\documentclass[sigconf]{acmart}

\copyrightyear{}
\acmYear{}
\acmDOI{}

\acmConference[MobiHoc'20]{MobiHoc'20: International Symposium on Theory, Algorithmic Foundations, and Protocol Design for Mobile Networks and Mobile Computing}{October 11--14, 2020}{Online}
\acmBooktitle{}
\acmPrice{}
\acmISBN{}

\usepackage{booktabs} 

\usepackage{amsmath,amssymb}
\usepackage{bbm}
\usepackage{bm}
\usepackage{subfigure}
\usepackage{algorithm,algpseudocode}
\floatname{algorithm}{Policy}

\newtheorem{theorem}{Theorem}[section]

\renewcommand*{\proofname}{Sketch of Proof}

\renewcommand\footnotetextcopyrightpermission[1]{}
\DeclareMathOperator*{\argmin}{\arg\min}

\begin{document}

\title{Learning to Price Vehicle Service with Unknown Demand}


\author{Haoran Yu}
\affiliation{
   \institution{Beijing Institute of Technology\\Beijing, China}
}
\email{yhrhawk@gmail.com}

\author{Ermin Wei}
\affiliation{
  \institution{Northwestern University\\Evanston, IL, USA}
}
\email{ermin.wei@northwestern.edu}

\author{Randall A. Berry}
\affiliation{
   \institution{Northwestern University\\Evanston, IL, USA}
}
\email{rberry@ece.northwestern.edu}



\begin{abstract}
It can be profitable for vehicle service providers to set service prices based on users' travel demand on different origin-destination pairs. 
The prior studies on the spatial pricing of vehicle service rely on the assumption that providers know users' demand. 
In this paper, we study a monopolistic provider who initially does not know users' demand and needs to learn it over time by observing the users' responses to the service prices. 
We design a pricing and vehicle supply policy, considering the tradeoff between \emph{exploration} (i.e., learning the demand) and \emph{exploitation} (i.e., maximizing the provider's short-term payoff). 
Considering that the provider needs to ensure the vehicle flow balance at each location, its pricing and supply decisions for different origin-destination pairs are tightly coupled. This makes it challenging to theoretically analyze the performance of our policy. 
We analyze the gap between the provider's expected time-average payoffs under our policy and a clairvoyant policy, which makes decisions based on complete information of the demand. 
We prove that after running our policy for $D$ days, the loss in the expected time-average payoff can be at most ${\mathcal O}\left({\left(\ln D\right)}^{\frac{1}{2}} D^{-\frac{1}{4}}\right)$, which decays to zero as $D$ approaches infinity.

\end{abstract}



\pagestyle{plain}

\maketitle

\section{Introduction}
Many vehicle service providers (e.g., taxi companies and ride-sharing platforms) charge users based on the users' origins and destinations as well as the travel distances \cite{Uberlocation}. 
This is because users' travel demand varies significantly across origin-destination pairs (hereafter referred to as links). 
A vehicle service provider needs to jointly optimize its service prices for different links. 
The reason is that a user takes the vehicle service if and only if it accepts the price and there is enough vehicle supply on the corresponding link. 
Since the provider should route vehicles across links, its vehicle supply and pricing decisions for different links are tightly coupled. 
This makes the pricing of vehicle service different from those of many other services (e.g., electric power service \cite{khezeli2017risk} and mobile data service \cite{ha2012tube}).

There have been some studies investigating providers' pricing and vehicle supply decisions \cite{bimpikis2016spatial,banerjee2015pricing,ma2018spatio}. 
A common assumption made in these studies is that for each link, the providers have complete information of users' aggregate demand as a function of the service price. 
In practice, the providers initially may not know the users' aggregate demand. Instead, since the users have similar demand patterns during the same time period of different days, the providers can learn the users' demand over days by testing different prices and observing the users' responses. Intuitively, the providers can test prices with a large variance to expedite the learning process. However, this may yield low payoffs to the providers in the short run. 
Therefore, the providers should carefully make their decisions to trade off the \emph{exploration} (i.e., learning the demand to improve the long-term decision making) and \emph{exploitation} (i.e., maximizing their short-term payoffs). 
As far as we know, none of the prior studies investigated this tradeoff in the pricing of vehicle service with unknown demand. This motivates our study in this work. 

\subsection{Our Work}\label{sec:ourwork}
We study a vehicle service provider's pricing and supply policy. On each day, the provider decides its service prices for all links, where the prices are measured in dollars per time slot.{\footnote{In practice, many providers charge users based on their travel distances instead of their travel times. Given the average vehicle velocity on a link, one can convert between the two measures of prices.}} Moreover, for each link, the provider decides the rate at which its vehicles depart from the origin to the destination.{\footnote{We assume that the provider has its own vehicle fleet and has full control over the supply. In our future work, we are interested in studying providers (e.g., ride-sharing platforms) who incentivize people to use private vehicles to offer service.}} The provider's vehicle supply decisions should ensure the \emph{vehicle flow balance}. In the system's steady state, the rate that the vehicles depart from a location (to other locations) should equal the rate that the vehicles arrive at this location. 
The vehicle flow balance constraint couples the provider's decisions for different links, and makes it challenging to design the pricing and supply policy. 

At the beginning of each day, the provider announces its prices. During the day, the aggregate demand on each link is realized and observed by the provider. The realized aggregate demand is a function of the price and a random demand shock. The demand shocks are different across links and days, and their values are not known by the provider. Our purpose is to design a policy that helps the provider estimate the parameters associated with the demand functions and achieve a high time-average payoff. 

To focus on the policy design with the unknown demand, we study a stationary model where the users' aggregate demand does not fluctuate during each considered time period. Similar stationary models have been considered in \cite{bimpikis2016spatial,lagos2000alternative,yu2019analyzing}. References \cite{banerjee2016pricing} and \cite{braverman2019empty} studied more sophisticated models, where the arrivals of user requests follow Poisson processes and the systems are modeled by closed-queueing networks. It is even more challenging to design and theoretically analyze learning and pricing policies for these models. As the first study in this direction, our work focuses on the stationary model, and our results may provide guidelines for the policy design in other more sophisticated models. 

We summarize our key contributions as follows.

{\bf I. Policy Design.} We design a pricing and supply policy that has different operations on odd and even days. On each odd day, the provider first estimates the parameters associated with the user demand functions. The estimation is based on the prices implemented on the prior days and the corresponding demand observed by the provider. Then, the provider makes the pricing and supply decisions to maximize its payoff \emph{as if} its estimation is correct. On each even day, the provider makes decisions by modifying its decisions on the last odd day. For example, it modifies its prices by adding \emph{offsets}. This induces a dispersion between the prices implemented on the current even day and the last odd day, which facilitates the provider's learning of the users' demand. The sizes of the offsets decay over days, and the provider can control the decay rate to balance the exploitation and exploration. 

{\bf II. Regret Analysis.} We compare our policy with a clairvoyant policy, where the provider is assumed to initially know the users' demand and makes decisions accordingly. We evaluate our policy by the \emph{time-average regret}, which is the difference between the provider's expected time-average payoffs under the clairvoyant policy and our policy. The theoretical analysis of the time-average regret in our problem is much more challenging than that in some prior work (e.g., \cite{khezeli2017risk,keskin2014dynamic}) which studied pricing services with unknown demand. The reason is that the vehicle flow balance considered in our problem complicates the provider's decision making and makes it difficult to derive closed forms for the pricing and supply decisions. To tackle the difficulty, we construct a resistor network given the traffic network (which is inspired by \cite{yu2019analyzing}). We leverage the notion of \emph{effective resistances} (defined based on the resistor network) to derive the closed forms for the provider's decisions. 
Then, we prove that our policy is a \emph{no-regret policy}, i.e., as time goes by, the provider's time-average payoff under our policy will converge to that under the clairvoyant policy.

\subsection{Related Work}\label{subsec:literature}

\subsubsection{Spatial Pricing of Vehicle Service}
There have been some studies analyzing providers' spatial pricing decisions, e.g., \cite{banerjee2015pricing,bimpikis2016spatial,banerjee2016pricing,ma2018spatio,yu2019analyzing}. 
Banerjee \emph{et al.} in \cite{banerjee2016pricing} used a continuous-time Markov chain to track the mass of vehicles at each location, and designed pricing policies with approximation guarantees. 
Bimpikis \emph{et al.} in \cite{bimpikis2016spatial} considered a stationary model with time-invariant user demand, and investigated the impacts of the network topology and demand pattern on the spatial pricing. 
Ma \emph{et al.} in \cite{ma2018spatio} studied a ride-sharing platform's problem of dispatching drivers and charging riders, considering the drivers' decisions of accepting the dispatching. 
In our prior work \cite{yu2019analyzing}, we analyzed the impact of location-based advertising on providers' spatial pricing, and investigated the providers' optimal collaboration with advertisers. 
None of the above studies considered the spatial pricing with unknown user demand, which is the focus in this work. 

\subsubsection{Pricing with Unknown Demand}
Our work is closely related to the stream of research that analyzes service providers' optimal pricing with unknown demand, e.g., \cite{besbes2009dynamic,broder2012dynamic,den2013simultaneously,keskin2014dynamic,khezeli2017risk}. Most of the related work assumed that users' demand functions belong to a parametric family and some parameters characterizing the functions are unknown. Service providers iterate between estimating the parameters and optimizing their prices based on the estimated models. In terms of the policy design, our work is most related to \cite{keskin2014dynamic} and \cite{khezeli2017risk}, where providers add offsets to prices to create price dispersions. As discussed before, it is difficult to derive closed forms for a vehicle service provider's decisions because of the vehicle flow balance. The theoretical analysis of our policy is more challenging than that in \cite{keskin2014dynamic} and \cite{khezeli2017risk}.

\subsubsection{Multi-Armed Bandit Problem}
Our work is also related to the studies on the multi-armed bandit problem, e.g., \cite{berry1985bandit,vermorel2005multi,kleinberg2005nearly,wang2018multi}. These studies also analyzed dynamic decision problems with uncertainty, and considered the exploitation-exploration tradeoff. Different from our work, these studies did not assume fixed parametric structures for objective functions. Moreover, most of them considered finite decision spaces for decision makers. Although a few studies considered infinite decision spaces, their solutions cannot be applied to our problem. For example, the solution in \cite{kleinberg2005nearly} requires a convex objective function, while the vehicle service provider's objective function in our problem is non-convex.{\footnote{As we will see in (\ref{equ:payoff}), maximizing the provider's expected payoff is a non-convex problem if we do not enforce the local supply-demand balance constraint. When using the solution in \cite{kleinberg2005nearly}, we cannot enforce this constraint. This is because enforcing the constraint requires the provider to estimate the demand model parameters, which is not included in the solution in \cite{kleinberg2005nearly}.}} 

\section{Model}
We consider a monopolistic provider offering vehicle service over multiple days, which are indexed by $d=1,\ldots,D$. In Section \ref{subsec:userdemand}, we model users' demand. In Section \ref{subsec:payoffpolicy}, we define the provider's decisions, payoff, and policies. In Section \ref{subsec:metric}, we introduce a metric for evaluating the provider's policies.

\subsection{Users' Demand}\label{subsec:userdemand}
We use ${\mathcal N}\triangleq \left\{1,\ldots,N\right\}$ to denote the set of locations, and assume that the time within each day $d$ is slotted. Let $p_{ij}^d$ denote the vehicle service price for link $\left(i,j\right)$ on day $d$, where $i\ne j$, $i,j\in{\mathcal N}$, and $p_{ij}^d$ is measured in dollars per time slot. If a user takes the vehicle service and travels from $i$ to $j$, its payment to the provider \emph{per time slot} is $p_{ij}^d$.

Given $p_{ij}^d$, we assume that the mass of users who want to travel from $i$ to $j$ via the vehicle service in each time slot during day $d$ is 
\begin{align}
\Psi_{ij}^d\left(p_{ij}^d,\epsilon_{ij}^d\right)=\alpha_{ij} - \beta_{ij} p_{ij}^d + \epsilon_{ij}^d, \forall i\ne j,i,j\in{\mathcal N}.\label{equ:demand}
\end{align}
Here, $\alpha_{ij}$ and $\beta_{ij}$ are the demand model parameters. We assume that $0<\alpha_{\min}\le \alpha_{ij} \le \alpha_{\max}$ and $0<\beta_{\min}\le \beta_{ij} \le \beta_{\max}$ for all $i\ne j,i,j\in{\mathcal N}$. \emph{The provider initially only knows $\alpha_{\min}$, $\alpha_{\max}$, $\beta_{\min}$, and $\beta_{\max}$, and needs to learn the values of $\alpha_{ij}$ and $\beta_{ij}$ over time.}

The random variable $\epsilon_{ij}^d\in\left[{\underline \epsilon},{\overline \epsilon}\right]$ captures the shock in the demand on day $d$. 
For each link $\left(i,j\right)$, we assume that $\left\{\epsilon_{ij}^d\right\}_{d=1,\ldots,D}$ is a set of independent and identically distributed random variables with a cumulative distribution function $F_{ij}\left(\cdot\right)$. We further assume that ${\mathbb E}\left\{\epsilon_{ij}^d\right\}=0$. The provider does not know $\epsilon_{ij}^d$, but knows $F_{ij}\left(\cdot\right)$.

We assume that there exists a maximum price $p_{\max}$ that the provider can charge, e.g., due to government regulations \cite{productivity2002regulation}. We further assume that $\alpha_{\min} - \beta_{\max} p_{\max} + {\underline \epsilon} \ge0$, which ensures the demand's non-negativity.

In (\ref{equ:demand}), the demand is linear in the price, and we consider an additive demand shock. References \cite{keskin2014dynamic} and \cite{khezeli2017risk} studied the pricing of products with unknown demand, and considered similar models. References \cite{bimpikis2016spatial} and \cite{yu2019analyzing} studied the pricing of vehicle service with \emph{known} demand, and also considered linear demand models. The linear demand model enables us to theoretically characterize the performance of our policy and shed light on the design of effective learning and pricing policies.{\footnote{A linear demand curve can be a reasonable approximation of some other demand curves. For example, using a linear curve to approximate the demand curve in \cite{fang2017prices} can achieve a small mean squared error.}} We can easily extend our policy to other demand models, e.g., the exponential demand model (note that the theoretical performance analysis will become even more challenging). 

An underlying assumption in (\ref{equ:demand}) is that the demand model parameters do not change within day $d$. This is to simplify the notations and presentation. In practice, users may have quite different demand patterns during different time periods (e.g., daytime and nighttime). We can easily generalize our model and solutions by considering different demand model parameters (e.g., $\alpha_{ij}^{\rm day}$, $\beta_{ij}^{\rm day}$, $\alpha_{ij}^{\rm night}$, and $\beta_{ij}^{\rm night}$) and pricing decisions (e.g., $p_{ij}^{d,{\rm day}}$ and $p_{ij}^{d,{\rm night}}$) for different time periods. For example, the provider can learn $\alpha_{ij}^{\rm day}$ and $\beta_{ij}^{\rm day}$ by choosing $\left\{p_{ij}^{d,{\rm day}}\right\}_{d=1,\ldots,D}$.

\subsection{Provider's Decisions, Payoff, and Policies}\label{subsec:payoffpolicy}
\subsubsection{Decisions}
At the beginning of each day $d$, the provider decides price $p_{ij}^d$ for each link $\left(i,j\right)$, and announces $p_{ij}^d$ to the users. Meanwhile, the provider decides the vehicle supply for each link. Specifically, we use $w_{ij}^d\ge0$ to denote the mass of vehicles departing from $i$ to $j$ ($i\ne j$) in each time slot during day $d$. Our work focuses on the system's steady state. Hence, when deciding $w_{ij}^d$, the provider should ensure the following \emph{vehicle flow balance} \cite{bimpikis2016spatial} \cite{lagos2000alternative}:
\begin{align}
\sum_{j\in{\mathcal N}\setminus\left\{i\right\}} w_{ij}^d = \sum_{j\in{\mathcal N}\setminus\left\{i\right\}} w_{ji}^d, \forall i\in{\mathcal N}.\label{equ:flowbalance}
\end{align}
For link $\left(i,j\right)$, $w_{ij}^d$ is the \emph{rate} that the vehicles depart $i$, and also equals the rate that the vehicles arrive at $j$. Considering all links, the vehicles' departure rate at $i$ is $\sum_{j\in{\mathcal N}\setminus\left\{i\right\}} w_{ij}^d$, and the arrival rate at $i$ is $\sum_{j\in{\mathcal N}\setminus\left\{i\right\}} w_{ji}^d$. Constraint (\ref{equ:flowbalance}) implies that these two rates should be equal.


During day $d$, the randomness in the demand (captured by $\epsilon_{ij}^d$) is realized. After observing the demand $\Psi_{ij}^d\left(p_{ij}^d,\epsilon_{ij}^d\right)$ for each link $\left(i,j\right)$, the provider can update its knowledge about $\alpha_{ij}$ and $\beta_{ij}$, and adjust its pricing and supply decisions on day $d+1$ (which will be discussed in later sections). 
In practice, the demand may fluctuate over time during a day, and we use the random variable $\epsilon_{ij}^d$ to approximate the average fluctuation.

\subsubsection{Payoff}
Next, we define the provider's time-average payoff on day $d$ in the system's steady state. We introduce some parameters. Let $\xi_{ij}>0$ denote the vehicle travel time on link $\left(i,j\right)$. It is defined as the number of time slots required for a vehicle to travel from $i$ to $j$. First, we assume that $\xi_{ij}$ is a fixed parameter and does not change with the users' demand for taking the provider's vehicle service (i.e., $\Psi_{ij}^d\left(p_{ij}^d,\epsilon_{ij}^d\right)$). When the provider increases $p_{ij}^d$, $\Psi_{ij}^d\left(p_{ij}^d,\epsilon_{ij}^d\right)$ will decrease, and some users will not take the provider's vehicle service. These users will travel to $j$ by other means (e.g., taking their own vehicles). Therefore, the impact of $\Psi_{ij}^d\left(p_{ij}^d,\epsilon_{ij}^d\right)$ on the traffic load and travel time on $\left(i,j\right)$ is negligible. Second, we assume that $\xi_{ij}$ does not change during a day. Similar to $\alpha_{ij}$ and $\beta_{ij}$, we can easily generalize the model by considering different travel times (e.g., $\xi_{ij}^{\rm day}$ and $\xi_{ij}^{\rm night}$) for different time periods. Third, we assume that $\xi_{ij}$ does not change \emph{over} days. This is to simplify the presentation, and our policy can be generalized to the day-variant travel time case. 

We use $c>0$ to denote the provider's cost of supplying a vehicle per time slot. The cost can include the provider's payment to the vehicle's driver and energy cost. Since the cost per time slot is normally independent of the vehicle's location, we consider a homogeneous cost $c$ for different links.{\footnote{Mathematically, it is easy to extend the model to the heterogeneous cost case.}} Our work focuses on the case where $c<p_{\max}$, i.e., the cost is smaller than the maximum price that the provider can charge. 

We use the function $\Pi\left({\bm p}^d,{\bm w}^d,{\bm \epsilon}^d\right)$ to denote the provider's payoff per time slot on day $d$ in the system's steady state. Here, we define ${\bm p}^d \triangleq \left(p_{ij}^d, \forall i\ne j, i,j\in{\mathcal N}\right)$ for $d=1,\ldots, D$, and ${\bm w}^d$ and ${\bm \epsilon}^d$ are defined similarly. Then, we define $\Pi\left({\bm p}^d,{\bm w}^d,{\bm \epsilon}^d\right)$ as follows:
\begin{align}
\nonumber
\Pi\left({\bm p}^d,{\bm w}^d,{\bm \epsilon}^d\right)\triangleq 
& \sum_{i\in{\mathcal N}} \sum_{j\in{\mathcal N}\setminus\left\{i\right\}} \xi_{ij} \min\left\{\Psi_{ij}^d\left(p_{ij}^d,\epsilon_{ij}^d\right),w_{ij}^d\right\} p_{ij}^d \\
& - \sum_{i\in{\mathcal N}} \sum_{j\in{\mathcal N}\setminus\left\{i\right\}} \xi_{ij} w_{ij}^d c.\label{equ:payoff}
\end{align}
The provider's payoff per time slot consists of two parts. The first part corresponds to the users' payments. Based on the definition of $\Psi_{ij}^d\left(p_{ij}^d,\epsilon_{ij}^d\right)$ in (\ref{equ:demand}), in each time slot, a continuum of users of mass $\Psi_{ij}^d\left(p_{ij}^d,\epsilon_{ij}^d\right)$ want to depart from $i$ to $j$ by taking the vehicle service. Given the provider's supply decision $w_{ij}^d$, the actual mass of users departing from $i$ to $j$ via the vehicle service per time slot is $\min\left\{\Psi_{ij}^d\left(p_{ij}^d,\epsilon_{ij}^d\right),w_{ij}^d\right\}$. Therefore, considering the travel time $\xi_{ij}$, the mass of users traveling on $\left(i,j\right)$ via the vehicle service (i.e., including the users traveling on the link but departing from $i$ in earlier slots) in any time slot is $\xi_{ij} \min\left\{\Psi_{ij}^d\left(p_{ij}^d,\epsilon_{ij}^d\right),w_{ij}^d\right\}$. Since the provider gets $p_{ij}^d$ by serving each of these users in this time slot, the first part on the right side of (\ref{equ:payoff}) captures the users' overall payment per time slot. The second part corresponds to the cost of supplying vehicles. In any time slot, the mass of vehicles traveling on $\left(i,j\right)$ is $\xi_{ij} w_{ij}^d$.{\footnote{When $w_{ij}^d> \Psi_{ij}^d\left(p_{ij}^d,\epsilon_{ij}^d\right)$, some vehicles traveling on $\left(i,j\right)$ are empty, i.e., do not carry users. The provider may intentionally route empty vehicles in the network to ensure the vehicle flow balance in (\ref{equ:flowbalance}).}} Therefore, the overall cost per time slot is $\sum_{i\in{\mathcal N}} \sum_{j\in{\mathcal N}\setminus\left\{i\right\}} \xi_{ij} w_{ij}^d c$.


\subsubsection{Policies}
At the beginning of day $d\ge2$, the provider knows the history of the realized demand and its decisions during the past $d-1$ days. We use $\left({{\bm \Psi}^1},{\bm p}^1,{\bm w}^1,\ldots,{{\bm \Psi}^{d-1}},{\bm p}^{d-1},{\bm w}^{d-1}\right)$ to denote this history, where ${{\bm \Psi}^1},\ldots,{{\bm \Psi}^{d-1}}$ represent the users' demand on all links during the first $d-1$ days. Note that the provider does not know the history of the demand shocks (i.e., ${\bm \epsilon}^1,\ldots,{\bm \epsilon}^{d-1}$).

We define a \emph{policy} ${\bm \pi}$ as a sequence of functions $\left(\pi^1,\ldots,\pi^D\right)$. Here, $\pi^1$ is a constant function, and $\pi^d$ ($d\ge2$) maps the vector $\left({{\bm \Psi}^1},{\bm p}^1,{\bm w}^1,\ldots,{{\bm \Psi}^{d-1}},{\bm p}^{d-1},{\bm w}^{d-1}\right)$ to the vector $\left({\bm p}^d,{\bm w}^d\right)$, i.e., it maps the history during the first $d-1$ days to the provider's decisions on day $d$. Note that $\left({\bm p}^d,{\bm w}^d\right)$ should satisfy $p_{ij}^d\le p_{\max}$ and $w_{ij}^d\ge0$ for all $\left(i,j\right)$, and ensure the vehicle flow balance in (\ref{equ:flowbalance}). The function $\pi^d$ ($d\ge2$) is assumed measurable with respect to the $\sigma$-algebra generated by $\left({{\bm \Psi}^1},{\bm p}^1,{\bm w}^1,\ldots,{{\bm \Psi}^{d-1}},{\bm p}^{d-1},{\bm w}^{d-1}\right)$.

Next, we define the provider's time-average payoff (i.e., its average payoff per time slot) during the first $D$ days. Recall that $\Pi\left({\bm p}^d,{\bm w}^d,{\bm \epsilon}^d\right)$ denotes the provider's time-average payoff on day $d$. Given a policy ${\bm \pi}$, the provider's expected time-average payoff during the first $D$ days is ${\mathbb E}^{\bm \pi}\left\{ \frac{1}{D} \sum_{d=1}^D \Pi\left({\bm p}^d,{\bm w}^d,{\bm \epsilon}^d\right) \right\}$, where the expectation is taken with respect to the random variables ${\bm \epsilon}^1,\ldots,{\bm \epsilon}^{D}$ and the (possible) randomness in the policy ${\bm \pi}$.


\subsection{Performance Metric}\label{subsec:metric}
Our target is to design policies for the provider, who initially does not know the demand model parameters. In order to evaluate the designed policies, we first assume that the provider knows the demand model parameters, and define a clairvoyant policy. Then, we will introduce a metric for evaluating the provider's policies based on the clairvoyant policy.

To facilitate the presentation, we define ${\bm \theta}_{ij}\triangleq \left(\alpha_{ij},\beta_{ij}\right)$ for each link $\left(i,j\right)$, and let ${\bm \theta}\triangleq \left({\bm \theta}_{ij}, \forall i\ne j, i,j\in{\mathcal N}\right)$. 
\subsubsection{Clairvoyant Policy}\label{subsubsec:clai}
When the provider knows ${\bm \theta}$, it does not need to adjust decisions over time to learn ${\bm \theta}$. Under the clairvoyant policy, the provider solves the following problem on each day $d$:
\begin{subequations}\label{problem:1}
\begin{align}
& \max {\mathbb E}_{{\bm \epsilon}^d} \left\{ \Pi\left({\bm p}^d,{\bm w}^d,{\bm \epsilon}^d\right) \right\} \label{equ:cla:obj} \\
& {\rm s.t.~~~} \sum_{j\in{\mathcal N}\setminus\left\{i\right\}} w_{ij}^d = \sum_{j\in{\mathcal N}\setminus\left\{i\right\}} w_{ji}^d, \forall i\in{\mathcal N},\label{equ:cla:con1} \\
& {~~}{~~}{~~}{~~}{~~}{~~}{~~}{~~}{~~}{~~} w_{ij}^d = {\mathbb E}_{\epsilon_{ij}^d} \left\{ \Psi_{ij}^d\left(p_{ij}^d,\epsilon_{ij}^d\right) \right\}, \forall i\ne j, i,j\in{\mathcal N},\label{equ:cla:con2} \\
& {\rm var.}  {~~}{~~}{~~}{~~} p_{ij}^d \le p_{\max}, w_{ij}^d\ge0, \forall i\ne j, i,j\in{\mathcal N}. \label{equ:cla:var}
\end{align}
\end{subequations}
As shown in (\ref{equ:cla:obj}), the provider makes the decisions to maximize its expected payoff per time slot on day $d$, where the expectation is taken with respect to ${\bm \epsilon}^d$. The constraint (\ref{equ:cla:con1}) ensures the \emph{vehicle flow balance}, as discussed in (\ref{equ:flowbalance}).{\footnote{Note that the travel time $\xi_{ij}$ does not appear in the flow balance constraint (\ref{equ:cla:con1}), since $\xi_{ij}$ does not affect the vehicles' departure rates and arrival rates.}} 
The constraint (\ref{equ:cla:con2}) captures the \emph{local supply-demand balance}, meaning that the provider chooses the vehicle supply (i.e., $w_{ij}^d$) to equal the users' expected demand (i.e., ${\mathbb E}_{\epsilon_{ij}^d} \left\{ \Psi_{ij}^d\left(p_{ij}^d,\epsilon_{ij}^d\right) \right\}$) on each link. This implies that we consider the vehicle service in a large city with thousands of links and the provider simply sets its supply to satisfy the local supply-demand balance, which simplifies its operation.{\footnote{In our future work, we plan to relax the constraint (\ref{equ:cla:con2}) and analyze the corresponding clairvoyant policy. In this case, the provider's operation is more complex. For example, even if the expected demand on $\left(i,j\right)$ is small, the provider may choose a large $w_{ij}^d$, which increases the mass of vehicles available at location $j$ and enables the provider to serve more users departing from $j$.}} Note that ${\bm \theta}$ (which includes the demand model parameters) appears in the expressions of both $\Pi\left({\bm p}^d,{\bm w}^d,{\bm \epsilon}^d\right)$ and $\Psi_{ij}^d\left(p_{ij}^d,\epsilon_{ij}^d\right)$. Therefore, the provider needs to know ${\bm \theta}$ to solve problem (\ref{problem:1}).

Recall that for each $\left(i,j\right)$, $\epsilon_{ij}^1,\ldots,\epsilon_{ij}^D$ are independent and identically distributed. As a result, the provider's optimal solutions of $\left({\bm p}^d,{\bm w}^d\right)$ to problem (\ref{problem:1}) for different $d$ are the same. We use $\left({\bm p}^*\left({\bm \theta}\right),{\bm w}^*\left({\bm \theta}\right)\right)$ to denote the optimal solution. We include ${\bm \theta}$ in the notation to indicate that the solution is derived based on the knowledge of ${\bm \theta}$. 
Under the clairvoyant policy, the provider's expected payoff per time slot on day $d$ is ${\mathbb E}_{{\bm \epsilon}^d} \left\{ \Pi\left({\bm p}^*\left({\bm \theta}\right),{\bm w}^*\left({\bm \theta}\right),{\bm \epsilon}^d\right) \right\}$.

\subsubsection{No-Regret Policies}
We intend to design policies for the provider who initially does not know ${\bm \theta}$ and achieve a time-average payoff that is close to ${\mathbb E}_{{\bm \epsilon}^d} \left\{ \Pi\left({\bm p}^*\left({\bm \theta}\right),{\bm w}^*\left({\bm \theta}\right),{\bm \epsilon}^d\right) \right\}$ in the long run.

We evaluate a policy ${\bm \pi}$ based on the \emph{time-average regret} during the first $D$ days, which is defined as follows:
\begin{align}
\!\Delta_D^{\bm \pi} \!\triangleq\! {\mathbb E}^{\bm \pi} \!\left\{ \frac{1}{D} \sum_{d=1}^D \!\Bigg( \Pi\left({\bm p}^*\left({\bm \theta}\right),{\bm w}^*\left({\bm \theta}\right),{\bm \epsilon}^d\right) \! -\! \Pi\left({\bm p}^d,{\bm w}^d,{\bm \epsilon}^d\right) \Bigg) \right\}. \label{equ:averegret}
\end{align}
The time-average regret $\Delta_D^{\bm \pi}$ captures the difference between the provider's expected time-average payoffs during the first $D$ days achieved under the clairvoyant policy and the policy ${\bm \pi}$. In (\ref{equ:averegret}), the expectation is taken with respect to ${\bm \epsilon}^1,\ldots,{\bm \epsilon}^{D}$ and the possible randomness in the policy ${\bm \pi}$. 

Our work focuses on designing \emph{no-regret policies}, which are defined as the policies with $\lim_{D\rightarrow \infty} \Delta_D^{\bm \pi}=0$, i.e., the time-average payoffs achieved under these policies converge to that achieved under the clairvoyant policy. We summarize the key notations (including those introduced in later sections) in Table \ref{table:notation}.

\begin{table}[t]
\caption{Key Notations.}\label{table:notation}
\vspace{-0.2cm}
\begin{tabular}{|p{2cm}p{5.6cm}|}
\hline
{$d=1,\ldots,D$} & {Index of days}\\
{$i,j\in {\mathcal N}$} & {Index of locations}\\
{$p_{ij}^d$} & {Provider's pricing decision for $\left(i,j\right)$ on day $d$}\\
{$w_{ij}^d$} & {Provider's supply decision for $\left(i,j\right)$ on day $d$}\\
{${\bm \pi}$} & {Provider's pricing and supply policy}\\
{${\bm \theta}_{ij}= \left(\alpha_{ij},\beta_{ij}\right)$} & {Demand model parameters for $\left(i,j\right)$}\\
{$\epsilon_{ij}^d\in\left[{\underline \epsilon},{\overline \epsilon}\right]$} & {Demand shock on $\left(i,j\right)$ on day $d$}\\
{$\Psi_{ij}^d\left(p_{ij}^d,\epsilon_{ij}^d\right)$} & {Demand (per time slot) on $\left(i,j\right)$ on day $d$}\\
{$\xi_{ij}$} & {Vehicle travel time on $\left(i,j\right)$}\\
{$c$} & {Cost of supplying a vehicle per time slot}\\
{$\Pi\left({\bm p}^d,{\bm w}^d,{\bm \epsilon}^d\right)$} & {Provider's payoff per time slot on day $d$}\\
{$\Delta_D^{\bm \pi}$} & {Time-average regret achieved by policy ${\bm \pi}$}\\
{${\hat{\bm \theta}}_{ij}^d= \left({\hat \alpha}_{ij}^d,{\hat \beta}_{ij}^d\right)$} & {Estimated demand model parameters for $\left(i,j\right)$ based on the history of the first $d$ days}\\
{$\rho,\eta$} & {Control parameters used in our policy}\\
\hline
\end{tabular}
\vspace{-0.5cm}
\end{table}


\section{Our Pricing and Supply Policy}
In this section, we introduce our \emph{No-Regret Pricing and Supply} (NRPS) policy. In Section \ref{subsec:policy:estimation}, we explain the method of estimating the demand model parameters. In Sections \ref{subsec:policy:procedure} and \ref{subsec:policy:complexity}, we show the procedure of our policy and discuss its complexity, respectively.

\subsection{Estimation of Demand Model Parameters}\label{subsec:policy:estimation}
At the beginning of day $d\ge2$, the information that the provider has includes the provider's decisions and the realized demand during the past $d-1$ days. Based on its past pricing decisions (i.e., ${{\bm p}^1},\ldots,{{\bm p}^{d-1}}$) and the corresponding realized demand (i.e., ${{\bm \Psi}^1},\ldots,{{\bm \Psi}^{d-1}}$), the provider can update its estimation of the demand model parameters (i.e., ${\bm \theta}$) and make the pricing and supply decisions for day $d$ accordingly. 

For each link $\left(i,j\right)$, we use ${\hat{\bm \theta}}_{ij}^{d-1}= \left({\hat \alpha}_{ij}^{d-1},{\hat \beta}_{ij}^{d-1}\right)$ to denote the provider's estimate of ${\bm \theta}_{ij}=\left(\alpha_{ij},\beta_{ij}\right)$ given the history of the first $d-1$ days ($d\ge2$). In our policy, the provider computes ${\hat{\bm \theta}}_{ij}^{d-1}$ based on the following \emph{least squares estimation}:
\begin{align}
{\tilde {\bm \theta}}_{ij}^{d-1} = \argmin_{\left({\bar \alpha}_{ij}, {\bar \beta}_{ij}\right)\in {\mathbb R}^2 } \sum_{\tau=1}^{d-1} \bigg(\Psi_{ij}^\tau\left(p_{ij}^\tau,\epsilon_{ij}^\tau\right) - \left( {\bar \alpha}_{ij} - {\bar \beta}_{ij} p_{ij}^\tau \right) \bigg)^2, \label{equ:least:1}
\end{align}
\begin{align}
\! {\hat {\bm \theta}}_{ij}^{d-1} \!\!=\!\! \Big(\!\max\!\left\{\min\left\{{\tilde \alpha}_{ij}^{d-1},\alpha_{\max} \right\},\alpha_{\min}\right\}\!, \!\max\!\left\{\min\left\{{\tilde \beta}_{ij}^{d-1},\beta_{\max} \right\},\beta_{\min}\right\}\!\Big). \label{equ:least:2}
\end{align}
In (\ref{equ:least:1}), the provider computes a vector $\left({\bar \alpha}_{ij}, {\bar \beta}_{ij}\right)$ that belongs to the set ${\mathbb R}^2$ and minimizes the sum of $\left(\Psi_{ij}^\tau\left(p_{ij}^\tau,\epsilon_{ij}^\tau\right) - \left( {\bar \alpha}_{ij} - {\bar \beta}_{ij} p_{ij}^\tau \right) \right)^2$ over $\tau=1,\ldots,d-1$. Here, $\Psi_{ij}^\tau\left(p_{ij}^\tau,\epsilon_{ij}^\tau\right)$ is the realized demand on link $\left(i,j\right)$ on day $\tau$, and ${\bar \alpha}_{ij} - {\bar \beta}_{ij} p_{ij}^\tau$ is the \emph{expected} demand on $\left(i,j\right)$ under $p_{ij}^\tau$ when the demand model parameters are ${\bar \alpha}_{ij}$ and ${\bar \beta}_{ij}$.{\footnote{According to (\ref{equ:demand}), when the demand model parameters are ${\bar \alpha}_{ij}$ and ${\bar \beta}_{ij}$, the demand on $\left(i,j\right)$ is ${\bar \alpha}_{ij} - {\bar \beta}_{ij} p_{ij}^\tau + \epsilon_{ij}^\tau$. We can compute the \emph{expected} demand using ${\mathbb E}\left\{\epsilon_{ij}^\tau\right\}=0$.}} We use ${\tilde {\bm \theta}}_{ij}^{d-1}$ to denote the solution vector $\left({\bar \alpha}_{ij}, {\bar \beta}_{ij}\right)$. 

Note that ${\tilde {\bm \theta}}_{ij}^{d-1}$ belongs to ${\mathbb R}^2$, while ${\bm \theta}_{ij}$ lies in the compact rectangle $\left[\alpha_{\min},\alpha_{\max}\right] \times \left[\beta_{\min},\beta_{\max}\right]$. Therefore, in (\ref{equ:least:2}), the provider projects ${\tilde {\bm \theta}}_{ij}^{d-1}$ onto the set $\left[\alpha_{\min},\alpha_{\max}\right] \times \left[\beta_{\min},\beta_{\max}\right]$ to get the estimate ${\hat{\bm \theta}}_{ij}^{d-1}$.

\subsection{Our NRPS Policy}\label{subsec:policy:procedure}
In Policy \ref{policy}, we show the complete procedure of our NRPS policy, which includes the parameter estimation introduced in Section \ref{subsec:policy:estimation}. 

\begin{algorithm}[t]
\begin{algorithmic}[1]
\caption{No-Regret Pricing and Supply (NRPS) Policy}\label{policy}
\State {{\textbf{Initialization}:}} For each link $\left(i,j\right)$, set ${\hat{\bm \theta}}_{ij}^0$ to be a vector randomly picked from set $\left[\alpha_{\min},\alpha_{\max}\right] \times \left[\beta_{\min},\beta_{\max}\right]$. Choose control parameters $\rho\in\left(0,\infty\right)$ and $\eta\in\left(0,\frac{1}{2}\right)$. 
\For {$d=1,\ldots, D$}
\If {$d$ is odd}
\State For each link $\left(i,j\right)$, compute ${\hat{\bm \theta}}_{ij}^{d-1}$ based on the least squares estimation method shown in equations (\ref{equ:least:1}) and (\ref{equ:least:2}). 
\State Solve problem (\ref{problem:2}) to obtain $\left({\bm p}^*\left({\hat{\bm \theta}}^{d-1}\right),{\bm w}^*\left({\hat{\bm \theta}}^{d-1}\right)\right)$.
\State For each link $\left(i,j\right)$, implement $p_{ij}^*\left({\hat{\bm \theta}}^{d-1}\right)$ as the pricing decision and $w_{ij}^*\left({\hat{\bm \theta}}^{d-1}\right)$ as the supply decision.
\Else
\State For each link $\left(i,j\right)$, implement $p_{ij}^*\left({\hat{\bm \theta}}^{d-2}\right)-\frac{\rho}{{\hat\beta}_{ij}^{d-2}}d^{-\eta}$ as the pricing decision and $w_{ij}^*\left({\hat{\bm \theta}}^{d-2}\right)+\rho d^{-\eta}$ as the supply decision.
\EndIf 
\EndFor
\end{algorithmic}
\end{algorithm}

\subsubsection{Initialization} 
In line 1, the provider chooses values for ${\hat{\bm \theta}}_{ij}^0$, $\rho$, and $\eta$. Let ${\hat{\bm \theta}}_{ij}^0$ denote the provider's estimate of ${\bm \theta}_{ij}$ \emph{without any history}. Recall that we assume that the provider initially does not have any prior knowledge of ${\bm \theta}_{ij}$ except the feasible region of ${\bm \theta}_{ij}$. Therefore, for each link $\left(i,j\right)$, the provider can choose ${\hat{\bm \theta}}_{ij}^0$ by randomly drawing a vector from $\left[\alpha_{\min},\alpha_{\max}\right] \times \left[\beta_{\min},\beta_{\max}\right]$ (according to an arbitrary distribution). 
We use $\rho\in\left(0,\infty\right)$ and $\eta\in\left(0,\frac{1}{2}\right)$ to denote two control parameters of our policy. As discussed later, the provider can tune $\rho$ and $\eta$ to improve the rate at which the time-average regret $\Delta_D^{\bm \pi}$ converges to zero. The concrete choices of $\rho$ and $\eta$ depend on the values of other parameters (e.g., ${\bm \theta}_{ij}$ and $c$). In Section \ref{sec:numerical}, we will numerically show the impacts of the control parameters on the policy's performance. 

\subsubsection{Operation on Odd Days} 
The provider's operation on each odd day is shown in lines 4-6 of Policy \ref{policy}. First, the provider computes ${\hat{\bm \theta}}_{ij}^{d-1}=\left({\hat \alpha}_{ij}^{d-1},{\hat \beta}_{ij}^{d-1}\right)$ (i.e., estimates demand model parameters) for each $\left(i,j\right)$ as described in Section \ref{subsec:policy:estimation}. Note that when $d=1$, ${\hat{\bm \theta}}_{ij}^{d-1}$ is simply ${\hat{\bm \theta}}_{ij}^0$, which has been chosen in the initialization. Second, the provider decides its pricing and supply based on the information of ${\hat{\bm \theta}}^{d-1} \triangleq \left({\hat{\bm \theta}}_{ij}^{d-1},\forall i\ne j, i,j\in {\mathcal N} \right)$. The provider's decision problem is formulated as follows:
\begin{subequations}\label{problem:2}
\begin{align}
\nonumber
& \max \sum_{i\in{\mathcal N}} \sum_{j\in{\mathcal N}\setminus\left\{i\right\}} \xi_{ij} {\mathbb E}_{\epsilon_{ij}^d} \left\{\min\left\{{\hat \alpha}_{ij}^{d-1} - {\hat \beta}_{ij}^{d-1} p_{ij}^d + \epsilon_{ij}^d ,w_{ij}^d\right\}\right\} p_{ij}^d  \\
& {~~}{~~}{~~}{~~}{~~}{~~}{~~}{~~} - \sum_{i\in{\mathcal N}} \sum_{j\in{\mathcal N}\setminus\left\{i\right\}} \xi_{ij} w_{ij}^d c \label{equ:our:obj}\\
& {\rm s.t.~~~} \sum_{j\in{\mathcal N}\setminus\left\{i\right\}} w_{ij}^d = \sum_{j\in{\mathcal N}\setminus\left\{i\right\}} w_{ji}^d, \forall i\in{\mathcal N}, \label{equ:our:con1}\\
& {~~}{~~}{~~}{~~}{~~}{~~}{~~}{~~}{~~}{~~} w_{ij}^d = {\hat \alpha}_{ij}^{d-1} - {\hat \beta}_{ij}^{d-1} p_{ij}^d, \forall i\ne j, i,j\in{\mathcal N}, \label{equ:our:con2}\\
& {\rm var.}  {~~}{~~}{~~}{~~} p_{ij}^d \le p_{\max}, w_{ij}^d\ge0, \forall i\ne j, i,j\in{\mathcal N}.\label{equ:our:var}
\end{align}
\end{subequations}
We get problem (\ref{problem:2}) by replacing $\alpha_{ij}$ and $\beta_{ij}$ in problem (\ref{problem:1}) with ${\hat \alpha}_{ij}^{d-1}$ and ${\hat \beta}_{ij}^{d-1}$, respectively. Specifically, both $\alpha_{ij}$ and $\beta_{ij}$ appear in two places of problem (\ref{problem:1}). First, they appear in the objective (\ref{equ:cla:obj}). According to (\ref{equ:demand}) and (\ref{equ:payoff}), $\alpha_{ij}$ and $\beta_{ij}$ affect the expression of $\Pi\left({\bm p}^d,{\bm w}^d,{\bm \epsilon}^d\right)$ in (\ref{equ:cla:obj}). We can replace $\alpha_{ij}$ and $\beta_{ij}$ in (\ref{equ:demand}) with ${\hat \alpha}_{ij}^{d-1}$ and ${\hat \beta}_{ij}^{d-1}$, plug the result into (\ref{equ:payoff}), and take an expectation with respect to ${\bm \epsilon}^d$. This leads to a new objective, i.e., (\ref{equ:our:obj}). Second, $\alpha_{ij}$ and $\beta_{ij}$ appear in the constraint (\ref{equ:cla:con2}), because they affect the expression of $\Psi_{ij}^d\left(p_{ij}^d,\epsilon_{ij}^d\right)$ in (\ref{equ:demand}). We can replace $\alpha_{ij}$ and $\beta_{ij}$ in (\ref{equ:demand}) with ${\hat \alpha}_{ij}^{d-1}$ and ${\hat \beta}_{ij}^{d-1}$, plug the result into (\ref{equ:cla:con2}), and utilize ${\mathbb E}\left\{\epsilon_{ij}^d\right\}=0$ to get a new constraint, i.e., (\ref{equ:our:con2}). In problem (\ref{problem:2}), (\ref{equ:our:con1}) and (\ref{equ:our:var}) are the same as (\ref{equ:cla:con1}) and (\ref{equ:cla:var}), respectively. 

Next, we explain the intuition behind the formulation of problem (\ref{problem:2}). Since we target to achieve a performance that is close to that under the clairvoyant policy, we formulate problem (\ref{problem:2}) to be analogous to problem (\ref{problem:1}). Because the provider only has the information of ${\hat{\bm \theta}}^{d-1}$, we get problem (\ref{problem:2}) by replacing all $\alpha_{ij}$ and $\beta_{ij}$ in problem (\ref{problem:1}) with ${\hat \alpha}_{ij}^{d-1}$ and ${\hat \beta}_{ij}^{d-1}$, respectively. We use $\left({\bm p}^*\left({\hat{\bm \theta}}^{d-1}\right),{\bm w}^*\left({\hat{\bm \theta}}^{d-1}\right)\right)$ to denote the optimal solution to problem (\ref{problem:2}). Here, we include ${\hat{\bm \theta}}^{d-1}$ in the notation to indicate that the solution is derived based on ${\hat{\bm \theta}}^{d-1}$. On an odd day $d$, the provider implements $\left({\bm p}^*\left({\hat{\bm \theta}}^{d-1}\right),{\bm w}^*\left({\hat{\bm \theta}}^{d-1}\right)\right)$ as its pricing and supply. 


\subsubsection{Operation on Even Days}\label{subsubsec:operationeven} 
The provider's operation on each even day is shown in line 8 of Policy \ref{policy}. Different from the operation on each odd day, the provider does not update its estimate of demand model parameters or solve an optimization problem on each even day. Instead, the provider decides its pricing and supply by modifying its decisions on the last odd day. Specifically, for an even day $d$, the decisions made on the last odd day (i.e., day $d-1$) are captured by $\left({\bm p}^*\left({\hat{\bm \theta}}^{d-2}\right),{\bm w}^*\left({\hat{\bm \theta}}^{d-2}\right)\right)$ (according to line 5). On the even day $d$, the provider implements $p_{ij}^*\left({\hat{\bm \theta}}^{d-2}\right)-\frac{\rho}{{\hat\beta}_{ij}^{d-2}}d^{-\eta}$ as its pricing and $w_{ij}^*\left({\hat{\bm \theta}}^{d-2}\right)+\rho d^{-\eta}$ as its supply for each link $\left(i,j\right)$. Recall that $\rho\in\left(0,\infty\right)$ and $\eta\in\left(0,\frac{1}{2}\right)$ are the control parameters chosen in the initialization phase, and ${\hat\beta}_{ij}^{d-2}$ is the provider's estimate of $\beta_{ij}$ given the history of the first $d-2$ days. Based on the feasibility of $\left({\bm p}^*\left({\hat{\bm \theta}}^{d-2}\right),{\bm w}^*\left({\hat{\bm \theta}}^{d-2}\right)\right)$, we can verify that the provider's decisions on each even day are feasible and the supply decisions ensure the flow balance (we leave the proof to Appendix \ref{app:sec:feas}).


Next, we explain the intuition behind the design. When setting the price for $\left(i,j\right)$, the provider adds an \emph{offset} (i.e., $-\frac{\rho}{{\hat\beta}_{ij}^{d-2}}d^{-\eta}$) to $p_{ij}^*\left({\hat{\bm \theta}}^{d-2}\right)$. This induces a dispersion between the prices implemented on the odd and even days, which facilitates the provider's learning of ${\bm \theta}_{ij}$. The size of the offset is affected by the control parameters $\rho$ and $\eta$, and decays to zero as $d$ approaches infinity. When $\eta$ is large, the offset decays at a high rate, which may lead to a slow learning of ${\bm \theta}_{ij}$. When $\eta$ is small, the offset decays at a low rate. As a result, the provider implements ``non-optimal'' prices on many even days, which may reduce the provider's expected time-average payoff. Therefore, the provider should tune the control parameters to achieve a good balance between the exploration (i.e., learning ${\bm \theta}_{ij}$) and exploitation (i.e., maximizing the payoff). We will show the impacts of the control parameters in Section \ref{sec:numerical}. When setting the supply for $\left(i,j\right)$, the provider increases the supply from $w_{ij}^*\left({\hat{\bm \theta}}^{d-2}\right)$ to $w_{ij}^*\left({\hat{\bm \theta}}^{d-2}\right)+\rho d^{-\eta}$. This is to accommodate the change in the demand caused by the offset to $p_{ij}^*\left({\hat{\bm \theta}}^{d-2}\right)$. 


In contrast with our NRPS policy, one can design a myopic pricing and supply policy, where the provider updates its estimate of ${\bm \theta}$ and solves problem (\ref{problem:2}) on each day $d$ (without adding offsets to the prices). In Section \ref{sec:numerical}, we will numerically show that the myopic policy can lead to an incomplete learning of ${\bm \theta}$ and achieve a worse performance than our policy.

\subsection{Complexity of Our Policy}\label{subsec:policy:complexity}
When implementing our policy, the provider computes ${\hat{\bm \theta}}^{d-1}$ and $\left({\bm p}^*\left({\hat{\bm \theta}}^{d-1}\right),{\bm w}^*\left({\hat{\bm \theta}}^{d-1}\right)\right)$ on each odd day $d$. First, computing ${\hat{\bm \theta}}^{d-1}$ mainly requires the provider to solve (\ref{equ:least:1}), which is a linear regression problem. In Appendix \ref{app:sec:least}, we show that solving (\ref{equ:least:1}) is simple, as it mainly includes a multiplication between a $2 \times 2$ matrix and a $2 \times 1$ vector. 
Second, computing $\left({\bm p}^*\left({\hat{\bm \theta}}^{d-1}\right),{\bm w}^*\left({\hat{\bm \theta}}^{d-1}\right)\right)$ requires the provider to solve problem (\ref{problem:2}). We can utilize constraint (\ref{equ:our:con2}) to transform problem (\ref{problem:2}) to a simpler form. Specifically, we can replace $w_{ij}^d$ in (\ref{problem:2}) with ${\hat \alpha}_{ij}^{d-1} - {\hat \beta}_{ij}^{d-1} p_{ij}^d$. Recall that $F_{ij}\left(\cdot\right)$ is the cumulative distribution function of $\epsilon_{ij}^d$. We define $\epsilon_{ij}^-$ as a non-positive parameter that equals $\int_{\underline \epsilon}^0 \epsilon_{ij}^d d F_{ij}\left(\epsilon_{ij}^d\right)$, and simplify the term ${\mathbb E}_{\epsilon_{ij}^d} \left\{\min\left\{{\hat \alpha}_{ij}^{d-1} - {\hat \beta}_{ij}^{d-1} p_{ij}^d + \epsilon_{ij}^d ,w_{ij}^d\right\}\right\}$ in (\ref{equ:our:obj}) as ${\hat \alpha}_{ij}^{d-1} - {\hat \beta}_{ij}^{d-1} p_{ij}^d + \epsilon_{ij}^-$. 
Then, we can transform problem (\ref{problem:2}) to the following problem:
\begin{subequations}\label{problem:2trans}
\begin{align}
\nonumber
& \max \sum_{i\in{\mathcal N}} \sum_{j\in{\mathcal N}\setminus\left\{i\right\}} \xi_{ij}  \left({\hat \alpha}_{ij}^{d-1} - {\hat \beta}_{ij}^{d-1} p_{ij}^d + \epsilon_{ij}^-\right) p_{ij}^d   \\
& {~~}{~~}{~~}{~~}{~~}{~~}{~~}{~~} - \sum_{i\in{\mathcal N}} \sum_{j\in{\mathcal N}\setminus\left\{i\right\}} \xi_{ij} \left({\hat \alpha}_{ij}^{d-1} - {\hat \beta}_{ij}^{d-1} p_{ij}^d\right) c \label{equ:opt:ourtrans:1}\\
& {\rm s.t.}\! \sum_{j\in{\mathcal N}\setminus\left\{i\right\}} \!\!\!\!\!\left({\hat \alpha}_{ij}^{d-1} -\! {\hat \beta}_{ij}^{d-1} p_{ij}^d \right) =\!\!\!\!\!\! \sum_{j\in{\mathcal N}\setminus\left\{i\right\}} \!\!\!\!\! \left({\hat \alpha}_{ji}^{d-1} -\! {\hat \beta}_{ji}^{d-1} p_{ji}^d\right), \!\forall i\in{\mathcal N}, \label{equ:opt:ourtrans:2}\\
& {\rm var.}  {~~}{~~} p_{ij}^d \le p_{\max}, \forall i\ne j, i,j\in{\mathcal N}. \label{equ:opt:ourtrans:3}
\end{align}
\end{subequations}
Since problem (\ref{problem:2trans}) has a quadratic and concave objective function and affine constraints, it is a convex problem. The provider can solve (\ref{problem:2trans}) by interior-point methods, and use $w_{ij}^d={\hat \alpha}_{ij}^{d-1} - {\hat \beta}_{ij}^{d-1} p_{ij}^d$ to determine the supply for each link. Recall that we assume that $\alpha_{\min} - \beta_{\max} p_{\max} + {\underline \epsilon} \ge0$. This ensures the non-negativity of the determined supply. 

\section{Performance of Our Policy}
In this section, we analyze the time-average regret $\Delta_D^{\bm \pi}$ achieved by our NRPS policy. 
In Section \ref{subsec:performance:est}, we analyze the error of the provider's estimation of ${\bm \theta}$. In Section \ref{subsec:performance:dif}, we discuss the main difficulty of analyzing the $\Delta_D^{\bm \pi}$ achieved by our policy. We propose a solution to tackle the difficulty in Section \ref{subsec:performance:res}, and characterize an upper bound on $\Delta_D^{\bm \pi}$ in Section \ref{subsec:performance:bou}. 

\subsection{Upper Bound on Squared Estimation Error}\label{subsec:performance:est}
Suppose that the provider implements our NRPS policy. Next, we show that the provider can gradually achieve an accurate estimation of the demand model parameters. 
At the beginning of each odd day $d$, the provider estimates ${\bm \theta}_{ij}$ for each $\left(i,j\right)$ based on the history of the first $d-1$ days, and the estimate is denoted by ${\hat{\bm \theta}}_{ij}^{d-1}$. Then, we can use ${\mathbb E}\left\{ || {\hat{\bm \theta}}_{ij}^{d-1} - {\bm \theta}_{ij} ||_2^2 \right\}$ to characterize the \emph{mean squared error} of the provider's estimate. Here, the expectation is taken with respect to ${\bm \epsilon}^1,\ldots,{\bm \epsilon}^{d-1}$ and the randomness in our NRPS policy (e.g., in the random setting of ${\hat{\bm \theta}}_{ij}^0$). 
In the following theorem, we characterize an upper bound on ${\mathbb E}\left\{ || {\hat{\bm \theta}}_{ij}^{d-1} - {\bm \theta}_{ij} ||_2^2 \right\}$.

\begin{theorem}\label{theorem:estimate}
Under the NRPS policy, there exists a function $\Phi_1\left(\rho,\eta\right)$ such that (i) it is finite and positive for all $\rho\in\left(0,\infty\right)$ and $\eta\in\left(0,\frac{1}{2}\right)$; and (ii) the following relation holds for all $d\ge5$ and all $\left(i,j\right)$:
\begin{align}
{\mathbb E}\left\{ || {\hat{\bm \theta}}_{ij}^{d-1} - {\bm \theta}_{ij} ||_2^2 \right\} < \Phi_1\left(\rho,\eta\right) \frac{\ln \left(d-1\right)}{\left(d-1\right)^{1-2\eta}}.\label{equ:bound:est}
\end{align}
For all $\left(i,j\right)$, ${\mathbb E}\left\{ || {\hat{\bm \theta}}_{ij}^{d-1} - {\bm \theta}_{ij} ||_2^2 \right\}$ approaches zero as $d$ goes to infinity. 
\end{theorem}
The concrete expression of $\Phi_1\left(\rho,\eta\right)$ is complicated and can be found in Appendix \ref{app:sec:proof:the1} (all the proofs of the results in the paper can also be found in our appendices). Theorem \ref{theorem:estimate} implies that implementing the NRPS policy can help the provider accurately estimate ${\bm \theta}$ as $d$ goes to infinity. As $d$ increases, we can see that under a large $\eta$, the rate at which the right side of (\ref{equ:bound:est}) converges to zero becomes low. This implies that increasing $\eta$ can reduce the rate of learning ${\bm \theta}$.

\begin{figure*}[t]
  \centering
  \includegraphics[scale=0.45]{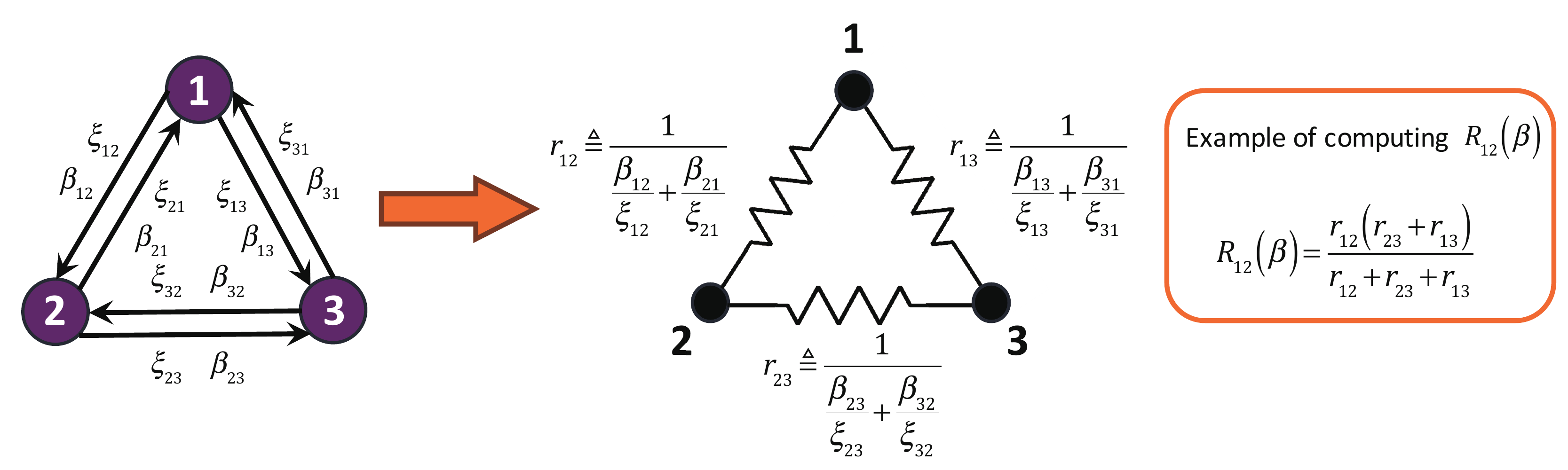}\\
  \vspace{-0.3cm}
  \caption{An Example of Constructing A Resistor Network for A Traffic Network (When $N=3$).}
  \label{fig:resistance}
  \vspace{-0.3cm}
\end{figure*}

\subsection{Difficulty of Regret Analysis}\label{subsec:performance:dif}
In this section, we discuss the difficulty of proving that our NRPS policy is a no-regret policy. To show that $\lim_{D\rightarrow \infty} \Delta_D^{\bm \pi}=0$ under the NRPS policy, we plan to first characterize an upper bound on $\Delta_D^{\bm \pi}$, and then prove that the upper bound converges to zero as $D$ goes to infinity. According to the definition of $\Delta_D^{\bm \pi}$ in (\ref{equ:averegret}), the key step of characterizing an upper bound on $\Delta_D^{\bm \pi}$ is to bound ${\mathbb E}^{\bm \pi}\left\{\Pi\left({\bm p}^*\left({\bm \theta}\right),{\bm w}^*\left({\bm \theta}\right),{\bm \epsilon}^d\right)-\Pi\left({\bm p}^d,{\bm w}^d,{\bm \epsilon}^d\right)\right\}$ for $d=1,\ldots,D$. Recall that $\left({\bm p}^*\left({\bm \theta}\right),{\bm w}^*\left({\bm \theta}\right)\right)$ is determined under the clairvoyant policy, and is the solution to problem (\ref{problem:1}). Under our NRPS policy, $\left({\bm p}^d,{\bm w}^d\right)$ is determined based on Policy \ref{policy}.  Next, we show that it is challenging to derive \emph{closed-form expressions} for both $\left({\bm p}^*\left({\bm \theta}\right),{\bm w}^*\left({\bm \theta}\right)\right)$ and $\left({\bm p}^d,{\bm w}^d\right)$.

Solving problem (\ref{problem:1}) gives $\left({\bm p}^*\left({\bm \theta}\right),{\bm w}^*\left({\bm \theta}\right)\right)$. Similar to the transformation from problem (\ref{problem:2}) to problem (\ref{problem:2trans}) (discussed in Section \ref{subsec:policy:complexity}), we can utilize constraint (\ref{equ:cla:con2}) to transform problem (\ref{problem:1}) to a simpler form. Based on (\ref{equ:cla:con2}), we have $w_{ij}^d = {\mathbb E}_{\epsilon_{ij}^d} \left\{ \Psi_{ij}^d\left(p_{ij}^d,\epsilon_{ij}^d\right) \right\}=\alpha_{ij} - \beta_{ij} p_{ij}^d$ for all $\left(i,j\right)$. Then, we can utilize this relation to transform problem (\ref{problem:1}) to the following problem:
\begin{subequations}\label{problem:1trans}
\begin{align}
\nonumber
& \max \sum_{i\in{\mathcal N}} \sum_{j\in{\mathcal N}\setminus\left\{i\right\}} \xi_{ij} \left(\alpha_{ij} - \beta_{ij} p_{ij}^d + \epsilon_{ij}^-\right) p_{ij}^d \\
& {~~}{~~}{~~}{~~}{~~}{~~}{~~} - \sum_{i\in{\mathcal N}} \sum_{j\in{\mathcal N}\setminus\left\{i\right\}} \xi_{ij} \left(\alpha_{ij} - \beta_{ij} p_{ij}^d\right) c  \\
& {\rm s.t.} \!\sum_{j\in{\mathcal N}\setminus\left\{i\right\}} \left(\alpha_{ij} - \beta_{ij} p_{ij}^d\right) \!=\!\!\!\! \sum_{j\in{\mathcal N}\setminus\left\{i\right\}} \left(\alpha_{ji} - \beta_{ji} p_{ji}^d\right), \forall i\in{\mathcal N}, \label{equ:obj:1trans:con}\\
& {\rm var.}  {~~} p_{ij}^d \le p_{\max}, \forall i\ne j, i,j\in{\mathcal N}. \label{equ:obj:1trans:var}
\end{align}
\end{subequations}
Recall that $\epsilon_{ij}^-$ is defined in Section \ref{subsec:policy:complexity} as $\int_{\underline \epsilon}^0 \epsilon_{ij}^d d F_{ij}\left(\epsilon_{ij}^d\right)$. In fact, problem (\ref{problem:1trans}) is similar to problem (\ref{problem:2trans}), except that problem (\ref{problem:1trans}) is formulated based on the actual demand model parameters (i.e., $\alpha_{ij}$ and $\beta_{ij}$). From (\ref{problem:1trans}), we can see that it is challenging to derive a closed-form expression for ${\bm p}^*\left({\bm \theta}\right)$. Given ${\bm p}^*\left({\bm \theta}\right)$, we can compute ${\bm w}^*\left({\bm \theta}\right)$ using $w_{ij}^*\left({\bm \theta}\right) = \alpha_{ij} - \beta_{ij} p_{ij}^*\left({\bm \theta}\right)$. Therefore, it is also hard to derive a closed-form expression for ${\bm w}^*\left({\bm \theta}\right)$.


Under the NRPS policy, when $d$ is odd, $\left({\bm p}^d,{\bm w}^d\right)$ is the solution to problem (\ref{problem:2}), denoted by $\left({\bm p}^*\left({\hat{\bm \theta}}^{d-1}\right),{\bm w}^*\left({\hat{\bm \theta}}^{d-1}\right)\right)$. 
Based on our discussion in Section \ref{subsec:policy:complexity}, we can transform problem (\ref{problem:2}) to problem (\ref{problem:2trans}), and solve problem (\ref{problem:2trans}) to get ${\bm p}^*\left({\hat{\bm \theta}}^{d-1}\right)$. From (\ref{problem:2trans}), we can see that it is also challenging to derive a closed-form expression for ${\bm p}^*\left({\hat{\bm \theta}}^{d-1}\right)$. Since ${\bm w}^*\left({\hat{\bm \theta}}^{d-1}\right)$ satisfies $w_{ij}^*\left({\hat{\bm \theta}}^{d-1}\right)={\hat \alpha}_{ij}^{d-1} - {\hat \beta}_{ij}^{d-1} p_{ij}^*\left({\hat{\bm \theta}}^{d-1}\right)$, it is hard to get a closed-form expression for ${\bm w}^*\left({\hat{\bm \theta}}^{d-1}\right)$. 
Under the NRPS policy, when $d$ is even, $\left({\bm p}^d,{\bm w}^d\right)$ is gotten by modifying the decisions on the last odd day. As a result, it is also hard to get a closed-form expression for $\left({\bm p}^d,{\bm w}^d\right)$ for an even $d$.


When we cannot use ${\bm \theta}$, ${\hat{\bm \theta}}^{d-1}$, and other parameters to represent $\left({\bm p}^*\left({\bm \theta}\right),{\bm w}^*\left({\bm \theta}\right)\right)$ and $\left({\bm p}^d,{\bm w}^d\right)$ in closed forms, it is difficult to utilize the bound on ${\mathbb E}\left\{ || {\hat{\bm \theta}}_{ij}^{d-1} - {\bm \theta}_{ij} ||_2^2 \right\}$ in Theorem \ref{theorem:estimate} to bound ${\mathbb E}^{\bm \pi}\left\{\Pi\left({\bm p}^*\left({\bm \theta}\right),{\bm w}^*\left({\bm \theta}\right),{\bm \epsilon}^d\right)-\Pi\left({\bm p}^d,{\bm w}^d,{\bm \epsilon}^d\right)\right\}$. 
Note that this difficulty does not exist in some earlier work that studied no-regret pricing policies \cite{khezeli2017risk,keskin2014dynamic}. For example, the service provider in \cite{khezeli2017risk} essentially sells a single item, and its decision under the pricing policy can be easily written in a closed form using the estimated demand model parameters. 
In our problem, the vehicle service provider makes the pricing and supply decisions for multiple links, and these decisions are coupled through the vehicle flow balance constraint. This makes it difficult to derive closed forms for the provider's decisions and further characterize a bound on $\Delta_D^{\bm \pi}$.{\footnote{We can see that without the flow balance constraints in problems (\ref{problem:1trans}) and (\ref{problem:2trans}) (i.e., constraints (\ref{equ:obj:1trans:con}) and (\ref{equ:opt:ourtrans:2})), one can easily derive the closed forms of the optimal solutions.}} 

\subsection{Effective Resistance-Based Solution}\label{subsec:performance:res}
In this section, we tackle the difficulty discussed in Section \ref{subsec:performance:dif}. The key idea is that we can construct a resistor network given the traffic network, and then utilize the notion of effective resistances to derive the closed forms of the provider's decisions. 

The idea is inspired by our prior work \cite{yu2019analyzing}, which studied a vehicle service provider's pricing in a \emph{complete information} setting with location-based advertising. Although the problem in \cite{yu2019analyzing} is quite different from the problem in this paper (as discussed in Section \ref{subsec:literature}), the problem in \cite{yu2019analyzing} also requires deriving closed forms for the optimal prices. Therefore, we use the technique proposed in \cite{yu2019analyzing} (with proper modification) to tackle the difficulty here. 

\subsubsection{Resistor Network and Effective Resistances}\label{subsub:ressolution:a}
Next, we focus on deriving the expression for ${\bm p}^*\left({\bm \theta}\right)$, which is the optimal solution to problem (\ref{problem:1trans}). First, we construct a resistor network based on the traffic network (an example is illustrated in Fig. \ref{fig:resistance}). We can replace the locations in the traffic network with nodes, and the two links between each pair of locations with a resistor. Specifically, for all $i,j\in{\mathcal N}$ with $i< j$, we replace the links $\left(i,j\right)$ and $\left(j,i\right)$ with a resistor, use $r_{ij}$ to denote its resistance, and let $r_{ij}\triangleq \frac{1}{\frac{\beta_{ij}}{\xi_{ij}}+\frac{\beta_{ji}}{\xi_{ji}}}$. 
Recall that $\beta_{ij}$, $\beta_{ji}$, $\xi_{ij}$, and $\xi_{ji}$ are defined under the traffic network (e.g., $\xi_{ij}$ is the vehicle travel time, and $\beta_{ij}$ is related to the slope of the demand curve for $\left(i,j\right)$).

In a resistor network, the \emph{effective resistance} between any two nodes $i$ and $j$ is defined as the voltage between $i$ and $j$ if a unit current is injected at $i$ and extracted from $j$ \cite{dorfler2018electrical}. We use $R_{ij}\left({\bm \beta}\right)$ to denote the effective resistance between nodes $i$ and $j$ in our constructed resistor network. 
Here, we include ${\bm \beta}\triangleq \left(\beta_{ij},\forall i\ne j,i,j\in{\mathcal N}\right)$ in the notation to indicate the dependence of $R_{ij}\left({\bm \beta}\right)$ on $\bm \beta$ and differentiate $R_{ij}\left({\bm \beta}\right)$ from $R_{ij}\left({\hat {\bm \beta}}^{d-1}\right)$, which will be introduced later. 
Note that we have $R_{ii}\left({\bm \beta}\right)=0$ and $R_{ij}\left({\bm \beta}\right)=R_{ji}\left({\bm \beta}\right)$ for all $i,j\in{\mathcal N}$ \cite{klein1993resistance}. 
In the example in Fig. \ref{fig:resistance}, we show the computation of $R_{12}\left({\bm \beta}\right)$. Readers can refer to \cite{klein1993resistance} for more properties of effective resistances.

\subsubsection{Closed Forms of $\left({\bm p}^*\left({\bm \theta}\right),{\bm w}^*\left({\bm \theta}\right)\right)$}\label{subsub:ressolution:b}
In the following proposition, we characterize ${\bm p}^*\left({\bm \theta}\right)$ (i.e., the solution to problem (\ref{problem:1trans})) using the effective resistances. Here, we use $\mu_{ij}^*$ to denote the optimal dual variable corresponding to the constraint $p_{ij}^d \le p_{\max}$ (i.e., (\ref{equ:obj:1trans:var})) for each $\left(i,j\right)$. 

\begin{proposition}\label{proposition:price}
When $\mu_{ij}^*=0$ for all $\left(i,j\right)$, we have
\begin{align}
p_{ij}^*\left({\bm \theta}\right)\!=\!\frac{c\beta_{ij} \!+\! \alpha_{ij} \!+\! \epsilon_{ij}^-}{2 \beta_{ij}} \!+\! \frac{1}{4\xi_{ij}} \!\sum_{k\in{\mathcal N}} \!\left(R_{jk}\left({\bm \beta}\right) \!-\! R_{ik}\left({\bm \beta}\right) \right) \!v_k\left({\bm \theta}\right),\label{equ:proposition:p}
\end{align}
where $v_k\left({\bm \theta}\right)$ is defined as follows:
\begin{align}
v_k\left({\bm \theta}\right)\triangleq \!\!\!\!\sum_{j\in{\mathcal N}\setminus\left\{k\right\}} \left(\alpha_{kj} - c\beta_{kj} - \epsilon_{kj}^-\right) - \!\!\! \sum_{j\in{\mathcal N}\setminus\left\{k\right\}}\left(\alpha_{jk} - c\beta_{jk} -\epsilon_{jk}^-\right).
\end{align}
\end{proposition}

We will conduct the regret analysis under the assumption that $\mu_{ij}^*=0$ for all $\left(i,j\right)$. Intuitively, under a large $p_{\max}$, the optimal solution to problem (\ref{problem:1trans}) can satisfy $p_{ij}^*\left({\bm \theta}\right)<p_{\max}$ for all $\left(i,j\right)$. In this case, we have $\mu_{ij}^*=0$ for all $\left(i,j\right)$ (based on the complementary slackness condition). In Appendix \ref{app:sec:sufficient}, we prove that the following condition ensures $\mu_{ij}^*=0$ for all $\left(i,j\right)$:
\begin{align}
\sum_{k\in{\mathcal N}} \left|v_k\left({\bm \theta}\right) \right| \le \!\!\!\min_{\left(i,j\right):i\ne j,i,j\in{\mathcal N}} \!\! 2 \left(\beta_{ij} + \frac{\xi_{ij}}{\xi_{ji}}\beta_{ji}\right) \left(2p_{\max} -c - \frac{\alpha_{ij} + \epsilon_{ij}^-}{\beta_{ij}}\right).
\end{align}
Given ${\bm p}^*\left({\bm \theta}\right)$ in (\ref{equ:proposition:p}), we can derive the closed form of ${\bm w}^*\left({\bm \theta}\right)$ using $w_{ij}^*\left({\bm \theta}\right) = \alpha_{ij} - \beta_{ij} p_{ij}^*\left({\bm \theta}\right)$.

\subsubsection{Closed Forms of $\left({\bm p}^d,{\bm w}^d\right)$ Under Our Policy}\label{subsub:discussmu}
Recall that problem (\ref{problem:2trans}) is similar to problem (\ref{problem:1trans}), except that problem (\ref{problem:2trans}) is formulated based on the estimated demand model parameters (i.e., ${\hat \alpha}_{ij}^{d-1}$ and ${\hat \beta}_{ij}^{d-1}$). Therefore, we can apply the same resistance-based approach to solve problem (\ref{problem:2trans}). Specifically, we can construct a new resistor network, and define the corresponding effective resistances, i.e., $R_{ij}\left(\hat {\bm \beta}^{d-1}\right)$. We use ${\hat \mu}_{ij}^{d-1,*}$ to denote the optimal dual variable associated with (\ref{equ:opt:ourtrans:3}) for each $\left(i,j\right)$. Similar to Proposition \ref{proposition:price}, when ${\hat \mu}_{ij}^{d-1,*}=0$ for all $\left(i,j\right)$, we can derive $p_{ij}^*\left({\hat{\bm \theta}}^{d-1}\right)$ using ${\hat \beta}_{ij}^{d-1}$, ${\hat \alpha}_{ij}^{d-1}$, $R_{jk}\left(\hat {\bm \beta}^{d-1}\right)$, $R_{ik}\left(\hat {\bm \beta}^{d-1}\right)$, and $v_k\left({\hat {\bm \theta}}^{d-1}\right)$. Then, we can characterize the closed forms for the provider's decisions on the odd and even days under our NRPS policy.

We will conduct the regret analysis under the assumption that ${\hat \mu}_{ij}^{d-1,*}=0$ for all $\left(i,j\right)$ and $d$. In fact, this assumption is not a necessary condition for proving that our policy is a no-regret policy (i.e., $\lim_{D\rightarrow \infty} \Delta_D^{\bm \pi}=0$). As shown in Section \ref{sec:numerical}, when $\mu_{ij}^*=0$ for all $\left(i,j\right)$, ${\hat \mu}_{ij}^{d-1,*}$ may be positive for some $\left(i,j\right)$ at the beginning. After several days, ${\hat \mu}_{ij}^{d-1,*}$ becomes zero for all $\left(i,j\right)$, and no longer changes. In this case, we can still prove that $\lim_{D\rightarrow \infty} \Delta_D^{\bm \pi}=0$. We explain the reason in Appendix \ref{app:sec:reason} to save space. 

Based on our discussion above, we can get the closed forms of the provider's decisions (e.g., $p_{ij}^*\left({\bm \theta}\right)$ and $p_{ij}^*\left({\hat{\bm \theta}}^{d-1}\right)$) using the effective resistances. Then, we can leverage the properties of the effective resistances to compare the provider's decisions under the clairvoyant policy and our policy. For example, we can utilize Theorem \ref{theorem:estimate} to bound $|R_{jk}\left({\bm \beta}\right)-R_{jk}\left(\hat {\bm \beta}^{d-1}\right)|$ and $|R_{ik}\left({\bm \beta}\right)-R_{ik}\left(\hat {\bm \beta}^{d-1}\right)|$, and further bound $|p_{ij}^*\left({\bm \theta}\right)-p_{ij}^*\left({\hat{\bm \theta}}^{d-1}\right)|$. This enables us to analyze the time-average regret $\Delta_D^{\bm \pi}$.

\subsection{Upper Bound on Time-Average Regret}\label{subsec:performance:bou}
In this section, we characterize an upper bound on $\Delta_D^{\bm \pi}$, and show that $\lim_{D\rightarrow \infty} \Delta_D^{\bm \pi}=0$. 
For both the clairvoyant policy and our policy, we can plug the closed forms of the provider's decisions into (\ref{equ:payoff}) to get the provider's payoff per time slot on day $d$ (i.e., $\Pi\left({\bm p}^*\left({\bm \theta}\right),{\bm w}^*\left({\bm \theta}\right),{\bm \epsilon}^d\right)$ and the $\Pi\left({\bm p}^d,{\bm w}^d,{\bm \epsilon}^d\right)$ under our policy). 
Then, we utilize the bound on ${\mathbb E}\left\{ || {\hat{\bm \theta}}_{ij}^{d-1} - {\bm \theta}_{ij} ||_2^2 \right\}$ in Theorem \ref{theorem:estimate} to bound ${\mathbb E}^{\bm \pi}\left\{\Pi\left({\bm p}^*\left({\bm \theta}\right),{\bm w}^*\left({\bm \theta}\right),{\bm \epsilon}^d\right)\!-\!\Pi\left({\bm p}^d,{\bm w}^d,{\bm \epsilon}^d\right)\right\}$. 
Intuitively, as the provider's estimate ${\hat {\bm \theta}}^{d-1}$ becomes closer to $\bm \theta$, the expected gap between $\Pi\left({\bm p}^*\left({\bm \theta}\right),{\bm w}^*\left({\bm \theta}\right),{\bm \epsilon}^d\right)$ and $\Pi\left({\bm p}^d,{\bm w}^d,{\bm \epsilon}^d\right)$ becomes smaller. 
We can get $\Delta_D^{\bm \pi}$ by taking the average of the expected gap over $d$. In the following theorem, we characterize an upper bound on $\Delta_D^{\bm \pi}$ (let $e$ denote the base of the natural logarithm). 


\begin{theorem}\label{theorem:key}
Under the NRPS policy, there exist functions $\Phi_2\left(\rho,\eta\right)$, $\Phi_3\left(\rho,\eta\right)$, and $\Phi_4\left(\rho,\eta\right)$ such that (i) they are finite and positive for all $\rho\in\left(0,\infty\right)$ and $\eta\in\left(0,\frac{1}{2}\right)$; and (ii) the following relation holds for all $D>4+e^{\frac{1}{1-2\eta}}$:
\begin{align}
\Delta_D^{\bm \pi} < \Phi_2\left(\rho,\eta\right) D^{-1} + \Phi_3\left(\rho,\eta\right) {\left(\ln D\right)}^{\frac{1}{2}} D^{\eta-\frac{1}{2}} + \Phi_4\left(\rho,\eta\right) D^{-\eta}.\label{equ:regretbound} 
\end{align}
\end{theorem}

\begin{figure*}[t]
  \centering
  \subfigure[Squared Estimation Errors Under Different Policies.]{
  \label{fig:simu:A2}
    \includegraphics[scale=0.39]{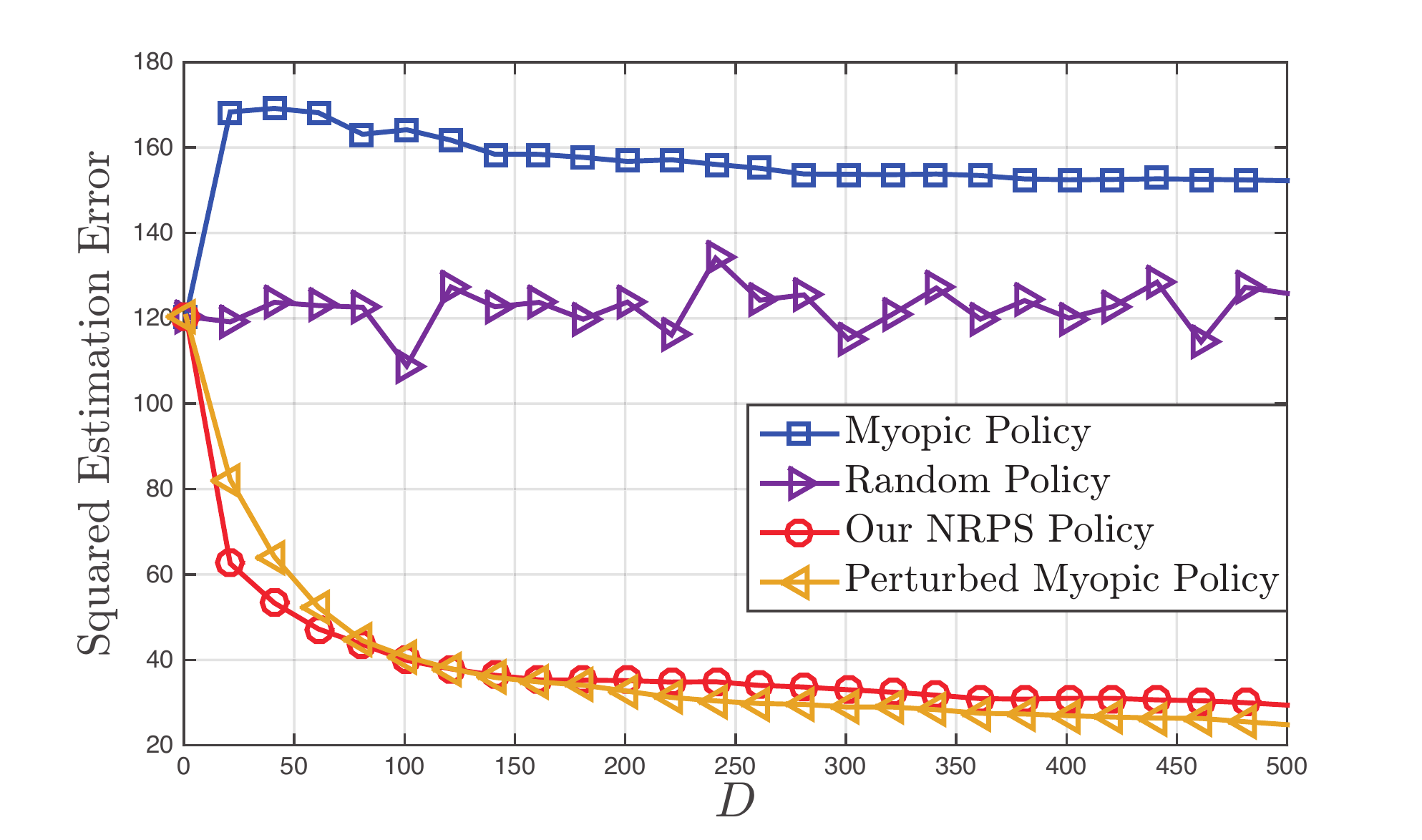}}   
  \vspace{-0.2cm}     
  \subfigure[Provider's Time-Average Payoffs During $D$ Days Under Different Policies.]{
  \label{fig:simu:A1}
    \includegraphics[scale=0.39]{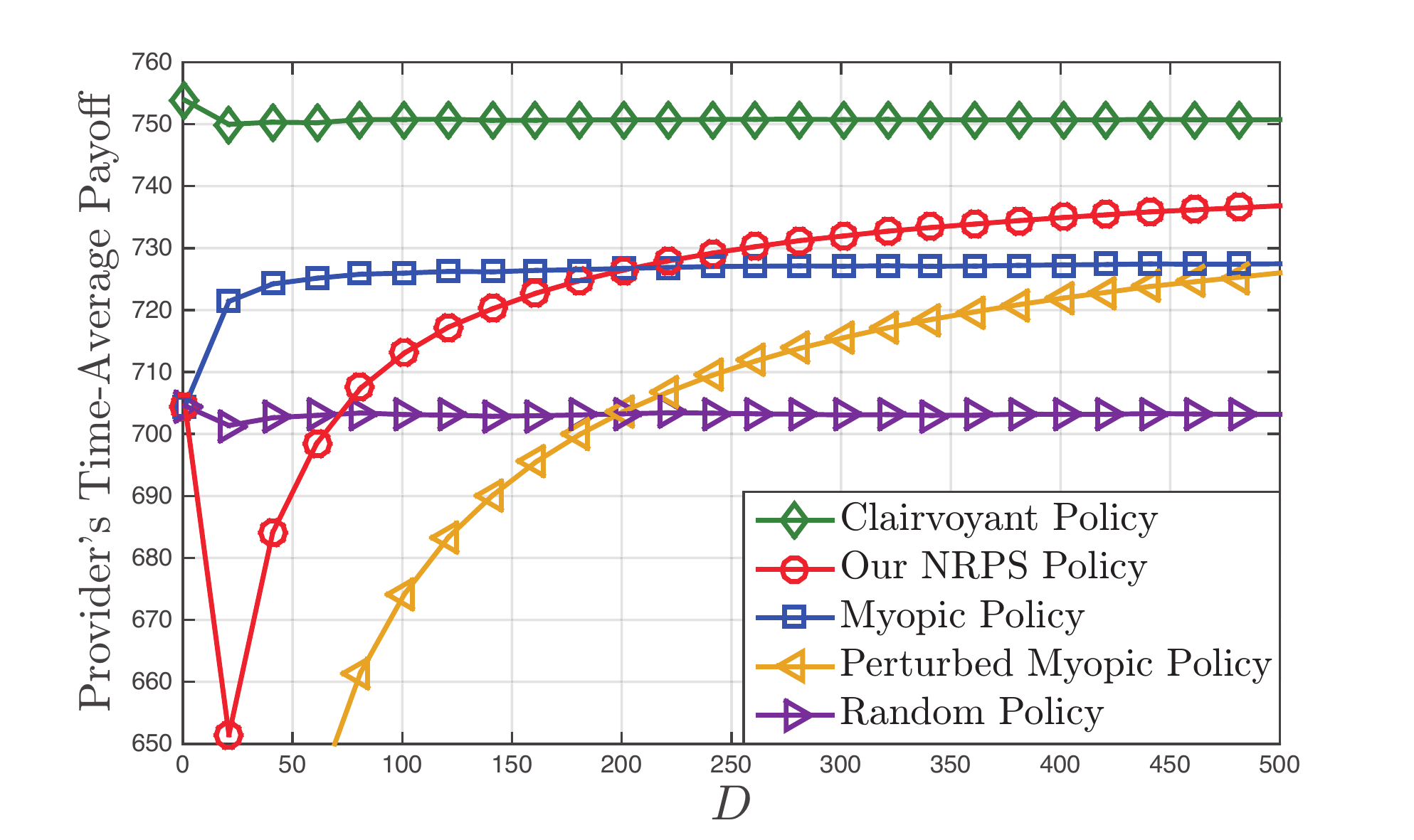}}
  \vspace{-0.2cm}
  \caption{Comparison Between Different Policies.}
  \label{fig:simu:A}
  \vspace{-0.4cm}
\end{figure*}

We can see that if $\eta\in\left(0,\frac{1}{4}\right)$, the upper bound of $\Delta_D^{\bm \pi}$ in (\ref{equ:regretbound}) is ${\mathcal O}\left(D^{-\eta}\right)$; if $\eta\in\left[\frac{1}{4},\frac{1}{2}\right)$, the upper bound is ${\mathcal O}\left({\left(\ln D\right)}^{\frac{1}{2}} D^{\eta-\frac{1}{2}}\right)$. Hence, we can choose $\eta=\frac{1}{4}$, which leads to an upper bound of ${\mathcal O}\left({\left(\ln D\right)}^{\frac{1}{2}} D^{-\frac{1}{4}}\right)$. 

Based on Theorem \ref{theorem:key}, we can get the following corollary.
\begin{corollary}\label{corollary:only}
The NRPS policy is a no-regret policy, i.e., the policy ensures that $\lim_{D\rightarrow \infty} \Delta_D^{\bm \pi}=0$.  
\end{corollary}


\section{Numerical Results}\label{sec:numerical}
In this section, we compare our policy with several other policies via numerical experiments, and investigate the impact of the control parameter $\eta$ on the performance of our policy. 


\subsection{Experiment Settings}
We compare our NRPS policy with the following four policies:
\begin{itemize}
\item \emph{Clairvoyant policy:} As introduced in Section \ref{subsubsec:clai}, the provider makes decisions based on the complete information of $\bm \theta$. 
\item \emph{Myopic policy:} As mentioned in Section \ref{subsubsec:operationeven}, the provider updates its estimate of ${\bm \theta}$ based on (\ref{equ:least:1}) and (\ref{equ:least:2}), and solves problem (\ref{problem:2}) on \emph{every} day (i.e., the operation under the myopic policy on each day is the same as the operation under our policy on each odd day).
\item \emph{Perturbed myopic policy:} It is similar to the myopic policy, except that the provider further adds offsets to its myopic optimal pricing and supply solutions. The sizes of the offsets are similar to those described in line 8 of Policy \ref{policy}.
\item \emph{Random policy:} The provider estimates ${\bm \theta}$ via random guessing and solves problem (\ref{problem:2}) on every day.{\footnote{Recall that the provider initially does not have any prior knowledge of $\bm \theta$ except the feasible region of $\bm \theta$. Under the random policy, the provider uniformly randomly picks an element from the feasible region of $\bm \theta$ as its estimate on each day.}}
\end{itemize}

We generate $\left\{\xi_{ij}\right\}_{i\ne j,i,j\in{\mathcal N}}$ using a real-world dataset from DiDi Chuxing (the largest ride-sharing platform in China), which contains information of the rides taken in November, 2016 in Chengdu, China \cite{DiDidata}. We focus on the rides whose (i) origins and destinations are within a $4.8\times 4.4 {\rm ~km}^2$ area and (ii) departure time and arrival time are between 8:30 pm and 11:30 pm on weekdays. We cluster the origins and destinations into $25$ locations (i.e., $N=25$), and set $\xi_{ij}$ to be the average travel time of the rides from $i$ to $j$. 

For each link $\left(i,j\right)$, we randomly generate $\alpha_{ij}$ and $\beta_{ij}$ according to truncated normal distributions (recall that $\alpha_{ij}$ and $\beta_{ij}$ are bounded). Specifically, we obtain the distribution of $\alpha_{ij}$ by truncating the normal distribution ${\mathcal N}\left(3.75,2.25\right)$ to interval $\left[3.5,4\right]$, and obtain the distribution of $\beta_{ij}$ by truncating ${\mathcal N}\left(2.5,2.25\right)$ to $\left[2,3\right]$. We randomly generate each $\epsilon_{ij}^d$ ($i\ne j,i,j\in{\mathcal N},d=1,\ldots,D$) according to a truncated normal distribution, which is obtained by truncating ${\mathcal N}\left(0,1\right)$ to $\left[-0.5,0.5\right]$.{\footnote{We plan to evaluate our policy under non-i.i.d. demand shocks in our future work.}} We set $p_{\max}=1$ and $c=0.1$. 

\subsection{Comparison Between Different Policies}
We use random guessing to get the initial estimate of $\bm \theta$ under our NRPS policy, myopic policy, perturbed myopic policy, and random policy. We choose $\rho=2$ and $\eta=0.45$, and show the comparison between different policies under one experiment in Fig. \ref{fig:simu:A}. 


\begin{figure*}[t]
  \centering
  \subfigure[Squared Estimation Errors Under Different $\eta$.]{
  \label{fig:simu:B2}
    \includegraphics[scale=0.39]{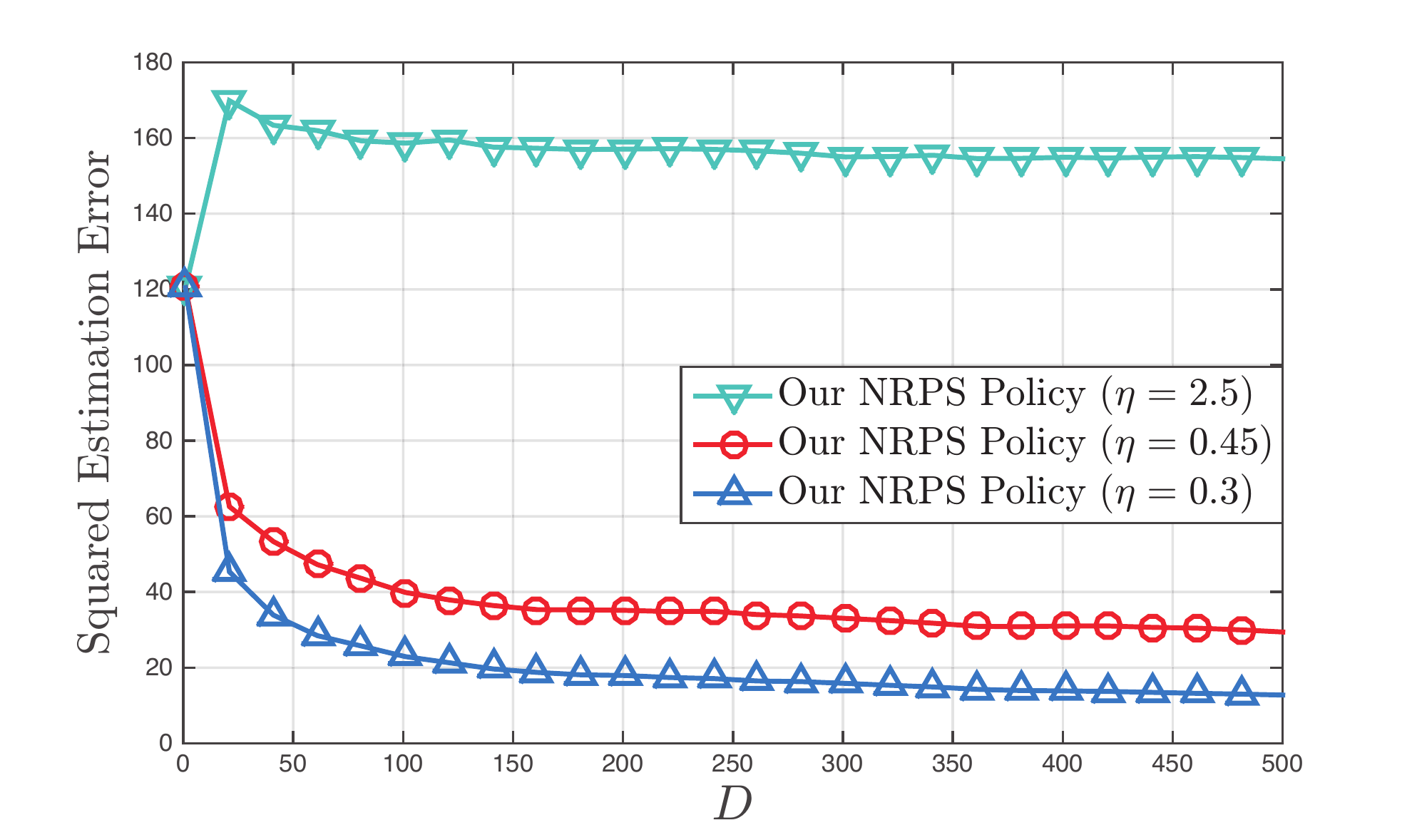}}    
  \vspace{-0.2cm}    
  \subfigure[Provider's Time-Average Payoffs During $D$ Days Under Different $\eta$.]{
  \label{fig:simu:B1}
    \includegraphics[scale=0.39]{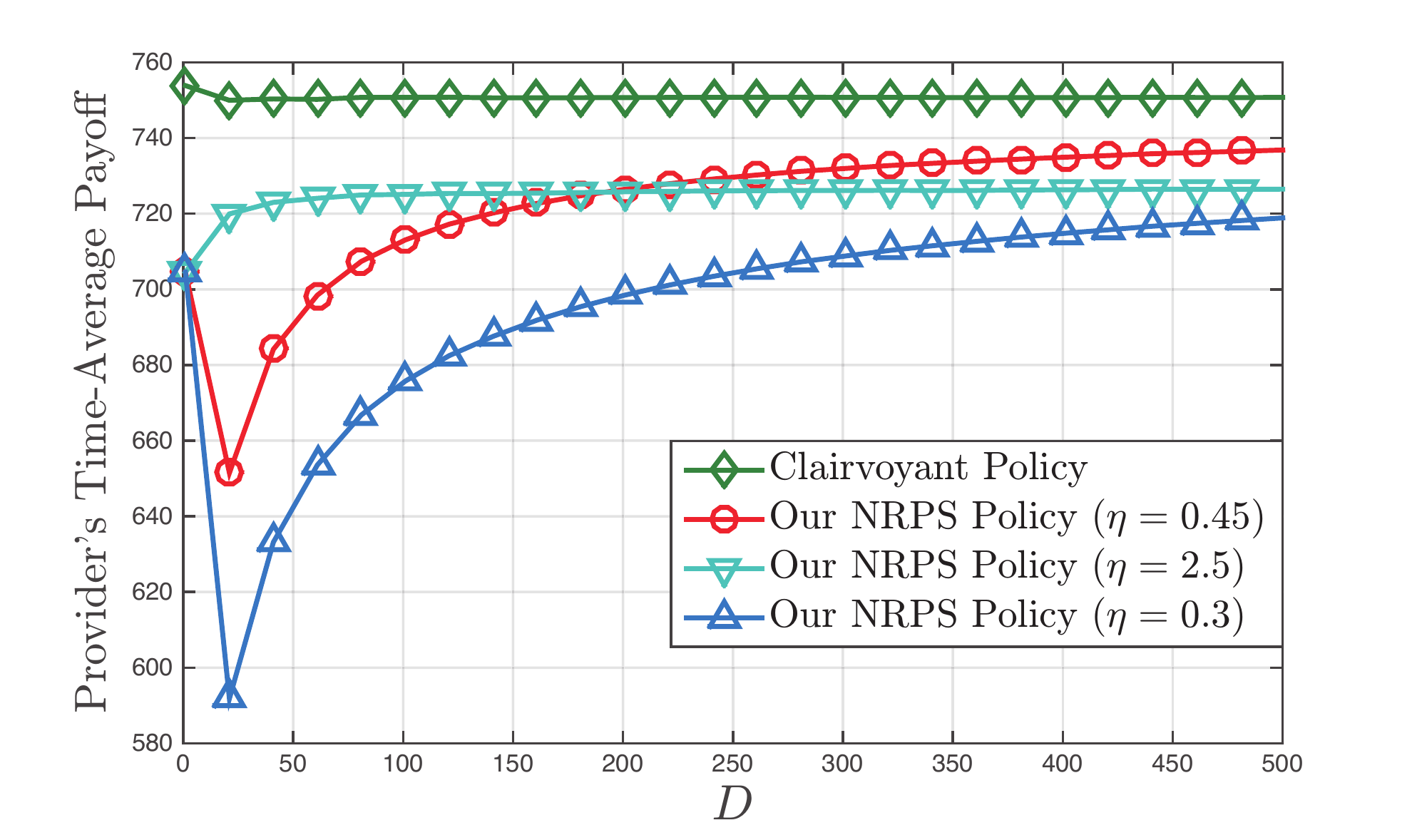}}
  \vspace{-0.2cm}
  \caption{Impact of Control Parameter $\eta$.}
  \label{fig:simu:B}
  \vspace{-0.4cm}
\end{figure*}

In Fig. \ref{fig:simu:A2}, we compare the accuracies of estimating $\bm \theta$, and plot the squared estimation errors (i.e., $\sum_{i\in{\mathcal N}}\sum_{j\in{\mathcal N}\setminus\left\{i\right\}}{|| {\hat{\bm \theta}}_{ij}^{D-1} - {\bm \theta}_{ij} ||_2^2}$) under different policies against $D$. As $D$ increases, our policy achieves a smaller squared estimation error (i.e., a better estimate of $\bm \theta$). The random policy keeps guessing $\bm\theta$ randomly, and hence its estimate does not improve over time. The myopic policy achieves an even worse estimate than the random policy. This is because the myopic policy does not explore sufficiently many prices and its estimate can get stuck at an incorrect value due to incomplete learning. The perturbed myopic policy achieves a similar estimate as our policy, since adding the offsets to the myopic optimal decisions leads to more exploration.

In Fig. \ref{fig:simu:A1}, we plot the provider's time-average payoffs during the first $D$ days under different policies (i.e., the average is taken over all the time slots during the first $D$ days). The time-average payoff under our policy first drops down to a low value when $D$ is small. This is because when $D$ is small, our policy adds large offsets to prices on even days to do exploration. Implementing these ``non-optimal'' prices reduces the time-average payoff. When $D$ is large, our policy outperforms the myopic policy and random policy, since our policy has a better estimate of $\bm \theta$ and the offsets added to prices have decayed to small values. When $D$ is large, the performance gap between our policy and the clairvoyant policy decreases with $D$.

In Fig. \ref{fig:simu:A1}, the myopic policy achieves a higher time-average payoff than the random policy. As shown in Fig. \ref{fig:simu:A2}, compared with the random policy, the myopic policy achieves a worse \emph{overall} squared estimation error (i.e., a larger $\sum_{i\in{\mathcal N}}\sum_{j\in{\mathcal N}\setminus\left\{i\right\}}{|| {\hat{\bm \theta}}_{ij}^{D-1} - {\bm \theta}_{ij} ||_2^2}$). However, the myopic policy can still well estimate ${\bm\theta}_{ij}$ for \emph{a few} links by learning from the history (although the estimation for the other links is much worse). This enables the myopic policy to make better decisions than the random policy, which estimates all ${\bm\theta}_{ij}$ only by random guessing. 

In Fig. \ref{fig:simu:A1}, the perturbed myopic policy has the worst performance under a small $D$, which is due to its frequent exploration of the solution space. Under a larger $D$, the perturbed myopic policy achieves a higher time-average payoff, which is because of its better estimate of ${\bm \theta}$ and the smaller values of the offsets.

In Section \ref{subsub:discussmu}, we claimed that when $\mu_{ij}^*=0$ for all $\left(i,j\right)$, we have ${\hat \mu}_{ij}^{D-1,*}=0$ for all $\left(i,j\right)$ if $D$ is large. In our experiment, we have checked that (i) $\mu_{ij}^*=0$ for all $\left(i,j\right)$, and (ii) ${\hat \mu}_{ij}^{D-1,*}=0$ for all $\left(i,j\right)$ and $D\ge473$, which verifies our claim. 

\vspace{-0.3cm}
\subsection{Impact of Control Parameter $\eta$}
In Fig. \ref{fig:simu:B}, we investigate the performance of our policy under three different $\eta$, i.e., $2.5$, $0.45$, and $0.3$. Recall that our theoretical results (e.g., Theorem \ref{theorem:key}) are derived under $\eta<0.5$, and $2.5$ is actually beyond the suggested region of $\eta$. We illustrate the performance under $\eta=2.5$ to show the problem of choosing a large $\eta$. In Fig. \ref{fig:simu:B}, the value of $\rho$ is fixed as $2$. 

In Fig. \ref{fig:simu:B2}, we can see that when $\eta=2.5$, the squared estimation error is large. Our policy adds offsets to prices on each even day $d$, and the sizes of the offsets are proportional to $d^{-\eta}$. When $\eta=2.5$, the offsets decay at a high rate, leading to less exploration and a worse estimation. Under a smaller value of $\eta$ (e.g., $0.45$ and $0.3$), our policy can better estimate $\bm \theta$. 

In Fig. \ref{fig:simu:B1}, when $D$ is small (e.g., $D\le181$), the time-average payoff under our policy increases with $\eta$. This is because under a larger $\eta$, our policy adds smaller offsets to prices on even days, and the negative impact of implementing ``non-optimal'' prices on the provider's payoff is smaller. When $D$ is large, all the offsets under different $\eta$ decay to small values. Meanwhile, our policy under a small $\eta$ (e.g., $0.45$ and $0.3$) achieves a better estimation of $\bm \theta$ as $D$ increases (as shown in Fig. \ref{fig:simu:B2}). In this case, the time-average payoffs under $\eta=0.45$ and $\eta=0.3$ increase with $D$. In particular, the time-average payoff under $\eta=0.45$ is greater than that under $\eta=2.5$ for $D\ge201$. 

\section{Conclusion and Future Extensions}
In this work, we studied a vehicle service provider's spatial pricing and supply with unknown demand. We proposed a policy that balances exploitation and exploration. To analyze the policy's performance, we leveraged the connection between the traffic network and a resistor network, and derived closed forms of the decisions under our policy. We proved that the time-average regret of our policy over $D$ days can be at most ${\mathcal O}\left({\left(\ln D\right)}^{\frac{1}{2}} D^{-\frac{1}{4}}\right)$. 

We considered a linear demand model in this work. However, we can extend our policy to other demand models. For example, given an exponential demand model \cite{fang2017prices}, we can estimate the demand model parameters using an exponential regression instead of the linear regression in (\ref{equ:least:1}). In this case, the design of the operations on odd and even days will be similar to those in our NRPS policy, except that the provider's optimization problem on each odd day will become non-convex. This makes it more difficult to theoretically analyze the policy's performance. 

There are some other interesting directions to extend our work. First, as discussed in Section \ref{sec:ourwork}, we could use a more sophisticated closed-queueing network to model users' stochastic demand, and design learning and pricing policies for the provider. Second, we could consider a ride-sharing platform, and study its spatial compensation to drivers. Besides learning user demand, the platform may need to learn drivers' willingness to work. 
Third, we are interested in analyzing the smallest achievable time-average regret for our problem. 
Fourth, it is interesting to consider multiple providers who compete for users and analyze their dynamic pricing strategies. 

\bibliographystyle{ACM-Reference-Format}
\bibliography{references}

\vspace{0.4cm}


\begin{centering}
{\LARGE{Appendices}}\\
\end{centering}
\vspace{0.4cm}

\begin{centering}
{{\large{\bf{Outline}}}}\\
\end{centering}
\vspace{0.2cm}

{\bf
{\ref{app:sec:feas}. Feasibility of Operation on Even Days}
\vspace{0.05cm}

{\ref{app:sec:least}. Solution to Equation (\ref{equ:least:1})}
\vspace{0.05cm}

{\ref{app:sec:proof:the1}. Proof of Theorem \ref{theorem:estimate}}
\vspace{0.05cm}

{\ref{app:sec:effective}. Proof of Proposition \ref{proposition:price}}
\vspace{0.05cm}

{\ref{app:sec:sufficient}. Sufficient Condition for $\mu_{ij}^*=0$}
\vspace{0.05cm}

{\ref{app:sec:reason}. Regret Analysis When ${\hat \mu}_{ij}^{d-1,*}\ne0$ at The Beginning}
\vspace{0.05cm}

{\ref{app:sec:keythe}. Proof of Theorem \ref{theorem:key}}
\vspace{0.05cm}

{\ref{app:sec:corollary}. Proof of Corollary \ref{corollary:only}}
\vspace{0.05cm}



}


\appendix

\section{Feasibility of Operation on Even Days}\label{app:sec:feas}
We prove that the provider's operation on even days under our NRPS policy is feasible and ensures the vehicle flow balance. 

First, we consider the pricing decisions. On an even day $d$, the provider's price for link $\left(i,j\right)$ is $p_{ij}^*\left({\hat{\bm \theta}}^{d-2}\right)-\frac{\rho}{{\hat\beta}_{ij}^{d-2}}d^{-\eta}$, where $p_{ij}^*\left({\hat{\bm \theta}}^{d-2}\right)$ is the provider's price on the last odd day. Based on the feasibility of the prices on the odd days, we have $p_{ij}^*\left({\hat{\bm \theta}}^{d-2}\right)\le p_{\max}$. Therefore, we also have $p_{ij}^*\left({\hat{\bm \theta}}^{d-2}\right)-\frac{\rho}{{\hat\beta}_{ij}^{d-2}}d^{-\eta}\le p_{\max}$, which shows the feasibility of the pricing decisions on each even day $d$.

Second, we consider the supply decisions. On an even day $d$, the provider's supply for link $\left(i,j\right)$ is $w_{ij}^*\left({\hat{\bm \theta}}^{d-2}\right)+\rho d^{-\eta}$, where $w_{ij}^*\left({\hat{\bm \theta}}^{d-2}\right)$ is the provider's supply on the last even day. Based on the feasibility of the supply decisions on the odd days, we have $w_{ij}^*\left({\hat{\bm \theta}}^{d-2}\right)\ge0$. Moreover, according to the flow balance achieved on the odd days, we have
\begin{align}
\sum_{j\in{\mathcal N}\setminus\left\{i\right\}} w_{ij}^*\left({\hat{\bm \theta}}^{d-2}\right) = \sum_{j\in{\mathcal N}\setminus\left\{i\right\}} w_{ji}^*\left({\hat{\bm \theta}}^{d-2}\right),\forall i\in{\mathcal N}.
\end{align}
Therefore, the supply decisions on the even day $d$ satisfy:
\begin{align}
& w_{ij}^*\left({\hat{\bm \theta}}^{d-2}\right)+\rho d^{-\eta}\ge0,\forall i\ne j, i,j\in{\mathcal N},\\
& \sum_{j\in{\mathcal N}\setminus\left\{i\right\}} \!\!\! \left( w_{ij}^*\left({\hat{\bm \theta}}^{d-2}\right)+\rho d^{-\eta} \right)\!=\!\!\!\!\! \sum_{j\in{\mathcal N}\setminus\left\{i\right\}} \!\!\! \left( w_{ji}^*\left({\hat{\bm \theta}}^{d-2}\right)+\rho d^{-\eta} \right),\forall i\in{\mathcal N}.
\end{align}
We can see that the supply decisions on each even day $d$ are feasible and ensure the vehicle flow balance.

\section{Solution to Equation (\ref{equ:least:1})}\label{app:sec:least}
We introduce the solution to (\ref{equ:least:1}). First, we show that $\sum_{\tau=1}^{d-1} \bigg(\Psi_{ij}^\tau\left(p_{ij}^\tau,\epsilon_{ij}^\tau\right) - \left( {\bar \alpha}_{ij} - {\bar \beta}_{ij} p_{ij}^\tau \right) \bigg)^2$ is a convex function of $\left({\bar \alpha}_{ij},{\bar \beta}_{ij}\right)$. We can derive the Hessian matrix of the function as
\begin{align}
\left[ {\begin{array}{cc}
   2\left(d-1\right) & -2\sum_{\tau=1}^{d-1} p_{ij}^\tau \\
   -2\sum_{\tau=1}^{d-1} p_{ij}^\tau & 2\sum_{\tau=1}^{d-1} \left(p_{ij}^\tau\right)^2 \\
  \end{array} } \right].\label{app:equ:hess}
\end{align}
In our NRPS policy, we run the least squares estimation on each odd day $d\ge3$ (when $d=1$, the estimation is given by the policy's initialization phase). Therefore, we have $d-1>0$ when solving (\ref{equ:least:1}). Next, we analyze the leading principal minor of the Hessian matrix of order $2$. We can easily derive the following relation:
\begin{align}
\left(d-1\right) \sum_{\tau=1}^{d-1} \left(p_{ij}^\tau\right)^2 - \left(\sum_{\tau=1}^{d-1} p_{ij}^\tau\right)^2 = \frac{1}{2} \sum_{\tau=1}^{d-1} \sum_{\nu=1}^{d-1} \left(p_{ij}^\tau-p_{ij}^\nu\right)^2.\label{app:equ:prin}
\end{align}
In our NRPS policy, we create a dispersion between the prices on odd and even days. For example, we have $p_{ij}^2-p_{ij}^1=-\frac{\rho}{{\hat \beta}_{ij}^0}2^{-\eta}<0$. Therefore, the value of $\frac{1}{2} \sum_{\tau=1}^{d-1} \sum_{\nu=1}^{d-1} \left(p_{ij}^\tau-p_{ij}^\nu\right)^2$ in (\ref{app:equ:prin}) is positive. Then, we can see that the Hessian matrix in (\ref{app:equ:hess}) is positive definite, which implies that the function $\sum_{\tau=1}^{d-1} \bigg(\Psi_{ij}^\tau\left(p_{ij}^\tau,\epsilon_{ij}^\tau\right) - \left( {\bar \alpha}_{ij} - {\bar \beta}_{ij} p_{ij}^\tau \right) \bigg)^2$ is convex. 

Based on the convexity of the function, $\left({{\tilde \alpha}_{ij}}^{d-1},{{\tilde \beta}_{ij}}^{d-1}\right)$ should satisfy the following equations:
\begin{align}
& -2 \sum_{\tau=1}^{d-1} \bigg(\Psi_{ij}^\tau\left(p_{ij}^\tau,\epsilon_{ij}^\tau\right) - \left( {{\tilde \alpha}_{ij}}^{d-1} - {{\tilde \beta}_{ij}}^{d-1} p_{ij}^\tau \right) \bigg) =0,\\
& 2 \sum_{\tau=1}^{d-1} p_{ij}^\tau \bigg(\Psi_{ij}^\tau\left(p_{ij}^\tau,\epsilon_{ij}^\tau\right) - \left( {{\tilde \alpha}_{ij}}^{d-1} - {{\tilde \beta}_{ij}}^{d-1} p_{ij}^\tau \right) \bigg) =0.
\end{align}
After rearrangement, we have
\begin{align}
\left[ {\begin{array}{cc}
 d-1 \!  & -\sum_{\tau=1}^{d-1}p_{ij}^\tau \! \\
 \sum_{\tau=1}^{d-1}p_{ij}^\tau  \! &  -\sum_{\tau=1}^{d-1} \left(p_{ij}^\tau\right)^2 \!\\
  \end{array} } \right]
\left[ {\begin{array}{c}
     {{\tilde \alpha}_{ij}}^{d-1} \\
     {{\tilde \beta}_{ij}}^{d-1}   \\
  \end{array} } \right]
\!\! = \!\!
\left[ {\begin{array}{c}
     \sum_{\tau=1}^{d-1} \Psi_{ij}^\tau\left(p_{ij}^\tau,\epsilon_{ij}^\tau\right) \\
     \sum_{\tau=1}^{d-1} \left( p_{ij}^\tau \Psi_{ij}^\tau\left(p_{ij}^\tau,\epsilon_{ij}^\tau\right)\right) \\
  \end{array} } \right].\label{app:equ:lineareq}
\end{align}
We can check that the coefficient matrix above is invertible. Specifically, we can compute the determinant as
\begin{align}
-\left( d-1\right) \sum_{\tau=1}^{d-1} \left(p_{ij}^\tau\right)^2 + \left(\sum_{\tau=1}^{d-1}p_{ij}^\tau\right)^2 = - \frac{1}{2} \sum_{\tau=1}^{d-1} \sum_{\nu=1}^{d-1} \left(p_{ij}^\tau-p_{ij}^\nu\right)^2.
\end{align}
According to our discussion for (\ref{app:equ:prin}), the determinant above is negative. Hence, the coefficient matrix in (\ref{app:equ:lineareq}) is invertible and its inverse can be computed as 
\begin{align}
\nonumber
& \left[ {\begin{array}{cc}
 d-1 \!  & -\sum_{\tau=1}^{d-1}p_{ij}^\tau \! \\
 \sum_{\tau=1}^{d-1}p_{ij}^\tau  \! &  -\sum_{\tau=1}^{d-1} \left(p_{ij}^\tau\right)^2 \!\\
  \end{array} } \right]^{-1} \\
& =  -\frac{2}{\sum_{\tau=1}^{d-1} \sum_{\nu=1}^{d-1} \left(p_{ij}^\tau-p_{ij}^\nu\right)^2}
\left[ {\begin{array}{cc}
  -\sum_{\tau=1}^{d-1} \left(p_{ij}^\tau\right)^2  & \sum_{\tau=1}^{d-1}p_{ij}^\tau  \\
 -\sum_{\tau=1}^{d-1}p_{ij}^\tau   &  d-1 \\
  \end{array} } \right].
\end{align}
Then, we can compute $\left[ {\begin{array}{c}
     {{\tilde \alpha}_{ij}}^{d-1} \\
     {{\tilde \beta}_{ij}}^{d-1}   \\
  \end{array} } \right]$ as follows:
\begin{align}
\nonumber
\left[ {\begin{array}{c}
     {{\tilde \alpha}_{ij}}^{d-1} \\
     {{\tilde \beta}_{ij}}^{d-1}   \\
  \end{array} } \right]  =  & -\frac{2}{\sum_{\tau=1}^{d-1} \sum_{\nu=1}^{d-1} \left(p_{ij}^\tau-p_{ij}^\nu\right)^2}
\left[ {\begin{array}{cc}
  -\sum_{\tau=1}^{d-1} \left(p_{ij}^\tau\right)^2  & \sum_{\tau=1}^{d-1}p_{ij}^\tau  \\
 -\sum_{\tau=1}^{d-1}p_{ij}^\tau   &  d-1 \\
  \end{array} } \right] \\
& \cdot \left[ {\begin{array}{c}
     \sum_{\tau=1}^{d-1} \Psi_{ij}^\tau\left(p_{ij}^\tau,\epsilon_{ij}^\tau\right) \\
     \sum_{\tau=1}^{d-1} \left( p_{ij}^\tau \Psi_{ij}^\tau\left(p_{ij}^\tau,\epsilon_{ij}^\tau\right)\right) \\
  \end{array} } \right].
\end{align}
Therefore, solving (\ref{equ:least:1}) mainly includes a multiplication between a $2 \times 2$ matrix and a $2 \times 1$ vector.

\section{Proof of Theorem \ref{theorem:estimate}}\label{app:sec:proof:the1}
Theorem \ref{theorem:estimate} characterizes the bound of ${\mathbb E}\left\{ || {\hat{\bm \theta}}_{ij}^{d-1} - {\bm \theta}_{ij} ||_2^2 \right\}$ under our policy. Recall that the provider gets ${\hat{\bm \theta}}_{ij}^{d-1}$ from (\ref{equ:least:1}) and (\ref{equ:least:2}). In (\ref{equ:least:1}), the provider gets ${\tilde {\bm \theta}}_{ij}^{d-1}$ via least squares estimation. In (\ref{equ:least:2}), the provider projects ${\tilde {\bm \theta}}_{ij}^{d-1}$ onto $\left[\alpha_{\min},\alpha_{\max}\right] \times \left[\beta_{\min},\beta_{\max}\right]$ to get ${\hat{\bm \theta}}_{ij}^{d-1}$. Since ${\bm \theta}_{ij}$ lies in $\left[\alpha_{\min},\alpha_{\max}\right] \times \left[\beta_{\min},\beta_{\max}\right]$, we have the following relation:
\begin{align}
{\mathbb E}\left\{ || {\hat{\bm \theta}}_{ij}^{d-1} - {\bm \theta}_{ij} ||_2^2 \right\} \le {\mathbb E}\left\{ || {\tilde{\bm \theta}}_{ij}^{d-1} - {\bm \theta}_{ij} ||_2^2 \right\}.
\end{align}
In the following, we characterize an upper bound on ${\mathbb E}\left\{ || {\tilde{\bm \theta}}_{ij}^{d-1} - {\bm \theta}_{ij} ||_2^2 \right\}$, and the upper bound will also be a bound on ${\mathbb E}\left\{ || {\hat{\bm \theta}}_{ij}^{d-1} - {\bm \theta}_{ij} ||_2^2 \right\}$. 
The proof in the following is similar to that in \cite{khezeli2017risk} (some details are different). Note that the proofs of other results in our paper (e.g., Theorem \ref{theorem:key}) are completely different from those in \cite{khezeli2017risk}.

{\bf Step 1:} We derive the expression for ${\mathbb E}\left\{ || {\tilde{\bm \theta}}_{ij}^{d-1} - {\bm \theta}_{ij} ||_2^2 \right\}$.

Based on (\ref{app:equ:lineareq}) in our analysis in Appendix \ref{app:sec:least}, we have the following relation:
\begin{align}
\nonumber
& \left[ {\begin{array}{cc}
 d-1   & \sum_{\tau=1}^{d-1}p_{ij}^\tau  \\
 \sum_{\tau=1}^{d-1}p_{ij}^\tau   &  \sum_{\tau=1}^{d-1} \left(p_{ij}^\tau\right)^2 \\
  \end{array} } \right]
\left[ {\begin{array}{c}
     {{\tilde \alpha}_{ij}}^{d-1} \\
     -{{\tilde \beta}_{ij}}^{d-1}   \\
  \end{array} } \right] \\
\nonumber  
&  = 
\left[ {\begin{array}{c}
     \sum_{\tau=1}^{d-1} \Psi_{ij}^\tau\left(p_{ij}^\tau,\epsilon_{ij}^\tau\right) \\
     \sum_{\tau=1}^{d-1} \left( p_{ij}^\tau \Psi_{ij}^\tau\left(p_{ij}^\tau,\epsilon_{ij}^\tau\right)\right) \\
  \end{array} } \right] \\
\nonumber   
& =
\left[ {\begin{array}{c}
     \sum_{\tau=1}^{d-1} \left(\alpha_{ij} - \beta_{ij} p_{ij}^\tau + \epsilon_{ij}^\tau\right) \\
     \sum_{\tau=1}^{d-1} \left( p_{ij}^\tau \left(\alpha_{ij} - \beta_{ij} p_{ij}^\tau + \epsilon_{ij}^\tau\right) \right) \\
  \end{array} } \right] \\
& =  \left[ {\begin{array}{cc}
 d-1   & \sum_{\tau=1}^{d-1}p_{ij}^\tau  \\
 \sum_{\tau=1}^{d-1}p_{ij}^\tau   &  \sum_{\tau=1}^{d-1} \left(p_{ij}^\tau\right)^2 \\
  \end{array} } \right]
\left[ {\begin{array}{c}
     \alpha_{ij} \\
     -\beta_{ij}   \\
  \end{array} } \right]
+    \left[ {\begin{array}{cc}
\sum_{\tau=1}^{d-1} \epsilon_{ij}^\tau \\
\sum_{\tau=1}^{d-1} \left(p_{ij}^\tau \epsilon_{ij}^\tau \right) \\
  \end{array} } \right].
\end{align}
After rearrangement, we have the following result:
\begin{align}
\nonumber
& \left[ {\begin{array}{c}
     {{\tilde \alpha}_{ij}}^{d-1} - \alpha_{ij}\\
      -{{\tilde \beta}_{ij}}^{d-1} + \beta_{ij}\\
  \end{array} } \right]  \\
&  = \left[ {\begin{array}{cc}
 d-1   & \sum_{\tau=1}^{d-1}p_{ij}^\tau  \\
 \sum_{\tau=1}^{d-1}p_{ij}^\tau   &  \sum_{\tau=1}^{d-1} \left(p_{ij}^\tau\right)^2 \\
  \end{array} } \right]^{-1} 
  \left[ {\begin{array}{cc}
\sum_{\tau=1}^{d-1} \epsilon_{ij}^\tau \\
\sum_{\tau=1}^{d-1} \left(p_{ij}^\tau \epsilon_{ij}^\tau \right) \\
  \end{array} } \right].
\end{align}
Note that in our analysis in Appendix \ref{app:sec:least}, we have proved that $\left(d-1\right) \sum_{\tau=1}^{d-1} \left(p_{ij}^\tau\right)^2 - \left(\sum_{\tau=1}^{d-1}p_{ij}^\tau\right)^2>0$ for any $d\ge3$. Hence, $\left[ {\begin{array}{cc}
 d-1   & \sum_{\tau=1}^{d-1}p_{ij}^\tau  \\
 \sum_{\tau=1}^{d-1}p_{ij}^\tau   &  \sum_{\tau=1}^{d-1} \left(p_{ij}^\tau\right)^2 \\
  \end{array} } \right]$ is invertible for any $d\ge3$. We define a matrix ${\bm \zeta}_{d-1}$ and a vector ${{\bm \varepsilon}}_{d-1}$ as follows (we omit the subscript $ij$ to simplify the notation):
\begin{align}
{\bm \zeta}_{d-1} \triangleq \left[ {\begin{array}{cc}
 d-1   & \sum_{\tau=1}^{d-1}p_{ij}^\tau  \\
 \sum_{\tau=1}^{d-1}p_{ij}^\tau   &  \sum_{\tau=1}^{d-1} \left(p_{ij}^\tau\right)^2 \\
  \end{array} } \right], {~} {{\bm \varepsilon}}_{d-1}\triangleq \left[ {\begin{array}{cc}
\sum_{\tau=1}^{d-1} \epsilon_{ij}^\tau \\
\sum_{\tau=1}^{d-1} \left(p_{ij}^\tau \epsilon_{ij}^\tau \right) \\
  \end{array} } \right].
\end{align}
Then, we can see that
\begin{align}
\nonumber
{\mathbb E}\left\{ || {\tilde{\bm \theta}}_{ij}^{d-1} - {\bm \theta}_{ij} ||_2^2 \right\} & =  
{\mathbb E}\left\{ || \left[ {{\tilde \alpha}_{ij}}^{d-1} - \alpha_{ij}, {{\tilde \beta}_{ij}}^{d-1}- \beta_{ij}\right] ||_2^2 \right\} \\
\nonumber
& =  
{\mathbb E}\left\{ || \left[ {{\tilde \alpha}_{ij}}^{d-1} - \alpha_{ij}, -{{\tilde \beta}_{ij}}^{d-1}+ \beta_{ij}\right] ||_2^2 \right\} \\
\nonumber
& =  
{\mathbb E}\left\{ || \left[ {{\tilde \alpha}_{ij}}^{d-1} - \alpha_{ij}, -{{\tilde \beta}_{ij}}^{d-1}+ \beta_{ij}\right]^T ||_2^2 \right\} \\
& = {\mathbb E}\left\{ || {\bm \zeta}_{d-1}^{-1}{{\bm \varepsilon}}_{d-1} ||_2^2 \right\}.\label{app:equ:normequs}
\end{align}
Furthermore, we can see that when $d\ge3$, the matrix ${\bm \zeta}_{d-1}$ is symmetric, positive definite, and invertible. 

{\bf Step 2:} We split ${\mathbb E}\left\{ || {\bm \zeta}_{d-1}^{-1}{{\bm \varepsilon}}_{d-1} ||_2^2 \right\}$ into two parts. 

Since ${\bm \zeta}_{d-1}$ is positive definite, it can be written as ${\bm Q} \left[ {\begin{array}{cc}
\lambda_1   &  0 \\
0   &  \lambda_2 \\
  \end{array} } \right] {\bm Q}^{-1}$, where $\lambda_1$ and $\lambda_2$ are the eigenvalues of ${\bm \zeta}_{d-1}$ and the columns of $\bm Q$ comprise an orthonormal basis of the eigenvectors of ${\bm \zeta}_{d-1}$. Furthermore, we have ${\bm Q}^T {\bm Q}={\bm Q} {\bm Q}^T ={\bm I}_2$, where ${\bm I}_2$ is the $2 \times 2$ identity matrix. The inverse of ${\bm \zeta}_{d-1}$ (i.e., ${\bm \zeta}_{d-1}^{-1}$) can be written as ${\bm Q} \left[ {\begin{array}{cc}
\frac{1}{\lambda_1}   &  0 \\
0   &  \frac{1}{\lambda_2} \\
  \end{array} } \right] {\bm Q}^{-1}$. 

We can further prove that ${\bm Q} \left[ {\begin{array}{cc}
\sqrt{\lambda_1}   &  0 \\
0   &  \sqrt{\lambda_2} \\
  \end{array} } \right] {\bm Q}^{-1}$ is positive definite and the following relation holds:
\begin{align}
{\bm Q} \left[ {\begin{array}{cc}
\sqrt{\lambda_1}   &  0 \\
0   &  \sqrt{\lambda_2} \\
  \end{array} } \right] {\bm Q}^{-1}
\left({\bm Q} \left[ {\begin{array}{cc}
\sqrt{\lambda_1}   &  0 \\
0   &  \sqrt{\lambda_2} \\
  \end{array} } \right] {\bm Q}^{-1}\right)
=  {\bm \zeta}_{d-1}.
\end{align}
Then, we can compute ${\bm \zeta}_{d-1}^{\frac{1}{2}}$ and ${\bm \zeta}_{d-1}^{-\frac{1}{2}}$ as 
\begin{align}
& {\bm \zeta}_{d-1}^{\frac{1}{2}}={\bm Q} \left[ {\begin{array}{cc}
\sqrt{\lambda_1}   &  0 \\
0   &  \sqrt{\lambda_2} \\
  \end{array} } \right] {\bm Q}^{-1},\\
& {\bm \zeta}_{d-1}^{-\frac{1}{2}}={\bm Q} \left[ {\begin{array}{cc}
\frac{1}{\sqrt{\lambda_1}}   &  0 \\
0   &  \frac{1}{\sqrt{\lambda_2}} \\
  \end{array} } \right] {\bm Q}^{-1}.\label{app:equ:zetasqrt}
\end{align}
Considering (\ref{app:equ:normequs}), we have the following relation:
\begin{align}
{\mathbb E}\left\{ || {\tilde{\bm \theta}}_{ij}^{d-1} - {\bm \theta}_{ij} ||_2^2 \right\} = {\mathbb E}\left\{ || {\bm \zeta}_{d-1}^{-1}{{\bm \varepsilon}}_{d-1} ||_2^2 \right\} = {\mathbb E}\left\{ || {\bm \zeta}_{d-1}^{-\frac{1}{2}} {\bm \zeta}_{d-1}^{-\frac{1}{2}} {{\bm \varepsilon}}_{d-1} ||_2^2 \right\}.\label{app:equ:s2l1}
\end{align}
Note that ${\bm \zeta}_{d-1}^{-\frac{1}{2}}$ is a $2 \times 2$ matrix and ${\bm \zeta}_{d-1}^{-\frac{1}{2}} {{\bm \varepsilon}}_{d-1}$ is a $2 \times 1$ vector. We use $|| {\bm \zeta}_{d-1}^{-\frac{1}{2}} ||_{\rm op}$ to denote the operator norm of the matrix ${\bm \zeta}_{d-1}^{-\frac{1}{2}}$. Then, we have
\begin{align}
|| {\bm \zeta}_{d-1}^{-\frac{1}{2}} ||_{\rm op} = \sup \left\{|| {\bm \zeta}_{d-1}^{-\frac{1}{2}} {\bm x}||_2: {\bm x}\in{\mathbb R}^{2\times1} {\rm ~with~} ||{\bm x}||_2=1\right\}.
\end{align}
Hence, for any given $\bm y \in{\mathbb R}^{2\times1}$ with $||{\bm y}||_2=1$, we have
\begin{align}
|| {\bm \zeta}_{d-1}^{-\frac{1}{2}} {\bm y}||_2 \le || {\bm \zeta}_{d-1}^{-\frac{1}{2}} ||_{\rm op}.
\end{align}
We can plug ${\bm y}=\frac{{\bm \zeta}_{d-1}^{-\frac{1}{2}} {{\bm \varepsilon}}_{d-1}}{|| {\bm \zeta}_{d-1}^{-\frac{1}{2}} {{\bm \varepsilon}}_{d-1} ||_2}$ into the above inequality and have 
\begin{align}
\frac{1}{{|| {\bm \zeta}_{d-1}^{-\frac{1}{2}} {{\bm \varepsilon}}_{d-1} ||_2}} || {\bm \zeta}_{d-1}^{-\frac{1}{2}} {{\bm \zeta}_{d-1}^{-\frac{1}{2}} {{\bm \varepsilon}}_{d-1}} ||_2 \le || {\bm \zeta}_{d-1}^{-\frac{1}{2}} ||_{\rm op}.
\end{align}
After rearrangement and taking the square on both sides, we get the following inequality:
\begin{align}
|| {\bm \zeta}_{d-1}^{-\frac{1}{2}} {\bm \zeta}_{d-1}^{-\frac{1}{2}} {{\bm \varepsilon}}_{d-1} ||_2^2 \le || {\bm \zeta}_{d-1}^{-\frac{1}{2}} ||_{\rm op}^2 || {\bm \zeta}_{d-1}^{-\frac{1}{2}} {{\bm \varepsilon}}_{d-1} ||_2^2.\label{app:equ:normope}
\end{align}
According to the property of the operator norm, $|| {\bm \zeta}_{d-1}^{-\frac{1}{2}} ||_{\rm op}^2$ equals the larger eigenvalue of $\left({\bm \zeta}_{d-1}^{-\frac{1}{2}}\right)^T {\bm \zeta}_{d-1}^{-\frac{1}{2}}$. Based on (\ref{app:equ:zetasqrt}) and the eigendecomposition of ${\bm \zeta}_{d-1}^{-1}$, we can see that $|| {\bm \zeta}_{d-1}^{-\frac{1}{2}} ||_{\rm op}^2$ also equals the larger eigenvalue of ${\bm \zeta}_{d-1}^{-1}$, which is the reciprocal of the smaller eigenvalue of ${\bm \zeta}_{d-1}$.

Considering (\ref{app:equ:s2l1}), (\ref{app:equ:normope}), and the above discussion, we have
\begin{align}
\nonumber
& {\mathbb E}\left\{ || {\tilde{\bm \theta}}_{ij}^{d-1} - {\bm \theta}_{ij} ||_2^2 \right\} = {\mathbb E}\left\{ || {\bm \zeta}_{d-1}^{-\frac{1}{2}} {\bm \zeta}_{d-1}^{-\frac{1}{2}} {{\bm \varepsilon}}_{d-1} ||_2^2 \right\} \\
& \le \frac{1}{{\rm the~smaller~eigenvalue~of~}{\bm \zeta}_{d-1}} {\mathbb E}\left\{ || {\bm \zeta}_{d-1}^{-\frac{1}{2}} {{\bm \varepsilon}}_{d-1} ||_2^2\right\}.\label{app:equ:s2split}
\end{align}
Hence, we have split the upper bound of ${\mathbb E}\left\{ || {\tilde{\bm \theta}}_{ij}^{d-1} - {\bm \theta}_{ij} ||_2^2 \right\}$ to two parts, i.e., $\frac{1}{{\rm the~smaller~eigenvalue~of~}{\bm \zeta}_{d-1}}$ and $ {\mathbb E}\left\{ || {\bm \zeta}_{d-1}^{-\frac{1}{2}} {{\bm \varepsilon}}_{d-1} ||_2^2\right\}$. We will characterize the upper bounds of them separately. 

{\bf Step 3:} In this step, we characterize a lower bound for the smaller eigenvalue of ${\bm \zeta}_{d-1}$, which will be an upper bound for $\frac{1}{{\rm the~smaller~eigenvalue~of~}{\bm \zeta}_{d-1}}$. 

Recall that ${\bm \zeta}_{d-1}$ is defined as $\left[ {\begin{array}{cc}
 d-1   & \sum_{\tau=1}^{d-1}p_{ij}^\tau  \\
 \sum_{\tau=1}^{d-1}p_{ij}^\tau   &  \sum_{\tau=1}^{d-1} \left(p_{ij}^\tau\right)^2 \\
  \end{array} } \right]$. Let $\lambda_L$ and $\lambda_S$ denote the larger and smaller eigenvalues of ${\bm \zeta}_{d-1}$, respectively. Then, $\lambda_L$ and $\lambda_S$ are the two solutions to the following equation:
\begin{align}
{\rm det} \left( \left[ {\begin{array}{cc}
 d-1-\lambda   & \sum_{\tau=1}^{d-1}p_{ij}^\tau  \\
 \sum_{\tau=1}^{d-1}p_{ij}^\tau   &  \sum_{\tau=1}^{d-1} \left(p_{ij}^\tau\right)^2-\lambda \\
  \end{array} } \right]\right)=0.
\end{align}
After rearrangement, we have
\begin{align}
\lambda^2 - \left(d-1+ \sum_{\tau=1}^{d-1} \left(p_{ij}^\tau\right)^2\right) \lambda + \left(d-1\right) \sum_{\tau=1}^{d-1} \left(p_{ij}^\tau\right)^2 - \left(\sum_{\tau=1}^{d-1}p_{ij}^\tau\right)^2=0.
\end{align}
Hence, $\lambda_L$ and $\lambda_S$ satisfy the following relations:
\begin{align}
& \lambda_L \lambda_S = \left(d-1\right) \sum_{\tau=1}^{d-1} \left(p_{ij}^\tau\right)^2 - \left(\sum_{\tau=1}^{d-1}p_{ij}^\tau\right)^2,\\
&  \lambda_L + \lambda_S =d-1+ \sum_{\tau=1}^{d-1} \left(p_{ij}^\tau\right)^2.
\end{align}
Next, we rearrange the expression for $\lambda_L \lambda_S$. We define ${\bar p}_{ij}^{d-1}\triangleq \frac{1}{d-1}\sum_{\tau=1}^{d-1} p_{ij}^\tau$, and can get the following relation:
\begin{align}
\lambda_L \lambda_S = \left(d-1\right) \sum_{\tau=1}^{d-1} \left(p_{ij}^\tau\right)^2 - \left(\sum_{\tau=1}^{d-1}p_{ij}^\tau\right)^2= \left(d-1\right) \sum_{\tau=1}^{d-1} \left(p_{ij}^\tau-{\bar p}_{ij}^{d-1}\right)^2.\label{app:equ:pbar}
\end{align}
Since ${\bm \zeta}_{d-1}$ is positive definite, both $\lambda_L$ and $\lambda_S$ are positive. Hence, $\lambda_L<d-1+ \sum_{\tau=1}^{d-1} \left(p_{ij}^\tau\right)^2$. We can derive the following result for $\lambda_S$:
\begin{align}
\nonumber
\lambda_S &=\frac{\left(d-1\right) \sum_{\tau=1}^{d-1} \left(p_{ij}^\tau-{\bar p}_{ij}^{d-1}\right)^2}{\lambda_L} \\
& >\frac{\left(d-1\right) \sum_{\tau=1}^{d-1} \left(p_{ij}^\tau-{\bar p}_{ij}^{d-1}\right)^2}{d-1+ \sum_{\tau=1}^{d-1} \left(p_{ij}^\tau\right)^2}\label{app:equ:lambdaS}
\end{align}
Next, we prove the existence of an upper bound of $|p_{ij}^\tau|$ for all $\left(i,j\right)$ and $\tau=1,\ldots,D$ under our policy. Under our policy, the prices on the odd days are determined by solving problem (\ref{problem:2}) (or equivalently, problem (\ref{problem:2trans})), and the prices on the even days are determined by modifying the prices on the odd days. It is easy to see that all the prices under our policy are upper-bounded by $p_{\max}$. From problem (\ref{problem:2trans}), we can see that the optimal prices obtained by solving problem (\ref{problem:2trans}) are also lower-bounded. We can prove this by contradiction. Suppose that the provider charges prices with negative infinite values on some links. The provider's overall payoff associated with these links has a negative infinite value. Since the provider's overall payoff associated with other links in the network is upper-bounded due to the quadratic shape of the payoff function, the provider's overall payoff has a negative infinite value. This implies that charging prices with negative infinite values is strictly dominated by charging prices that generate finite payoff values. In other words, the optimal prices obtained by solving problem (\ref{problem:2trans}) are lower-bounded. Then, we can easily see that the prices on odd and even days under our policy are lower-bounded. Because the prices under our policy are both lower-bounded and upper-bounded, we can use $p_{\rm up}$ to denote an upper bound of $|p_{ij}^\tau|$ for all $\left(i,j\right)$ and $\tau=1,\ldots,D$. 

Based on (\ref{app:equ:lambdaS}) and the definition of $p_{\rm up}$, we have
\begin{align}
\lambda_S>\frac{\left(d-1\right) \sum_{\tau=1}^{d-1} \left(p_{ij}^\tau-{\bar p}_{ij}^{d-1}\right)^2}{d-1+ \sum_{\tau=1}^{d-1} \left(p_{ij}^\tau\right)^2} >\frac{\sum_{\tau=1}^{d-1} \left(p_{ij}^\tau-{\bar p}_{ij}^{d-1}\right)^2}{1+p_{\rm up}^2}.
\end{align}
Next, we derive a lower bound of $\sum_{\tau=1}^{d-1} \left(p_{ij}^\tau-{\bar p}_{ij}^{d-1}\right)^2$. When $d\ge3$, we can derive the lower bound as follows:
\begin{align}
\nonumber
\sum_{\tau=1}^{d-1} \left(p_{ij}^\tau-{\bar p}_{ij}^{d-1}\right)^2 & \ge \sum_{\tau=1}^{\lfloor\frac{d-1}{2}\rfloor} \left(\left(p_{ij}^{2\tau-1}-{\bar p}_{ij}^{d-1}\right)^2 + \left(p_{ij}^{2\tau}-{\bar p}_{ij}^{d-1}\right)^2\right) \\
\nonumber
& \ge \sum_{\tau=1}^{\lfloor\frac{d-1}{2}\rfloor} \frac{1}{2} \left(p_{ij}^{2\tau-1}-{\bar p}_{ij}^{d-1} -p_{ij}^{2\tau}+{\bar p}_{ij}^{d-1}\right)^2 \\
& = \sum_{\tau=1}^{\lfloor\frac{d-1}{2}\rfloor} \frac{1}{2} \left(p_{ij}^{2\tau-1} -p_{ij}^{2\tau}\right)^2.
\end{align}
Under our policy, we have $p_{ij}^{2\tau-1} -p_{ij}^{2\tau}=\frac{\rho}{{\hat \beta}_{ij}^{2\tau-2}}\left(2\tau\right)^{-\eta}$ for $\tau=1,\ldots$. Hence, we can further derive the following inequality:
\begin{align}
\nonumber
\sum_{\tau=1}^{d-1} \left(p_{ij}^\tau-{\bar p}_{ij}^{d-1}\right)^2  & \ge \frac{\rho^2 2^{-2\eta}}{2 {\beta}_{\max}^2} \sum_{\tau=1}^{\lfloor\frac{d-1}{2}\rfloor} \tau^{-2\eta} \ge \frac{\rho^2 2^{-2\eta}}{2 {\beta}_{\max}^2} \int_1^{\lfloor\frac{d-1}{2}\rfloor+1} z^{-2\eta} dz \\
&= \frac{\rho^2 2^{-2\eta}}{2 {\beta}_{\max}^2} \frac{1}{1-2\eta} \left(\left(\lfloor\frac{d-1}{2}\rfloor+1\right)^{1-2\eta} - 1\right).
\end{align}
Recall that $\eta\in\left(0,\frac{1}{2}\right)$ and $d\ge5$ (which is the condition of Theorem \ref{theorem:estimate}). We can further derive the following result:
\begin{align}
\nonumber
\sum_{\tau=1}^{d-1} \left(p_{ij}^\tau-{\bar p}_{ij}^{d-1}\right)^2 & > \frac{\rho^2 2^{-2\eta}}{2 {\beta}_{\max}^2} \frac{1}{1-2\eta} \left(\left(\lfloor\frac{d-1}{2}\rfloor\right)^{1-2\eta} - 1\right) \\
\nonumber
& \ge \frac{\rho^2 2^{-2\eta}}{2 {\beta}_{\max}^2} \frac{1}{1-2\eta} \left(\left(\lfloor\frac{d-1}{2}\rfloor\right)^{1-2\eta} - \frac{\left(\lfloor\frac{d-1}{2}\rfloor\right)^{1-2\eta}}{2^{1-2\eta}} \right)\\
& =\frac{\rho^2 2^{-2\eta}}{2 {\beta}_{\max}^2} \frac{1}{1-2\eta}\left(\lfloor\frac{d-1}{2}\rfloor\right)^{1-2\eta} \left(1 - 2^{2\eta-1} \right).
\end{align}
To conclude, the smaller eigenvalue of ${\bm \zeta}_{d-1}$ is lower-bounded by
\begin{align}
\lambda_S>\frac{1}{1+p_{\rm up}^2} \frac{\rho^2 2^{-2\eta}}{2 {\beta}_{\max}^2} \frac{1}{1-2\eta}\left(\lfloor\frac{d-1}{2}\rfloor\right)^{1-2\eta} \left(1 - 2^{2\eta-1} \right).
\end{align}

{\bf Step 4:} We derive an upper bound for ${\mathbb E}\left\{ || {\bm \zeta}_{d-1}^{-\frac{1}{2}} {{\bm \varepsilon}}_{d-1} ||_2^2\right\}$. 

Recall that ${{\bm \varepsilon}}_{d-1}$ is defined as ${{\bm \varepsilon}}_{d-1}= \left[ {\begin{array}{cc}
\sum_{\tau=1}^{d-1} \epsilon_{ij}^\tau \\
\sum_{\tau=1}^{d-1} \left(p_{ij}^\tau \epsilon_{ij}^\tau \right) \\
  \end{array} } \right]$. When $d\ge3$, we have the following relation:
\begin{align}
{{\bm \varepsilon}}_{d-1} = {{\bm \varepsilon}}_{d-2} + \left[ {\begin{array}{cc}
 \epsilon_{ij}^{d-1} \\
 p_{ij}^{d-1} \epsilon_{ij}^{d-1} \\
  \end{array} } \right].
\end{align}
Then, we can expand the expression of ${\mathbb E}\left\{ || {\bm \zeta}_{d-1}^{-\frac{1}{2}} {{\bm \varepsilon}}_{d-1} ||_2^2\right\}$ as follows:
\begin{align}
\nonumber
& {\mathbb E}\left\{ || {\bm \zeta}_{d-1}^{-\frac{1}{2}} {{\bm \varepsilon}}_{d-1} ||_2^2\right\}= {\mathbb E}\left\{ \left({\bm \zeta}_{d-1}^{-\frac{1}{2}} {{\bm \varepsilon}}_{d-1}\right)^T {\bm \zeta}_{d-1}^{-\frac{1}{2}} {{\bm \varepsilon}}_{d-1} \right\} \\
\nonumber
& = {\mathbb E}\left\{ {{\bm \varepsilon}}_{d-1}^T \left({\bm \zeta}_{d-1}^{-\frac{1}{2}}\right)^T  {\bm \zeta}_{d-1}^{-\frac{1}{2}} {{\bm \varepsilon}}_{d-1} \right\} \overset{(a)}{=} {\mathbb E}\left\{ {{\bm \varepsilon}}_{d-1}^T {\bm \zeta}_{d-1}^{-1} {{\bm \varepsilon}}_{d-1} \right\}\\
\nonumber
& = {\mathbb E}\left\{ \left( {{\bm \varepsilon}}_{d-2}^T + \left[ {\begin{array}{cc}
 \epsilon_{ij}^{d-1} \\
 p_{ij}^{d-1} \epsilon_{ij}^{d-1} \\
  \end{array} } \right]^T\right) {\bm \zeta}_{d-1}^{-1} \left( {{\bm \varepsilon}}_{d-2} + \left[ {\begin{array}{cc}
 \epsilon_{ij}^{d-1} \\
 p_{ij}^{d-1} \epsilon_{ij}^{d-1} \\
  \end{array} } \right]\right) \right\} \\
\nonumber
& = {\mathbb E}\left\{{{\bm \varepsilon}}_{d-2}^T {\bm \zeta}_{d-1}^{-1} {{\bm \varepsilon}}_{d-2} \right\} + {\mathbb E}\left\{\left[ {\begin{array}{cc}
 \epsilon_{ij}^{d-1} \\
 p_{ij}^{d-1} \epsilon_{ij}^{d-1} \\
  \end{array} } \right]^T {\bm \zeta}_{d-1}^{-1}  \left[ {\begin{array}{cc}
 \epsilon_{ij}^{d-1} \\
 p_{ij}^{d-1} \epsilon_{ij}^{d-1} \\
  \end{array} } \right] \right\}\\
\nonumber  
& + {\mathbb E}\left\{{{\bm \varepsilon}}_{d-2}^T {\bm \zeta}_{d-1}^{-1} \left[ {\begin{array}{cc}
 1 \\
 p_{ij}^{d-1}  \\
  \end{array} } \right] \epsilon_{ij}^{d-1}\right\}
+ {\mathbb E}\left\{\epsilon_{ij}^{d-1} \left[ {\begin{array}{cc}
 1 \\
 p_{ij}^{d-1}  \\
  \end{array} } \right]^T {\bm \zeta}_{d-1}^{-1} {{\bm \varepsilon}}_{d-2}   \right\}\\
& \overset{(b)}{=} {\mathbb E}\left\{{{\bm \varepsilon}}_{d-2}^T {\bm \zeta}_{d-1}^{-1} {{\bm \varepsilon}}_{d-2} \right\} + {\mathbb E}\left\{\left[ {\begin{array}{cc}
 \epsilon_{ij}^{d-1} \\
 p_{ij}^{d-1} \epsilon_{ij}^{d-1} \\
  \end{array} } \right]^T {\bm \zeta}_{d-1}^{-1}  \left[ {\begin{array}{cc}
 \epsilon_{ij}^{d-1} \\
 p_{ij}^{d-1} \epsilon_{ij}^{d-1} \\
  \end{array} } \right] \right\}.
\end{align}
Here, the equality (a) is based on the decompositions of ${\bm \zeta}_{d-1}^{-\frac{1}{2}}$ and ${\bm \zeta}_{d-1}^{-1}$. The equality (b) is based on the fact that $\epsilon_{ij}^{d-1}$ is an independent and identically distributed random variable with ${\mathbb E}\left\{\epsilon_{ij}^{d-1}\right\}=0$. We can further rearrange the expression of ${\mathbb E}\left\{ || {\bm \zeta}_{d-1}^{-\frac{1}{2}} {{\bm \varepsilon}}_{d-1} ||_2^2\right\}$ as follows:
\begin{align}
\nonumber
& {\mathbb E}\left\{ || {\bm \zeta}_{d-1}^{-\frac{1}{2}} {{\bm \varepsilon}}_{d-1} ||_2^2\right\} \\
& = {\mathbb E}\left\{{{\bm \varepsilon}}_{d-2}^T {\bm \zeta}_{d-1}^{-1} {{\bm \varepsilon}}_{d-2} \right\} + {\mathbb E}\left\{\left[ {\begin{array}{cc}
 1 \\
 p_{ij}^{d-1}  \\
  \end{array} } \right]^T {\bm \zeta}_{d-1}^{-1}  \left[ {\begin{array}{cc}
 1 \\
 p_{ij}^{d-1}  \\
  \end{array} } \right] \right\} {\mathbb E}\left\{ \left(\epsilon_{ij}^{d-1}\right)^2 \right\}.
\end{align}
In the following, we derive upper bounds for the two components of ${\mathbb E}\left\{ || {\bm \zeta}_{d-1}^{-\frac{1}{2}} {{\bm \varepsilon}}_{d-1} ||_2^2\right\}$. 

({\bf Step 4-A}) First, we derive an upper bound for ${\mathbb E}\left\{{{\bm \varepsilon}}_{d-2}^T {\bm \zeta}_{d-1}^{-1} {{\bm \varepsilon}}_{d-2} \right\}$. Recall that ${\bm \zeta}_{d-1}$ is defined as $ \left[ {\begin{array}{cc}
 d-1   & \sum_{\tau=1}^{d-1}p_{ij}^\tau  \\
 \sum_{\tau=1}^{d-1}p_{ij}^\tau   &  \sum_{\tau=1}^{d-1} \left(p_{ij}^\tau\right)^2 \\
  \end{array} } \right]$. Then, we can derive the following expression:
\begin{align}
\nonumber
{\bm \zeta}_{d-1}^{-1}= & \left[ {\begin{array}{cc}
 d-1   & \sum_{\tau=1}^{d-1}p_{ij}^\tau  \\
 \sum_{\tau=1}^{d-1}p_{ij}^\tau   &  \sum_{\tau=1}^{d-1} \left(p_{ij}^\tau\right)^2 \\
  \end{array} } \right]^{-1} \\
\nonumber
=& \left({\bm \zeta}_{d-2} +  \left[ {\begin{array}{cc}
1   & p_{ij}^{d-1}  \\
p_{ij}^{d-1}   &  \left(p_{ij}^{d-1}\right)^2 \\
  \end{array} } \right]\right)^{-1} \\
\nonumber
=& \left({\bm \zeta}_{d-2} +  \left[ {\begin{array}{cc}
1  \\
p_{ij}^{d-1} \\
  \end{array} } \right]
 \left[ {\begin{array}{cc} 1 & p_{ij}^{d-1} \end{array}}\right]
\right)^{-1}  \\
=& {\bm \zeta}_{d-2} ^{-1} - \frac{{\bm \zeta}_{d-2}^{-1}    \left[ {\begin{array}{cc}
1  \\
p_{ij}^{d-1} \\
  \end{array} } \right]
 \left[ {\begin{array}{cc} 1 & p_{ij}^{d-1} \end{array}}\right]  {\bm \zeta}_{d-2}^{-1} }{1+ \left[ {\begin{array}{cc} 1 & p_{ij}^{d-1} \end{array}}\right] {\bm \zeta}_{d-2} ^{-1} \left[ {\begin{array}{cc}
1  \\
p_{ij}^{d-1} \\
  \end{array} } \right]}.
\end{align}
Here, the last equality is based on the Sherman-Morrison formula. Then, we can derive an upper bound for ${\mathbb E}\left\{{{\bm \varepsilon}}_{d-2}^T {\bm \zeta}_{d-1}^{-1} {{\bm \varepsilon}}_{d-2} \right\}$ as follows:
\begin{align}
\nonumber
& {\mathbb E}\left\{{{\bm \varepsilon}}_{d-2}^T {\bm \zeta}_{d-1}^{-1} {{\bm \varepsilon}}_{d-2} \right\} \\
\nonumber
= & {\mathbb E}\left\{ {{\bm \varepsilon}}_{d-2}^T {\bm \zeta}_{d-2}^{-1} {{\bm \varepsilon}}_{d-2}  \right\}\!-\! {\mathbb E}\!\left\{\! \! \frac{{{\bm \varepsilon}}_{d-2}^T{\bm \zeta}_{d-2}^{-1}    \left[ {\begin{array}{cc}
1  \\
p_{ij}^{d-1} \\
  \end{array} } \right]
 \left[ {\begin{array}{cc} 1 & p_{ij}^{d-1} \end{array}}\right]  {\bm \zeta}_{d-2}^{-1} {{\bm \varepsilon}}_{d-2}}{1+ \left[ {\begin{array}{cc} 1 & p_{ij}^{d-1} \end{array}}\right] {\bm \zeta}_{d-2} ^{-1} \left[ {\begin{array}{cc}
1  \\
p_{ij}^{d-1} \\
  \end{array} } \right]} \!\right\} \\
= & {\mathbb E}\left\{ {{\bm \varepsilon}}_{d-2}^T {\bm \zeta}_{d-2}^{-1} {{\bm \varepsilon}}_{d-2}  \right\}- {\mathbb E}\left\{  \frac{
 ||\left[ {\begin{array}{cc} 1 & p_{ij}^{d-1} \end{array}}\right]  {\bm \zeta}_{d-2}^{-1} {{\bm \varepsilon}}_{d-2} ||_2^2}{1+ \left[ {\begin{array}{cc} 1 & p_{ij}^{d-1} \end{array}}\right] {\bm \zeta}_{d-2} ^{-1} \left[ {\begin{array}{cc}
1  \\
p_{ij}^{d-1} \\
  \end{array} } \right]}  \right\}.
\end{align}
Since ${\bm \zeta}_{d-2} ^{-1} $ is positive definite, we have $\left[ {\begin{array}{cc} 1 & p_{ij}^{d-1} \end{array}}\right] {\bm \zeta}_{d-2} ^{-1} \left[ {\begin{array}{cc}
1  \\
p_{ij}^{d-1} \\
  \end{array} } \right]>0$. Therefore, we can derive the following relation:
\begin{align}
{\mathbb E}\left\{{{\bm \varepsilon}}_{d-2}^T {\bm \zeta}_{d-1}^{-1} {{\bm \varepsilon}}_{d-2} \right\} \le {\mathbb E}\left\{ {{\bm \varepsilon}}_{d-2}^T {\bm \zeta}_{d-2}^{-1} {{\bm \varepsilon}}_{d-2}  \right\}.
\end{align}

({\bf Step 4-B}) Second, we derive an upper bound for the expression ${\mathbb E}\left\{\left[ {\begin{array}{cc}
 1 \\
 p_{ij}^{d-1}  \\
  \end{array} } \right]^T {\bm \zeta}_{d-1}^{-1}  \left[ {\begin{array}{cc}
 1 \\
 p_{ij}^{d-1}  \\
  \end{array} } \right] \right\} {\mathbb E}\left\{ \left(\epsilon_{ij}^{d-1}\right)^2 \right\}$. Recall that ${\bm \zeta}_{d-1}$ is defined as $\left[ {\begin{array}{cc}
 d-1   & \sum_{\tau=1}^{d-1}p_{ij}^\tau  \\
 \sum_{\tau=1}^{d-1}p_{ij}^\tau   &  \sum_{\tau=1}^{d-1} \left(p_{ij}^\tau\right)^2 \\
  \end{array} } \right]$. We can compute ${\bm \zeta}_{d-1}^{-1}$ as follows:
\begin{align}
{\bm \zeta}_{d-1}^{-1} \!=\! \frac{1}{\left(d\!-\!1\right)\sum_{\tau=1}^{d-1} \left(p_{ij}^\tau\right)^2\!-\! \left(\sum_{\tau=1}^{d-1}p_{ij}^\tau\right)^2}\! \left[ {\begin{array}{cc}
 \sum_{\tau=1}^{d-1} \left(p_{ij}^\tau\right)^2  & -\sum_{\tau=1}^{d-1}p_{ij}^\tau  \\
 -\sum_{\tau=1}^{d-1}p_{ij}^\tau   &  d-1  \\
  \end{array} } \!\right].\label{app:equ:inversezeta}
\end{align}
We define $J_{d-1}\triangleq \sum_{\tau=1}^{d-1} \left(p_{ij}^\tau\right)^2- \frac{1}{d-1}\left(\sum_{\tau=1}^{d-1}p_{ij}^\tau\right)^2$. According to our prior proof, we have 
\begin{align}
J_{d-1}=\sum_{\tau=1}^{d-1} \left(p_{ij}^\tau\right)^2 - \frac{1}{d-1}\left(\sum_{\tau=1}^{d-1}p_{ij}^\tau\right)^2= \sum_{\tau=1}^{d-1} \left(p_{ij}^\tau-{\bar p}_{ij}^{d-1}\right)^2.
\end{align}
Note that ${\bar p}_{ij}^{d-1}$ is defined before (\ref{app:equ:pbar}). Since $p_{ij}^2$ is different from $p_{ij}^1$, the value of $J_{d-1}$ is positive for any $d\ge3$. By plugging the expression of  ${\bm \zeta}_{d-1}^{-1}$ and rearranging the result using $J_{d-1}$, we can rewrite ${\mathbb E}\left\{\left[ {\begin{array}{cc}
 1 \\
 p_{ij}^{d-1}  \\
  \end{array} } \right]^T {\bm \zeta}_{d-1}^{-1}  \left[ {\begin{array}{cc}
 1 \\
 p_{ij}^{d-1}  \\
  \end{array} } \right] \right\} {\mathbb E}\left\{ \left(\epsilon_{ij}^{d-1}\right)^2 \right\}$ as follows:
\begin{align}
\nonumber
& {\mathbb E}\left\{\left[ {\begin{array}{cc}
 1 \\
 p_{ij}^{d-1}  \\
  \end{array} } \right]^T {\bm \zeta}_{d-1}^{-1}  \left[ {\begin{array}{cc}
 1 \\
 p_{ij}^{d-1}  \\
  \end{array} } \right] \right\} {\mathbb E}\left\{ \left(\epsilon_{ij}^{d-1}\right)^2 \right\} \\
=  & {\mathbb E}\left\{ \frac{\left(d-1\right)\left(p_{ij}^{d-1}-{\bar p}_{ij}^{d-1}\right)^2+J_{d-1}}{\left(d-1\right) J_{d-1}}\right\}{\mathbb E}\left\{ \left(\epsilon_{ij}^{d-1}\right)^2 \right\}.
\end{align}
Next, we prove that $\left(p_{ij}^{d-1}-{\bar p}_{ij}^{d-1}\right)^2 \le J_{d-1}-J_{d-2}$. We can compute $J_{d-1}-J_{d-2}-\left(p_{ij}^{d-1}-{\bar p}_{ij}^{d-1}\right)^2$ as follows:
\begin{align}
\nonumber
& J_{d-1}-J_{d-2}-\left(p_{ij}^{d-1}-{\bar p}_{ij}^{d-1}\right)^2 \\
\nonumber
= & \left(p_{ij}^{d-1}-{\bar p}_{ij}^{d-1}\right)^2 + \sum_{\tau=1}^{d-2} \left(p_{ij}^\tau-{\bar p}_{ij}^{d-1}\right)^2 \\
\nonumber
& - \sum_{\tau=1}^{d-2} \left(p_{ij}^\tau-{\bar p}_{ij}^{d-2}\right)^2 -\left(p_{ij}^{d-1}-{\bar p}_{ij}^{d-1}\right)^2 \\
\nonumber
=& \sum_{\tau=1}^{d-2} \left(p_{ij}^\tau-{\bar p}_{ij}^{d-1}\right)^2- \sum_{\tau=1}^{d-2} \left(p_{ij}^\tau-{\bar p}_{ij}^{d-2}\right)^2 \\
\nonumber
=& \left({\bar p}_{ij}^{d-2}-{\bar p}_{ij}^{d-1}\right) \sum_{\tau=1}^{d-2} \left(2p_{ij}^\tau-{\bar p}_{ij}^{d-1}-{\bar p}_{ij}^{d-2}\right) \\
= & \left(d-2\right)\left({\bar p}_{ij}^{d-2}-{\bar p}_{ij}^{d-1}\right)^2\ge0.\label{app:equ:s4JJ}
\end{align}
Hence, we have $\left(p_{ij}^{d-1}-{\bar p}_{ij}^{d-1}\right)^2 \le J_{d-1}-J_{d-2}$. This implies the following relation:
\begin{align}
\nonumber
& {\mathbb E}\left\{\left[ {\begin{array}{cc}
 1 \\
 p_{ij}^{d-1}  \\
  \end{array} } \right]^T {\bm \zeta}_{d-1}^{-1}  \left[ {\begin{array}{cc}
 1 \\
 p_{ij}^{d-1}  \\
  \end{array} } \right] \right\} {\mathbb E}\left\{ \left(\epsilon_{ij}^{d-1}\right)^2 \right\} \\
\nonumber  
\le & {\mathbb E}\left\{ \frac{\left(d-1\right)\left(J_{d-1}-J_{d-2}\right)+J_{d-1}}{\left(d-1\right) J_{d-1}}\right\}{\mathbb E}\left\{ \left(\epsilon_{ij}^{d-1}\right)^2 \right\} \\
= & {\mathbb E}\left\{ 1-\frac{J_{d-2}}{J_{d-1}} + \frac{1}{d-1} \right\}{\mathbb E}\left\{ \left(\epsilon_{ij}^{d-1}\right)^2 \right\}.
\end{align}

({\bf Step 4-C}) Third, we combine the results derived in {\bf Step 4-A} and {\bf Step 4-B}. Recall that ${\mathbb E}\left\{ || {\bm \zeta}_{d-1}^{-\frac{1}{2}} {{\bm \varepsilon}}_{d-1} ||_2^2\right\}$ equals ${\mathbb E}\left\{ {{\bm \varepsilon}}_{d-1}^T {\bm \zeta}_{d-1}^{-1} {{\bm \varepsilon}}_{d-1} \right\}$, and it includes the following two terms: ${\mathbb E}\left\{{{\bm \varepsilon}}_{d-2}^T {\bm \zeta}_{d-1}^{-1} {{\bm \varepsilon}}_{d-2} \right\}$ and ${\mathbb E}\left\{\left[ {\begin{array}{cc}
 1 \\
 p_{ij}^{d-1}  \\
  \end{array} } \right]^T {\bm \zeta}_{d-1}^{-1}  \left[ {\begin{array}{cc}
 1 \\
 p_{ij}^{d-1}  \\
  \end{array} } \right] \right\} {\mathbb E}\left\{ \left(\epsilon_{ij}^{d-1}\right)^2 \right\}$. 
According to {\bf Step 4-A}, we have
\begin{align} 
{\mathbb E}\left\{{{\bm \varepsilon}}_{d-2}^T {\bm \zeta}_{d-1}^{-1} {{\bm \varepsilon}}_{d-2} \right\} \le {\mathbb E}\left\{ {{\bm \varepsilon}}_{d-2}^T {\bm \zeta}_{d-2}^{-1} {{\bm \varepsilon}}_{d-2}  \right\}. 
\end{align}
According to {\bf Step 4-B}, we have
\begin{align}
\nonumber
& {\mathbb E}\left\{\left[ {\begin{array}{cc}
 1 \\
 p_{ij}^{d-1}  \\
  \end{array} } \right]^T {\bm \zeta}_{d-1}^{-1}  \left[ {\begin{array}{cc}
 1 \\
 p_{ij}^{d-1}  \\
  \end{array} } \right] \right\} {\mathbb E}\left\{ \left(\epsilon_{ij}^{d-1}\right)^2 \right\} \\
\le & {\mathbb E}\left\{ 1-\frac{J_{d-2}}{J_{d-1}} + \frac{1}{d-1} \right\}{\mathbb E}\left\{ \left(\epsilon_{ij}^{d-1}\right)^2 \right\}.
\end{align}
Therefore, we have
{\small
\begin{align}
{\mathbb E}\left\{ {{\bm \varepsilon}}_{d-1}^T {\bm \zeta}_{d-1}^{-1} {{\bm \varepsilon}}_{d-1} \right\} \le {\mathbb E}\left\{ {{\bm \varepsilon}}_{d-2}^T {\bm \zeta}_{d-2}^{-1} {{\bm \varepsilon}}_{d-2}  \right\} + {\mathbb E}\left\{ 1-\frac{J_{d-2}}{J_{d-1}} + \frac{1}{d-1} \right\}{\mathbb E}\left\{ \left(\epsilon_{ij}^{d-1}\right)^2 \right\}.
\end{align}}
We can repeatedly apply similar inequalities:
{\small
\begin{align}
\nonumber
{\mathbb E}\left\{ {{\bm \varepsilon}}_{d-2}^T {\bm \zeta}_{d-2}^{-1} {{\bm \varepsilon}}_{d-2} \right\} & \le {\mathbb E}\left\{ {{\bm \varepsilon}}_{d-3}^T {\bm \zeta}_{d-3}^{-1} {{\bm \varepsilon}}_{d-3}  \right\} + {\mathbb E}\left\{ 1-\frac{J_{d-3}}{J_{d-2}} + \frac{1}{d-2} \right\}{\mathbb E}\left\{ \left(\epsilon_{ij}^{d-2}\right)^2 \right\},\\
\nonumber
& \ldots \\
\nonumber
{\mathbb E}\left\{ {{\bm \varepsilon}}_{3}^T {\bm \zeta}_{3}^{-1} {{\bm \varepsilon}}_{3} \right\} & \le {\mathbb E}\left\{ {{\bm \varepsilon}}_{2}^T {\bm \zeta}_{2}^{-1} {{\bm \varepsilon}}_{2}  \right\} + {\mathbb E}\left\{ 1-\frac{J_{2}}{J_{3}} + \frac{1}{3} \right\}{\mathbb E}\left\{ \left(\epsilon_{ij}^{3}\right)^2 \right\}.
\end{align}}
We can take the summation over the right sides and left sides of all these inequalities, and get the following inequality:
\begin{align}
{\mathbb E}\left\{ {{\bm \varepsilon}}_{d-1}^T {\bm \zeta}_{d-1}^{-1} {{\bm \varepsilon}}_{d-1} \right\} \le {\mathbb E}\left\{ {{\bm \varepsilon}}_{2}^T {\bm \zeta}_{2}^{-1} {{\bm \varepsilon}}_{2}  \right\} + \sum_{k=3}^{d-1} {\mathbb E}\left\{ 1-\frac{J_{k-1}}{J_{k}} + \frac{1}{k} \right\}{\mathbb E}\left\{ \left(\epsilon_{ij}^{k}\right)^2 \right\}.\label{app:equ:s4inequa}
\end{align}
Note that ${{\bm \varepsilon}}_{2}$ is defined as ${{\bm \varepsilon}}_{2}= \left[ {\begin{array}{cc}
\epsilon_{ij}^1+\epsilon_{ij}^2 \\
p_{ij}^1 \epsilon_{ij}^1 + p_{ij}^2 \epsilon_{ij}^2  \\
  \end{array} } \right]$ and ${\bm \zeta}_{2}^{-1}$ equals the following expression (based on (\ref{app:equ:inversezeta})):
\begin{align}
{\bm \zeta}_{2}^{-1} \!=\! \frac{1}{2\sum_{\tau=1}^{2} \left(p_{ij}^\tau\right)^2\!-\! \left(\sum_{\tau=1}^{2}p_{ij}^\tau\right)^2}\! \left[ {\begin{array}{cc}
 \sum_{\tau=1}^{2} \left(p_{ij}^\tau\right)^2  & -\sum_{\tau=1}^{2}p_{ij}^\tau  \\
 -\sum_{\tau=1}^{2}p_{ij}^\tau   &  2  \\
  \end{array} } \!\right].
\end{align}
We can verify that ${\mathbb E}\left\{ {{\bm \varepsilon}}_{2}^T {\bm \zeta}_{2}^{-1} {{\bm \varepsilon}}_{2}  \right\}={\mathbb E}\left\{\left(\epsilon_{ij}^1\right)^2 + \left(\epsilon_{ij}^2\right)^2\right\}$. Recall that $\left\{\epsilon_{ij}^k\right\}_{k=1,\ldots,D}$ is a set of independent and identically distributed random variables with $\epsilon_{ij}^k\in\left[{\underline \epsilon},{\overline \epsilon}\right]$ for all $\left(i,j\right)$ and $d$. We can see that ${\mathbb E}\left\{ \left(\epsilon_{ij}^{k}\right)^2 \right\}$ has the same value for different $\left(i,j\right)$ and $k$. Furthermore, since ${\mathbb E} \left\{ \epsilon_{ij}^{k} \right\}=0$, we can see that
\begin{align}
{\mathbb E}\left\{ \left(\epsilon_{ij}^{k}\right)^2 \right\} = {\rm Var} \left\{ \epsilon_{ij}^{k} \right\} + \left({\mathbb E} \left\{ \epsilon_{ij}^{k} \right\}\right)^2 = {\rm Var} \left\{ \epsilon_{ij}^{k} \right\} \le\frac{1}{4}\left({\overline \epsilon}-{\underline \epsilon}\right)^2,
\end{align}
where the last inequality is based on the Popoviciu's inequality. Therefore, we can derive the following result based on (\ref{app:equ:s4inequa}):
\begin{align}
{\mathbb E}\left\{ {{\bm \varepsilon}}_{d-1}^T {\bm \zeta}_{d-1}^{-1} {{\bm \varepsilon}}_{d-1} \right\} \le \frac{1}{2}\left({\overline \epsilon}-{\underline \epsilon}\right)^2 + \frac{1}{4}\left({\overline \epsilon}-{\underline \epsilon}\right)^2 \sum_{k=3}^{d-1} {\mathbb E}\left\{ 1-\frac{J_{k-1}}{J_{k}} + \frac{1}{k} \right\}.
\end{align}
Next, we derive upper bounds for $\sum_{k=3}^{d-1} {\mathbb E}\left\{ 1-\frac{J_{k-1}}{J_{k}} \right\}$ and $\sum_{k=3}^{d-1} {\mathbb E}\left\{\frac{1}{k} \right\}$. First, we can see that
\begin{align}
\sum_{k=3}^{d-1} {\mathbb E}\left\{ 1-\frac{J_{k-1}}{J_{k}} \right\} = \sum_{k=3}^{d-1} {\mathbb E}\left\{ \frac{J_k-J_{k-1}}{J_{k}} \right\} = \sum_{k=3}^{d-1} \int_{J_{k-1}}^{J_k} \frac{dz}{J_k}.
\end{align}
From (\ref{app:equ:s4JJ}), we can see that $J_{k}-J_{k-1}\ge \left(p_{ij}^{k}-{\bar p}_{ij}^{k}\right)^2\ge0$. Hence, we can further derive the following result:
\begin{align}
\sum_{k=3}^{d-1} {\mathbb E}\left\{ 1-\frac{J_{k-1}}{J_{k}} \right\} \le \sum_{k=3}^{d-1} \int_{J_{k-1}}^{J_k} \frac{dz}{z} = \int_{J_{2}}^{J_{d-1}} \frac{dz}{z} = \ln {J_{d-1}} - \ln{J_2}.
\end{align}
Note that $J_{d-1}=\sum_{\tau=1}^{d-1} \left(p_{ij}^\tau\right)^2 - \frac{1}{d-1}\left(\sum_{\tau=1}^{d-1}p_{ij}^\tau\right)^2\le \sum_{\tau=1}^{d-1} \left(p_{ij}^\tau\right)^2 \le \left(d-1\right) p_{\rm up}^2$ and $J_2=\left(p_{ij}^1\right)^2+\left(p_{ij}^2\right)^2-\frac{1}{2}\left(p_{ij}^1+p_{ij}^2\right)^2=\frac{1}{2}\left(p_{ij}^1-p_{ij}^2\right)^2 = \frac{1}{2} \left(\frac{\rho}{{\hat \beta}_{ij}^0}2^{-\eta}\right)^2=\frac{1}{2}\frac{\rho^2}{\left({\hat \beta}_{ij}^0\right)^2} 2^{-2\eta}$. We can get the following inequality:
\begin{align}
\nonumber
\sum_{k=3}^{d-1} {\mathbb E}\left\{ 1-\frac{J_{k-1}}{J_{k}} \right\} & \le \ln \left(\left(d-1\right) p_{\rm up}^2\right) - \ln \left(\frac{1}{2}\frac{\rho^2}{\left({\hat \beta}_{ij}^0\right)^2} 2^{-2\eta}\right) \\
& \le \ln \left(\left(d-1\right) p_{\rm up}^2\right) - \ln \left(\frac{1}{2}\frac{\rho^2}{{\beta}_{\max}^2} 2^{-2\eta}\right).
\end{align}
Second, we derive an upper bound for $\sum_{k=3}^{d-1} {\mathbb E}\left\{\frac{1}{k} \right\}$. We can see that the following result holds:
\begin{align}
\nonumber
\sum_{k=3}^{d-1} {\mathbb E}\left\{\frac{1}{k} \right\}& =\sum_{k=3}^{d-1} \int_k^{k+1} \frac{dz}{k} \le \sum_{k=3}^{d-1} \int_k^{k+1} \frac{dz}{z-1} \\
& = \int_3^d \frac{dz}{z-1} = \ln\left(d-1\right) -\ln2 < \ln\left(d-1\right). 
\end{align}
According to the above results and the fact that $1<\ln\left(d-1\right)$ when $d\ge5$, we can derive an upper bound on ${\mathbb E}\left\{ || {\bm \zeta}_{d-1}^{-\frac{1}{2}} {{\bm \varepsilon}}_{d-1} ||_2^2\right\}$ as follows:
\begin{align}
\nonumber
& {\mathbb E}\left\{ || {\bm \zeta}_{d-1}^{-\frac{1}{2}} {{\bm \varepsilon}}_{d-1} ||_2^2\right\}\le 
\frac{1}{2}\left({\overline \epsilon}-{\underline \epsilon}\right)^2 \\
\nonumber
& + \frac{1}{4}\left({\overline \epsilon}-{\underline \epsilon}\right)^2 \left(\ln \left(\left(d-1\right) p_{\rm up}^2\right) - \ln \left(\frac{1}{2}\frac{\rho^2}{{\beta}_{\max}^2} 2^{-2\eta}\right)+\ln\left(d-1\right)\right)\\
\nonumber
& = \frac{1}{4}\left({\overline \epsilon}-{\underline \epsilon}\right)^2 \left(2+\ln \left(p_{\rm up}^2\right) - \ln \left(\frac{\rho^2}{{\beta}_{\max}^2} 2^{-2\eta-1}\right)+2\ln\left(d-1\right)\right)\\
\nonumber
& = \frac{1}{2}\left({\overline \epsilon}-{\underline \epsilon}\right)^2 \left(1+\ln{\frac{p_{\rm up}{\beta}_{\max}2^{\eta+0.5}}{\rho} }+\ln\left(d-1\right)\right)\\
\nonumber
& \le \frac{1}{2}\left({\overline \epsilon}-{\underline \epsilon}\right)^2 \left(\left(1+\left|\ln{\frac{p_{\rm up}{\beta}_{\max}2^{\eta+0.5}}{\rho} }\right|\right)\ln\left(d-1\right)+\ln\left(d-1\right)\right)\\
& = \frac{1}{2}\left({\overline \epsilon}-{\underline \epsilon}\right)^2 \left(2+\left|\ln{\frac{p_{\rm up}{\beta}_{\max}2^{\eta+0.5}}{\rho} }\right|\right)\ln\left(d-1\right).
\end{align}

{\bf Step 5:} We derive an upper bound for ${\mathbb E}\left\{ || {\tilde{\bm \theta}}_{ij}^{d-1} - {\bm \theta}_{ij} ||_2^2 \right\}$. 

According to (\ref{app:equ:s2split}) of {\bf Step 2}, we can split an upper bound of ${\mathbb E}\left\{ || {\tilde{\bm \theta}}_{ij}^{d-1} - {\bm \theta}_{ij} ||_2^2 \right\}$ into two parts:
\begin{align}
{\mathbb E}\left\{ || {\tilde{\bm \theta}}_{ij}^{d-1} - {\bm \theta}_{ij} ||_2^2 \right\} \le \frac{1}{\lambda_S} {\mathbb E}\left\{ || {\bm \zeta}_{d-1}^{-\frac{1}{2}} {{\bm \varepsilon}}_{d-1} ||_2^2\right\}.
\end{align}

According to {\bf Step 3}, we have
\begin{align}
\nonumber
\frac{1}{\lambda_S} &<\left({1+p_{\rm up}^2}\right) \frac{2 {\beta}_{\max}^2}{\rho^2 2^{-2\eta}} \left({1-2\eta}\right)\left(\lfloor\frac{d-1}{2}\rfloor\right)^{2\eta-1} \frac{1}{1 - 2^{2\eta-1}} \\
& =\left({1+p_{\rm up}^2}\right) \frac{ {\beta}_{\max}^2}{\rho^2 } \left({1-2\eta}\right)\left(\frac{1}{\lfloor\frac{d-1}{2}\rfloor}\right)^{1-2\eta} \frac{4}{2^{1-2\eta} -1}.
\end{align}
Since $\lfloor\frac{d-1}{2}\rfloor\ge\frac{d-1}{4}$ for $d\ge5$ and $\eta<\frac{1}{2}$, we further have the following inequality:
\begin{align}
\frac{1}{\lambda_S} < \left({1+p_{\rm up}^2}\right) \frac{ {\beta}_{\max}^2}{\rho^2 } \left({1-2\eta}\right)\left(\frac{4}{{d-1}}\right)^{1-2\eta} \frac{4}{2^{1-2\eta} -1}.
\end{align}

According to {\bf Step 4}, we have
\begin{align}
\nonumber
& {\mathbb E}\left\{ || {\bm \zeta}_{d-1}^{-\frac{1}{2}} {{\bm \varepsilon}}_{d-1} ||_2^2\right\}\le 
\frac{1}{2}\left({\overline \epsilon}-{\underline \epsilon}\right)^2 \left(2+\left|\ln{\frac{p_{\rm up}{\beta}_{\max}2^{\eta+0.5}}{\rho} }\right|\right)\ln\left(d-1\right).
\end{align}
Then, we can derive an upper bound for ${\mathbb E}\left\{ || {\tilde{\bm \theta}}_{ij}^{d-1} - {\bm \theta}_{ij} ||_2^2 \right\}$ as:
\begin{align}
\nonumber
{\mathbb E}\left\{ || {\tilde{\bm \theta}}_{ij}^{d-1} - {\bm \theta}_{ij} ||_2^2 \right\}
< & \left({1+p_{\rm up}^2}\right) \frac{ {\beta}_{\max}^2}{\rho^2 } \left({1-2\eta}\right)\left(\frac{4}{{d-1}}\right)^{1-2\eta} \frac{4}{2^{1-2\eta} -1} \\
\nonumber
& \cdot \frac{1}{2}\left({\overline \epsilon}-{\underline \epsilon}\right)^2 \left(2+\left|\ln{\frac{p_{\rm up}{\beta}_{\max}2^{\eta+0.5}}{\rho} }\right|\right)\ln\left(d-1\right) \\
\nonumber
= & \left({1+p_{\rm up}^2}\right) \frac{ {\beta}_{\max}^2}{\rho^2 } \left({1-2\eta}\right) \frac{ 4^{1.5-2\eta}}{2^{1-2\eta} -1} \\
& \cdot \left({\overline \epsilon}-{\underline \epsilon}\right)^2 \left(2+\left|\ln{\frac{p_{\rm up}{\beta}_{\max}2^{\eta+0.5}}{\rho} }\right|\right)\frac{\ln\left(d-1\right)}{\left(d-1\right)^{1-2\eta}}.
\end{align}
Therefore, we can define $\Phi_1\left(\rho,\eta\right)$ as follows:
\begin{align}
\nonumber
\Phi_1\left(\rho,\eta\right)\triangleq &\left({1+p_{\rm up}^2}\right) \frac{ {\beta}_{\max}^2}{\rho^2 } \left({1-2\eta}\right) \frac{ 4^{1.5-2\eta}}{2^{1-2\eta} -1} \\
& \cdot \left({\overline \epsilon}-{\underline \epsilon}\right)^2 \left(2+\left|\ln{\frac{p_{\rm up}{\beta}_{\max}2^{\eta+0.5}}{\rho} }\right|\right).
\end{align}
We can see that $\Phi_1\left(\rho,\eta\right)$ is finite and positive for all $\rho\in\left(0,\infty\right)$ and $\eta\in\left(0,\frac{1}{2}\right)$. Furthermore, $\Phi_1\left(\rho,\eta\right)\frac{\ln\left(d-1\right)}{\left(d-1\right)^{1-2\eta}}$ is an upper bound on ${\mathbb E}\left\{ || {\tilde{\bm \theta}}_{ij}^{d-1} - {\bm \theta}_{ij} ||_2^2 \right\}$. Based on our discussion at the beginning of our proof in this section, $\Phi_1\left(\rho,\eta\right)\frac{\ln\left(d-1\right)}{\left(d-1\right)^{1-2\eta}}$ is also an upper bound on ${\mathbb E}\left\{ || {\hat{\bm \theta}}_{ij}^{d-1} - {\bm \theta}_{ij} ||_2^2 \right\}$. 

According to L'Hospital's rule, we can see that as $d$ goes to infinity, we have
\begin{align}
\nonumber
& \lim_{d\rightarrow \infty} \Phi_1\left(\rho,\eta\right)\frac{\ln\left(d-1\right)}{\left(d-1\right)^{1-2\eta}} = \lim_{d\rightarrow \infty} \Phi_1\left(\rho,\eta\right) \frac{\frac{1}{d-1}}{\left(1-2\eta\right) \left(d-1\right)^{-2\eta}} \\
& = \lim_{d\rightarrow \infty} \Phi_1\left(\rho,\eta\right) \frac{1}{\left(1-2\eta\right) \left(d-1\right)^{1-2\eta}}=0.
\end{align}
Since $\Phi_1\left(\rho,\eta\right)\frac{\ln\left(d-1\right)}{\left(d-1\right)^{1-2\eta}}$ is an upper bound on ${\mathbb E}\left\{ || {\hat{\bm \theta}}_{ij}^{d-1} - {\bm \theta}_{ij} ||_2^2 \right\}$, we can see that ${\mathbb E}\left\{ || {\hat{\bm \theta}}_{ij}^{d-1} - {\bm \theta}_{ij} ||_2^2 \right\}$ approaches zero as $d$ goes to infinity. This completes our proof. 

\section{Proof of Proposition \ref{proposition:price}}\label{app:sec:effective}
Recall that problem (\ref{problem:1trans}) is as follows:
\begin{subequations}
\begin{align}
\nonumber
& \max \sum_{i\in{\mathcal N}} \sum_{j\in{\mathcal N}\setminus\left\{i\right\}} \xi_{ij} \left(\alpha_{ij} - \beta_{ij} p_{ij}^d + \epsilon_{ij}^-\right) p_{ij}^d \\
& {~~}{~~}{~~}{~~}{~~}{~~}{~~} - \sum_{i\in{\mathcal N}} \sum_{j\in{\mathcal N}\setminus\left\{i\right\}} \xi_{ij} \left(\alpha_{ij} - \beta_{ij} p_{ij}^d\right) c  \\
& {\rm s.t.} \!\sum_{j\in{\mathcal N}\setminus\left\{i\right\}} \left(\alpha_{ij} - \beta_{ij} p_{ij}^d\right) \!=\!\!\!\! \sum_{j\in{\mathcal N}\setminus\left\{i\right\}} \left(\alpha_{ji} - \beta_{ji} p_{ji}^d\right), \forall i\in{\mathcal N}, \\
& {\rm var.}  {~~} p_{ij}^d \le p_{\max}, \forall i\ne j, i,j\in{\mathcal N}. 
\end{align}
\end{subequations}

First, we can see that the problem is a convex problem. Specifically, the objective function is a quadratic and concave function of the pricing decisions, and the flow balance constraints are affine. Hence, the KKT conditions are sufficient and necessary for optimality. Recall that we use $\mu_{ij}^*$ to denote the optimal dual variable associated with $p_{ij}^d \le p_{\max}$ for each $\left(i,j\right)$. We further use $\sigma_i^*$ to denote the optimal dual variable associated with $\sum_{j\in{\mathcal N}\setminus\left\{i\right\}} \left(\alpha_{ij} - \beta_{ij} p_{ij}^d\right) = \sum_{j\in{\mathcal N}\setminus\left\{i\right\}} \left(\alpha_{ji} - \beta_{ji} p_{ji}^d\right)$ for each $i$.

Second, according to the stationarity condition, we have the following relation:
\begin{multline}
-\xi_{ij} \left(\alpha_{ij} - \beta_{ij} p_{ij}^*\left({\bm \theta}\right) + \epsilon_{ij}^-\right) + \xi_{ij} \beta_{ij} p_{ij}^*\left({\bm \theta}\right) - \xi_{ij}\beta_{ij} c \\
- \sigma_i^* \beta_{ij} + \sigma_j^* \beta_{ij} + \mu_{ij}^* =0.
\end{multline}
After rearrangement, we can get the following result for each $\left(i,j\right)$:
\begin{align}
p_{ij}^*\left({\bm \theta}\right) 
=\frac{\xi_{ij} \alpha_{ij} + \xi_{ij} \epsilon_{ij}^- + \xi_{ij}\beta_{ij} c + \left(\sigma_i^*- \sigma_j^*\right) \beta_{ij} - \mu_{ij}^*}{2\xi_{ij}\beta_{ij}}.
\end{align}
When $\mu_{ij}^*=0$ for all $\left(i,j\right)$, we have
\begin{align}
\nonumber
p_{ij}^*\left({\bm \theta}\right) 
& =\frac{\xi_{ij} \alpha_{ij} + \xi_{ij} \epsilon_{ij}^- + \xi_{ij}\beta_{ij} c + \left(\sigma_i^*- \sigma_j^*\right) \beta_{ij} }{2\xi_{ij}\beta_{ij}}\\
& = \frac{\alpha_{ij} + \epsilon_{ij}^- + \beta_{ij} c }{2\beta_{ij}} + \frac{\sigma_i^*- \sigma_j^*}{2\xi_{ij}}.\label{app:equ:pexpre}
\end{align}

Third, we prove that $\left\{\sigma_i^*\right\}_{i\in{\mathcal N}}$ satisfies a system of linear equations, whose coefficient matrix is a Laplacian matrix. Note that $p_{ij}^*\left({\bm \theta}\right)$ satisfies the flow balance constraints. Hence, we can plug the expression of $p_{ij}^*\left({\bm \theta}\right)$ in (\ref{app:equ:pexpre}) into the flow balance constraints, and get the following relation:
\begin{align}
\nonumber
& \sum_{j\in{\mathcal N}\setminus\left\{i\right\}} \left(\alpha_{ij} - \beta_{ij} \left(\frac{\alpha_{ij} + \epsilon_{ij}^- + \beta_{ij} c }{2\beta_{ij}} + \frac{\sigma_i^*- \sigma_j^*}{2\xi_{ij}}\right)\right) \\
& = \sum_{j\in{\mathcal N}\setminus\left\{i\right\}} \left(\alpha_{ji} - \beta_{ji} \left(\frac{\alpha_{ji} + \epsilon_{ji}^- + \beta_{ji} c }{2\beta_{ji}} + \frac{\sigma_j^*- \sigma_i^*}{2\xi_{ji}}\right)\right), \forall i\in{\mathcal N}.
\end{align}
After rearrangement, we can get the following result for all $i\in{\mathcal N}$:
\begin{align}
\nonumber
& \sum_{j\in{\mathcal N}\setminus\left\{i\right\}} \left(\frac{\beta_{ij}}{\xi_{ij}}+\frac{\beta_{ji}}{\xi_{ji}}\right) \left(\sigma_i^*-\sigma_j^*\right) \\
\nonumber
& =  \sum_{j\in{\mathcal N}\setminus\left\{i\right\}} \left(\alpha_{ij} - \epsilon_{ij}^- - \beta_{ij}c\right) - \sum_{j\in{\mathcal N}\setminus\left\{i\right\}} \left(\alpha_{ji} - \epsilon_{ji}^- - \beta_{ji}c\right)\\
& = v_i\left({\bm \theta}\right).\label{app:equ:linearequa}
\end{align}
Recall that function $v_i\left({\bm \theta}\right)$ is defined in Proposition \ref{proposition:price}. Then, we define a matrix $\bm L$, whose $ij$-th entry is defined as
\begin{align}
l_{ij} \triangleq \left\{ {\begin{array}{*{20}{l}}
{\sum_{k\in{\mathcal N}\setminus\left\{i\right\}} \left(\frac{\beta_{ik}}{\xi_{ik}}+\frac{\beta_{ki}}{\xi_{ki}}\right),}&{{\rm if~}i=j,}\\
{-\frac{\beta_{ij}}{\xi_{ij}}-\frac{\beta_{ji}}{\xi_{ji}},}&{{\rm if~}i\ne j.}\\
\end{array}} \right.
\end{align}
We further define ${\bm \sigma}^*\triangleq \left(\sigma_i^*,\forall i\in{\mathcal N}\right)^T$ and ${\bm v}\left({\bm \theta}\right) \triangleq \left(v_i\left({\bm \theta}\right),\forall i\in{\mathcal N}\right)^T$, which are two column vectors. We can rewrite (\ref{app:equ:linearequa}) as follows:
\begin{align}
{\bm L} {\bm \sigma}^* = {\bm v}\left({\bm \theta}\right).\label{app:equ:simplele}
\end{align}
Note that $\bm L$ is the Laplacian matrix of a weighted undirected graph. Specifically, there exists an edge $\left(i,j\right)$ between any two different nodes $i$ and $j$, and edge $\left(i,j\right)$ is associated with a weight, which is $\frac{\beta_{ij}}{\xi_{ij}}+\frac{\beta_{ji}}{\xi_{ji}}$. 

Fourth, we show that the generalized inverse of the Laplacian matrix $\bm L$ has a strong connection with a resistor network. 
Since $\bm L$ is an $N\times N$ Laplacian matrix, the rank of ${\bm L}$ is $N-1$, and ${\bm L}$ is non-invertible. As a substitute for the inverse, we can consider the generalized inverse of ${\bm L}$ \cite{dorfler2018electrical}, and denote it by ${\bm L}^+$. Using the notion of ${\bm L}^+$, we can prove that the solution space of (\ref{app:equ:simplele}) is as follows:
\begin{align}
\left\{{\bm \sigma}: {\bm \sigma}={\bm L}^+ {\bm v}\left({\bm \theta}\right)   + \gamma \left(1,1,\ldots,1\right)^T, \gamma\in{\mathbb R}\right\}.
\end{align}
The concrete proof of the above result is the same as the proof of Proposition 3.2 of our prior work \cite{yu2019analyzing}. Hence, we skip the concrete proof procedure here. 

Based on the solution space of (\ref{app:equ:simplele}), we can easily see that
\begin{align}
\sigma_i^*-\sigma_j^*=\sum_{k\in{\mathcal N}} \left(l_{ik}^+ - l_{jk}^+\right) v_k\left({\bm \theta}\right),\forall i\ne j,i,j\in{\mathcal N}.\label{app:equ:sigmasigma}
\end{align}
Next, we show that the matrix ${\bm L}^+$ has a strong connection with a resistor network. As introduced in Section \ref{subsub:ressolution:a}, we can construct a resistor network based on the traffic network. Specifically, we replace the links between locations with resistors. For all $i,j\in{\mathcal N}$ with $i< j$, we replace the links $\left(i,j\right)$ and $\left(j,i\right)$ with a resistor, whose resistance is given by $r_{ij}= \frac{1}{\frac{\beta_{ij}}{\xi_{ij}}+\frac{\beta_{ji}}{\xi_{ji}}}$. Recall that we use $R_{ij}\left({\bm \beta}\right)$ to denote the effective resistance between nodes $i$ and $j$ in the constructed resistor network. 

The effective resistances in the resistor network have the following relation with ${\bm L}^+$:
\begin{align}
R_{ij}\left({\bm \beta}\right)=l_{ii}^+ + l_{jj}^+ - 2l_{ij}^+,\forall i,j\in{\mathcal N}.\label{app:equ:RandL}
\end{align}
Readers can refer to the following paper for more details of such a connection between the effective resistances and the generalized inverse of the Laplacian matrix: Gyan Ranjan, Zhi-Li Zhang, and Daniel Boley. 2014. Incremental computation of pseudo-inverse of {L}aplacian. In \emph{Proc. of COCOA}. Wailea, HI, USA, 729--749.

According to (\ref{app:equ:RandL}), we have the following relations for all $i,j,k\in{\mathcal N}$:
\begin{align}
& l_{ik}^+ =\frac{l_{ii}^+ + l_{kk}^+ - R_{ik}\left({\bm \beta}\right)}{2},\\
& l_{jk}^+ =\frac{l_{jj}^+ + l_{kk}^+ - R_{jk}\left({\bm \beta}\right)}{2}.
\end{align}
We further utilize (\ref{app:equ:sigmasigma}) and the above two equalities to get the following result:
\begin{align}
\nonumber
\sigma_i^*-\sigma_j^*=&\sum_{k\in{\mathcal N}} \left(\frac{l_{ii}^+ + l_{kk}^+ - R_{ik}\left({\bm \beta}\right)}{2} - \frac{l_{jj}^+ + l_{kk}^+ - R_{jk}\left({\bm \beta}\right)}{2}\right) v_k\left({\bm \theta}\right) \\
=& \sum_{k\in{\mathcal N}} \left(\frac{l_{ii}^+ - l_{jj}^+ + R_{jk}\left({\bm \beta}\right) - R_{ik}\left({\bm \beta}\right)}{2} \right) v_k\left({\bm \theta}\right).
\end{align}
Note that $\sum_{k\in{\mathcal N}} v_k\left({\bm \theta}\right)=0$. Hence, we have $\sum_{k\in{\mathcal N}} l_{ii}^+ v_k\left({\bm \theta}\right)=0$ and $\sum_{k\in{\mathcal N}} l_{jj}^+ v_k\left({\bm \theta}\right)=0$. Therefore, we can get the following result:
\begin{align}
\sigma_i^*-\sigma_j^*= \frac{1}{2}\sum_{k\in{\mathcal N}} \left({R_{jk}\left({\bm \beta}\right) - R_{ik}\left({\bm \beta}\right)} \right) v_k\left({\bm \theta}\right).\label{app:equ:diffsigma}
\end{align}

Last, we can utilize our results in (\ref{app:equ:pexpre}) and (\ref{app:equ:diffsigma}) to get the expression of $p_{ij}^*\left({\bm \theta}\right)$:
\begin{align}
\nonumber
p_{ij}^*\left({\bm \theta}\right) = &\frac{\alpha_{ij} + \epsilon_{ij}^- + \beta_{ij} c }{2\beta_{ij}} + \frac{\sigma_i^*- \sigma_j^*}{2\xi_{ij}} \\
=& \frac{\alpha_{ij} + \epsilon_{ij}^- + \beta_{ij} c }{2\beta_{ij}} + \frac{1}{4\xi_{ij}}\sum_{k\in{\mathcal N}} \left({R_{jk}\left({\bm \beta}\right) - R_{ik}\left({\bm \beta}\right)} \right) v_k\left({\bm \theta}\right).
\end{align}
This completes our proof of Proposition \ref{proposition:price}.

\section{Sufficient Condition for $\mu_{ij}^*=0$}\label{app:sec:sufficient}
In this section, we prove that when the following sufficient condition holds, we have $\mu_{ij}^*=0$ for all $\left(i,j\right)$:
\begin{align}
\sum_{k\in{\mathcal N}} \left|v_k\left({\bm \theta}\right) \right| \le \min_{\left(i,j\right):i\ne j,i,j\in{\mathcal N}} 2 \left(\beta_{ij} + \frac{\xi_{ij}}{\xi_{ji}}\beta_{ji}\right) \left(2p_{\max} -c - \frac{\alpha_{ij} + \epsilon_{ij}^-}{\beta_{ij}}\right).\label{app:equ:suffcondition}
\end{align}

According to the properties of effective resistances, the effective resistances satisfy the triangle inequality. Hence, we have $R_{jk}\left({\bm \beta}\right)-R_{ik}\left({\bm \beta}\right)\le R_{ij}\left({\bm \beta}\right)$ and $R_{ik}\left({\bm \beta}\right)-R_{jk}\left({\bm \beta}\right)\le R_{ij}\left({\bm \beta}\right)$ for any $i,j,k\in{\mathcal N}$. Furthermore, since the effective resistance between two locations is no greater than the resistance of the resistor between them, we have $R_{ij}\left({\bm \beta}\right)\le r_{ij}=\frac{1}{\frac{\beta_{ij}}{\xi_{ij}}+\frac{\beta_{ji}}{\xi_{ji}}}$. Then, we can get the following inequality for any $i,j,k\in{\mathcal N}$:
\begin{align}
\left| R_{jk}\left({\bm \beta}\right)-R_{ik}\left({\bm \beta}\right) \right| \le \frac{1}{\frac{\beta_{ij}}{\xi_{ij}}+\frac{\beta_{ji}}{\xi_{ji}}}.
\end{align}

By using the above inequality, we can show the following relation:
\begin{align}
\nonumber 
& \frac{\alpha_{ij} + \epsilon_{ij}^- + \beta_{ij} c }{2\beta_{ij}} + \frac{1}{4\xi_{ij}}\sum_{k\in{\mathcal N}} \left({R_{jk}\left({\bm \beta}\right) - R_{ik}\left({\bm \beta}\right)} \right) v_k\left({\bm \theta}\right) \\
\nonumber 
& \le \frac{\alpha_{ij} + \epsilon_{ij}^- + \beta_{ij} c }{2\beta_{ij}} + \frac{1}{4\xi_{ij}}\sum_{k\in{\mathcal N}} \left|{R_{jk}\left({\bm \beta}\right) - R_{ik}\left({\bm \beta}\right)} \right| \left|v_k\left({\bm \theta}\right) \right| \\
\nonumber
& \le \frac{\alpha_{ij} + \epsilon_{ij}^- + \beta_{ij} c }{2\beta_{ij}} + \frac{1}{4\xi_{ij}}\frac{1}{\frac{\beta_{ij}}{\xi_{ij}}+\frac{\beta_{ji}}{\xi_{ji}}}\sum_{k\in{\mathcal N}} \left|v_k\left({\bm \theta}\right) \right|. 
\end{align}
When the condition in (\ref{app:equ:suffcondition}), we can further derive the following relation:
\begin{align}
\nonumber 
& \frac{\alpha_{ij} + \epsilon_{ij}^- + \beta_{ij} c }{2\beta_{ij}} + \frac{1}{4\xi_{ij}}\sum_{k\in{\mathcal N}} \left({R_{jk}\left({\bm \beta}\right) - R_{ik}\left({\bm \beta}\right)} \right) v_k\left({\bm \theta}\right) \\
\nonumber 
& \le \frac{\alpha_{ij} + \epsilon_{ij}^- + \beta_{ij} c }{2\beta_{ij}} \\
\nonumber
& + \frac{1}{2\xi_{ij}}\frac{1}{\frac{\beta_{ij}}{\xi_{ij}}+\frac{\beta_{ji}}{\xi_{ji}}} \min_{\left({\tilde i},{\tilde j}\right):{\tilde i}\ne {\tilde j},{\tilde i},{\tilde j}\in{\mathcal N}} \left(\beta_{{\tilde i}{\tilde j}} + \frac{\xi_{{\tilde i}{\tilde j}}}{\xi_{{\tilde j}{\tilde i}}}\beta_{{\tilde j}{\tilde i}}\right) \left(2p_{\max} -c - \frac{\alpha_{{\tilde i}{\tilde j}} + \epsilon_{{\tilde i}{\tilde j}}^-}{\beta_{{\tilde i}{\tilde j}}}\right) \\
\nonumber 
& \le \frac{\alpha_{ij} + \epsilon_{ij}^- + \beta_{ij} c }{2\beta_{ij}} \\
\nonumber
& + \frac{1}{2}\frac{1}{\beta_{ij}+\frac{\xi_{ij}\beta_{ji}}{\xi_{ji}}} \left(\beta_{ij} + \frac{\xi_{ij}}{\xi_{ji}}\beta_{ji}\right) \left(2p_{\max} -c - \frac{\alpha_{ij} + \epsilon_{ij}^-}{\beta_{ij}}\right)\\
\nonumber
& \le \frac{\alpha_{ij} + \epsilon_{ij}^- + \beta_{ij} c }{2\beta_{ij}} + \left(p_{\max} -\frac{1}{2}c - \frac{\alpha_{ij} + \epsilon_{ij}^-}{2\beta_{ij}}\right)  \\
& \le p_{\max}.
\end{align}

We let $p_{ij}=\frac{\alpha_{ij} + \epsilon_{ij}^- + \beta_{ij} c }{2\beta_{ij}} + \frac{1}{4\xi_{ij}}\sum_{k\in{\mathcal N}} \left({R_{jk}\left({\bm \beta}\right) - R_{ik}\left({\bm \beta}\right)} \right) v_k\left({\bm \theta}\right)$ and $\mu_{ij}=0$ for all $\left(i,j\right)$, and let $\bm \sigma$ be the solution to ${\bm L}{\bm \sigma}={\bm v}\left({\bm \theta}\right)$. From our analysis above, we can see that $p_{ij}\le p_{\max}$ for all $\left(i,j\right)$. We can also verify that $\left\{p_{ij},\mu_{ij}\right\}_{i,j\in{\mathcal N},i\ne j}$ and $\bm \sigma$ satisfy the KKT conditions. This implies that they constitute an optimal solution to the optimization problem, which completes our proof.  

\section{Regret Analysis When ${\hat \mu}_{ij}^{d-1,*}\ne0$ at The Beginning}\label{app:sec:reason}
Recall that in Section \ref{subsub:discussmu}, we mention that when $\mu_{ij}^*=0$ for all $\left(i,j\right)$, ${\hat \mu}_{ij}^{d-1,*}$ may be positive for some $\left(i,j\right)$ at the beginning and will become zero for all $\left(i,j\right)$ after several days. We claim that in this case, we can still prove that $\lim_{D\rightarrow \infty} \Delta_D^{\bm \pi}=0$. We explain the reason in this section. 

Recall that $\Delta_D^{\bm \pi}$ is defined as follows:
\begin{align}
\Delta_D^{\bm \pi}= {\mathbb E}^{\bm \pi} \!\left\{ \frac{1}{D} \sum_{d=1}^D \!\Bigg( \Pi\left({\bm p}^*\left({\bm \theta}\right),{\bm w}^*\left({\bm \theta}\right),{\bm \epsilon}^d\right) \! -\! \Pi\left({\bm p}^d,{\bm w}^d,{\bm \epsilon}^d\right) \Bigg) \right\},
\end{align}
where ${\bm p}^d$ and ${\bm w}^d$ are the decisions under the policy $\bm \pi$. 

When $\mu_{ij}^*=0$ for all $\left(i,j\right)$, ${\hat \mu}_{ij}^{d-1,*}$ becomes zero for all $\left(i,j\right)$ after several days and no longer changes (as shown in Section \ref{sec:numerical}). We use $d_{\rm Th}\in\left\{1,2,\ldots\right\}$ to denote the \emph{threshold day}, which is defined as follows:
\begin{align}
d_{\rm Th} \triangleq \min\left\{{\tilde d}: {\hat \mu}_{ij}^{d-1,*}=0 {\rm ~for~all~}\left(i,j\right){\rm~and~all~}d\ge{\tilde d}\right\}.
\end{align}

When $D$ approaches infinity, we can rewrite $\Delta_D^{\bm \pi}$ as follows:
\begin{align}
\nonumber
\Delta_D^{\bm \pi}= & {\mathbb E}^{\bm \pi} \!\left\{ \frac{1}{D} \sum_{d=1}^{d_{\rm Th}-1} \!\Bigg( \Pi\left({\bm p}^*\left({\bm \theta}\right),{\bm w}^*\left({\bm \theta}\right),{\bm \epsilon}^d\right) \! -\! \Pi\left({\bm p}^d,{\bm w}^d,{\bm \epsilon}^d\right) \Bigg) \right\} \\
& + {\mathbb E}^{\bm \pi} \!\left\{ \frac{1}{D} \sum_{d=d_{\rm Th}}^D \!\Bigg( \Pi\left({\bm p}^*\left({\bm \theta}\right),{\bm w}^*\left({\bm \theta}\right),{\bm \epsilon}^d\right) \! -\! \Pi\left({\bm p}^d,{\bm w}^d,{\bm \epsilon}^d\right) \Bigg) \right\}.
\end{align}
We can see that as $D$ approaches infinity, the first term on the right side will become zero. Then, we have the following relation:
{\small
\begin{align}
\nonumber
\lim_{D\rightarrow\infty}\Delta_D^{\bm \pi}= \lim_{D\rightarrow\infty} {\mathbb E}^{\bm \pi} \!\left\{ \frac{1}{D} \sum_{d=d_{\rm Th}}^D \!\Bigg( \Pi\left({\bm p}^*\left({\bm \theta}\right),{\bm w}^*\left({\bm \theta}\right),{\bm \epsilon}^d\right) \! -\! \Pi\left({\bm p}^d,{\bm w}^d,{\bm \epsilon}^d\right) \Bigg) \right\}.
\end{align}}
That is to say, although ${\hat \mu}_{ij}^{d-1,*}\ne0$ for some $\left(i,j\right)$ and some $d< d_{\rm Th}$, it does not affect our analysis of $\lim_{D\rightarrow\infty}\Delta_D^{\bm \pi}$. 
Based on the definition of $d_{\rm Th}$, we have ${\hat \mu}_{ij}^{d-1,*}=0$ for all $\left(i,j\right)$ and all $d\ge d_{\rm Th}$. Then, we can still apply our proofs for Theorem \ref{theorem:key} and Corollary \ref{corollary:only} to show that as $D$ approaches infinity, the time-average regret for the days from $d=d_{\rm Th}$ to $d=D$ is zero:
{\small
\begin{align}
\lim_{D\rightarrow\infty} {\mathbb E}^{\bm \pi} \!\left\{ \frac{1}{D\!-\!d_{\rm Th}\!+\!1}\!\!\! \sum_{d=d_{\rm Th}}^D \!\Bigg( \Pi\left({\bm p}^*\left({\bm \theta}\right),{\bm w}^*\left({\bm \theta}\right),{\bm \epsilon}^d\right) \! -\! \Pi\left({\bm p}^d,{\bm w}^d,{\bm \epsilon}^d\right) \Bigg) \right\}\!=\!0.
\end{align}}
Then, we can easily prove that $\lim_{D\rightarrow\infty}\Delta_D^{\bm \pi}=0$.

\section{Proof of Theorem \ref{theorem:key}}\label{app:sec:keythe}
In this section, we derive an upper bound for $\Delta_D^{\bm \pi}$. We conduct the derivation by the following steps.

{\bf Step 1:} We analyze an upper bound for $\left|R_{jk}\left({\bm \beta}\right)-R_{jk}\left(\hat {\bm \beta}^{d-1}\right)\right|$.

Recall that $R_{jk}\left({\bm \beta}\right)$ is the effective resistance between nodes $j$ and $k$ when the resistor network is defined based on ${\bm \beta}$, and $R_{jk}\left(\hat {\bm \beta}^{d-1}\right)$ is the effective resistance between $j$ and $k$ when the resistor network is defined based on $\hat {\bm \beta}^{d-1}$. According to our discussion in Section \ref{subsub:discussmu}, we can derive $p_{ij}^*\left({\hat{\bm \theta}}^{d-1}\right)$ using ${\hat \beta}_{ij}^{d-1}$, ${\hat \alpha}_{ij}^{d-1}$, $R_{jk}\left(\hat {\bm \beta}^{d-1}\right)$, $R_{ik}\left(\hat {\bm \beta}^{d-1}\right)$, and $v_k\left({\hat {\bm \theta}}^{d-1}\right)$. In order to analyze $\left|p_{ij}^*\left({\bm \theta}\right) - p_{ij}^*\left({\hat{\bm \theta}}^{d-1}\right)\right|$, we first analyze $\left|R_{jk}\left({\bm \beta}\right)-R_{jk}\left(\hat {\bm \beta}^{d-1}\right)\right|$ in this step. 

Next, we prove the following relation:
\begin{align}
\nonumber
& \left|R_{jk}\left({\bm \beta}\right)-R_{jk}\left(\hat {\bm \beta}^{d-1}\right)\right| \\ 
\le 
& \frac{1}{2} \sum_{m\in{\mathcal N}} \sum_{n\in{\mathcal N}\setminus \left\{m\right\}} \left| \frac{1}{\frac{\beta_{mn}}{\xi_{mn}}+\frac{\beta_{nm}}{\xi_{nm}}} - \frac{1}{\frac{{\hat \beta_{mn}}^{d-1}}{\xi_{mn}}+\frac{{\hat\beta_{nm}}^{d-1}}{\xi_{nm}}} \right|.
\end{align}
Recall that in the resistor networks defined by ${\bm \beta}$ and $\hat {\bm \beta}^{d-1}$, the \emph{resistances} of the resistor between any two different nodes $m$ and $n$ are $\frac{1}{\frac{\beta_{mn}}{\xi_{mn}}+\frac{\beta_{nm}}{\xi_{nm}}}$ and $\frac{1}{\frac{{\hat \beta_{mn}}^{d-1}}{\xi_{mn}}+\frac{{\hat\beta_{nm}}^{d-1}}{\xi_{nm}}}$, respectively. We focus on the resistor network defined by ${\bm \beta}$ and $R_{jk}\left({\bm \beta}\right)$ with $j\ne k$. If we change the resistance of the resistor between two particular nodes $m$ and $n$ (where $m$ and $n$ can be any two different nodes including $j$ and $k$) from $\frac{1}{\frac{\beta_{mn}}{\xi_{mn}}+\frac{\beta_{nm}}{\xi_{nm}}}$ to $\frac{1}{\frac{{\hat \beta_{mn}}^{d-1}}{\xi_{mn}}+\frac{{\hat\beta_{nm}}^{d-1}}{\xi_{nm}}}$, then the \emph{effective resistance} between $j$ and $k$ will deviate from $R_{jk}\left({\bm \beta}\right)$ by at most $ \left| \frac{1}{\frac{\beta_{mn}}{\xi_{mn}}+\frac{\beta_{nm}}{\xi_{nm}}} - \frac{1}{\frac{{\hat \beta_{mn}}^{d-1}}{\xi_{mn}}+\frac{{\hat\beta_{nm}}^{d-1}}{\xi_{nm}}} \right|$ (we can prove this using Thomson's principle). We can repeat the above analysis. After changing the resistance of all the resistors in the network (i.e., for \emph{each} pair $\left(m,n\right)$, the resistance is changed from $\frac{1}{\frac{\beta_{mn}}{\xi_{mn}}+\frac{\beta_{nm}}{\xi_{nm}}}$ to $\frac{1}{\frac{{\hat \beta_{mn}}^{d-1}}{\xi_{mn}}+\frac{{\hat\beta_{nm}}^{d-1}}{\xi_{nm}}}$), we can show that the effective resistance between $j$ and $k$ will deviate from $R_{jk}\left({\bm \beta}\right)$ by at most $\sum_{m\in{\mathcal N}} \sum_{n>m,n\in{\mathcal N}} \left| \frac{1}{\frac{\beta_{mn}}{\xi_{mn}}+\frac{\beta_{nm}}{\xi_{nm}}} - \frac{1}{\frac{{\hat \beta_{mn}}^{d-1}}{\xi_{mn}}+\frac{{\hat\beta_{nm}}^{d-1}}{\xi_{nm}}} \right|$. Note that we consider the condition $n>m$ in the inner summation to avoid counting each pair $\left(m,n\right)$ twice. Formally, we get the following relation:
\begin{align}
\nonumber
& \left|R_{jk}\left({\bm \beta}\right)-R_{jk}\left(\hat {\bm \beta}^{d-1}\right)\right| \\ 
\nonumber
\le & 
\sum_{m\in{\mathcal N}} \sum_{n>m,n\in{\mathcal N}\setminus \left\{m\right\}} \left| \frac{1}{\frac{\beta_{mn}}{\xi_{mn}}+\frac{\beta_{nm}}{\xi_{nm}}} - \frac{1}{\frac{{\hat \beta_{mn}}^{d-1}}{\xi_{mn}}+\frac{{\hat\beta_{nm}}^{d-1}}{\xi_{nm}}} \right| \\
= & \frac{1}{2} \sum_{m\in{\mathcal N}} \sum_{n\in{\mathcal N}\setminus \left\{m\right\}} \left| \frac{1}{\frac{\beta_{mn}}{\xi_{mn}}+\frac{\beta_{nm}}{\xi_{nm}}} - \frac{1}{\frac{{\hat \beta_{mn}}^{d-1}}{\xi_{mn}}+\frac{{\hat\beta_{nm}}^{d-1}}{\xi_{nm}}} \right|.
\end{align}
We can further derive the following result:
\begin{align}
\nonumber
& \left|R_{jk}\left({\bm \beta}\right)\!-\!R_{jk}\left(\hat {\bm \beta}^{d-1}\right)\right| \! \le \! \frac{1}{2} \sum_{m\in{\mathcal N}}\! \sum_{n\in{\mathcal N}\setminus \left\{m\right\}} \left| \frac{1}{\frac{\beta_{mn}}{\xi_{mn}}+\frac{\beta_{nm}}{\xi_{nm}}} - \frac{1}{\frac{{\hat \beta_{mn}}^{d-1}}{\xi_{mn}}+\frac{{\hat\beta_{nm}}^{d-1}}{\xi_{nm}}} \right| \\
\nonumber
= & \frac{1}{2} \sum_{m\in{\mathcal N}} \sum_{n\in{\mathcal N}\setminus \left\{m\right\}} \left| \frac{\xi_{mn}\xi_{nm}}{{\beta_{mn}}{\xi_{nm}}+{\beta_{nm}}{\xi_{mn}}} - \frac{\xi_{mn}\xi_{nm}}{{{\hat \beta_{mn}}^{d-1}}{\xi_{nm}}+{{\hat\beta_{nm}}^{d-1}}{\xi_{mn}}} \right| \\
\nonumber
= & \frac{1}{2} \sum_{m\in{\mathcal N}} \sum_{n\in{\mathcal N}\setminus \left\{m\right\}} \frac{\left| \xi_{mn}\xi_{nm}^2 \left({\hat \beta_{mn}}^{d-1} - \beta_{mn}\right) + \xi_{mn}^2 \xi_{nm} \left({\hat\beta_{nm}}^{d-1}-\beta_{nm}\right) \right|}{\left({\beta_{mn}}{\xi_{nm}}+{\beta_{nm}}{\xi_{mn}}\right)\left({{\hat \beta_{mn}}^{d-1}}{\xi_{nm}}+{{\hat\beta_{nm}}^{d-1}}{\xi_{mn}}\right)} \\
\nonumber
\le & \frac{1}{2} \sum_{m\in{\mathcal N}} \sum_{n\in{\mathcal N}\setminus \left\{m\right\}} \frac{\left| \xi_{mn}\xi_{nm}^2 \left({\hat \beta_{mn}}^{d-1} - \beta_{mn}\right) \right| + \left| \xi_{mn}^2 \xi_{nm} \left({\hat\beta_{nm}}^{d-1}-\beta_{nm}\right) \right|}{\beta_{\min}^2 \left(\xi_{mn}+\xi_{nm}\right)^2} \\
\le & \frac{1}{2} \sum_{m\in{\mathcal N}} \sum_{n\in{\mathcal N}\setminus \left\{m\right\}} \frac{\xi_{mn}\xi_{nm} \left(\xi_{nm} || {\hat{\bm \theta}}_{mn}^{d-1} - {\bm \theta}_{mn} ||_2 + \xi_{mn} || {\hat{\bm \theta}}_{nm}^{d-1} - {\bm \theta}_{nm} ||_2\right)}{\beta_{\min}^2 \left(\xi_{mn}+\xi_{nm}\right)^2}. \label{app:equ:gapofRR}
\end{align}
Note that we can do the following rearrangement:
\begin{align}
\nonumber
& \sum_{m\in{\mathcal N}} \sum_{n\in{\mathcal N}\setminus \left\{m\right\}} \frac{\xi_{mn}\xi_{nm} \xi_{mn} || {\hat{\bm \theta}}_{nm}^{d-1} - {\bm \theta}_{nm} ||_2}{\beta_{\min}^2 \left(\xi_{mn}+\xi_{nm}\right)^2} \\
\nonumber
= & \sum_{n\in{\mathcal N}} \sum_{m\in{\mathcal N}\setminus \left\{n\right\}} \frac{\xi_{nm}\xi_{mn} \xi_{nm} || {\hat{\bm \theta}}_{mn}^{d-1} - {\bm \theta}_{mn} ||_2}{\beta_{\min}^2 \left(\xi_{nm}+\xi_{mn}\right)^2} \\
= & \sum_{m\in{\mathcal N}} \sum_{n\in{\mathcal N}\setminus \left\{m\right\}} \frac{\xi_{nm}\xi_{mn} \xi_{nm} || {\hat{\bm \theta}}_{mn}^{d-1} - {\bm \theta}_{mn} ||_2}{\beta_{\min}^2 \left(\xi_{nm}+\xi_{mn}\right)^2} .
\end{align}
Considering (\ref{app:equ:gapofRR}), we can further get the following result:
\begin{align}
\nonumber
\left|R_{jk}\left({\bm \beta}\right)\!-\!R_{jk}\left(\hat {\bm \beta}^{d-1}\right)\right| \!  &\le \sum_{m\in{\mathcal N}} \sum_{n\in{\mathcal N}\setminus \left\{m\right\}} \frac{\xi_{mn}\xi_{nm}^2 || {\hat{\bm \theta}}_{mn}^{d-1} - {\bm \theta}_{mn} ||_2 }{\beta_{\min}^2 \left(\xi_{mn}+\xi_{nm}\right)^2} \\
& \le \frac{1}{\beta_{\min}^2}\sum_{m\in{\mathcal N}} \sum_{n\in{\mathcal N}\setminus \left\{m\right\}} \xi_{mn}|| {\hat{\bm \theta}}_{mn}^{d-1} - {\bm \theta}_{mn} ||_2.\label{app:equ:gapRRresult}
\end{align}
The result above shows an upper bound of $\left|R_{jk}\left({\bm \beta}\right)\!-\!R_{jk}\left(\hat {\bm \beta}^{d-1}\right)\right|$. 

{\bf Step 2:} We analyze an upper bound for $\left| p_{ij}^*\left({\bm \theta}\right) - p_{ij}^*\left({\hat{\bm \theta}}^{d-1}\right) \right|$.

Recall that $p_{ij}^*\left({\bm \theta}\right)$ and $p_{ij}^*\left({\hat{\bm \theta}}^{d-1}\right)$ correspond to the optimal solutions to problems (\ref{problem:1trans}) and (\ref{problem:2trans}), respectively. Based on Proposition \ref{proposition:price}, we have the following relation:
\begin{align}
p_{ij}^*\left({\bm \theta}\right)\!=\!\frac{c\beta_{ij} \!+\! \alpha_{ij} \!+\! \epsilon_{ij}^-}{2 \beta_{ij}} \!+\! \frac{1}{4\xi_{ij}} \!\sum_{k\in{\mathcal N}} \!\left(R_{jk}\left({\bm \beta}\right) \!-\! R_{ik}\left({\bm \beta}\right) \right) \!v_k\left({\bm \theta}\right),
\end{align}
where $v_k\left({\bm \theta}\right)$ is given by
\begin{align}
v_k\left({\bm \theta}\right)= \!\!\!\!\sum_{j\in{\mathcal N}\setminus\left\{k\right\}} \left(\alpha_{kj} - c\beta_{kj} - \epsilon_{kj}^-\right) - \!\!\! \sum_{j\in{\mathcal N}\setminus\left\{k\right\}}\left(\alpha_{jk} - c\beta_{jk} -\epsilon_{jk}^-\right).
\end{align}
It is easy to see that we can rearrange the expression of $p_{ij}^*\left({\bm \theta}\right)$ as follows:
{\small
\begin{align}
\nonumber
& p_{ij}^*\left({\bm \theta}\right)= \frac{c\beta_{ij} + \alpha_{ij} + \epsilon_{ij}^-}{2 \beta_{ij}} \\
& \!+\! \frac{1}{4\xi_{ij}}\!\! \!\sum_{k\in{\mathcal N}} \!\sum_{m\in{\mathcal N}\setminus\left\{k\right\}} \!\!\!\left(R_{jk}\left({\bm \beta}\right)\!-\!R_{ik}\left({\bm \beta}\right)\!-\!R_{jm}\left({\bm \beta}\right)\!+\!R_{im}\left({\bm \beta}\right)\right) \!\left(\alpha_{km} \!-\! c\beta_{km} \!-\!  \epsilon_{km}^-\right).
\label{app:equ:plong:1}
\end{align}}

Similarly, we have the following relation for $p_{ij}^*\left({\hat{\bm \theta}}^{d-1}\right)$:
{\scriptsize
\begin{align}
\nonumber
& p_{ij}^*\left({\hat{\bm \theta}}^{d-1}\right)= \frac{c{\hat\beta}_{ij}^{d-1} + {\hat\alpha}_{ij}^{d-1} + \epsilon_{ij}^-}{2 {\hat\beta}_{ij}^{d-1}} \\
& \!\!\!\!\!+\!\! \frac{1}{4\xi_{ij}}\! \!\!\sum_{k\in{\mathcal N}} \!\!\sum_{m\in{\mathcal N}\setminus\left\{k\right\}} \!\!\!\!\left(R_{jk}\left({\hat{\bm \beta}}^{d-1}\right)\!-\!R_{ik}\left({\hat{\bm \beta}}^{d-1}\right)\!-\!R_{jm}\left({\hat{\bm \beta}}^{d-1}\right)\!+\!R_{im}\left({\hat{\bm \beta}}^{d-1}\right)\right) \!\left({\hat\alpha}_{km}^{d-1} \!-\! c{\hat\beta}_{km}^{d-1} \!-\!  \epsilon_{km}^-\right).\label{app:equ:plong:2}
\end{align}}
Next, we derive an upper bound for $\left| p_{ij}^*\left({\bm \theta}\right) - p_{ij}^*\left({\hat{\bm \theta}}^{d-1}\right) \right|$.

{\bf (Step 2-A)} In this part, we derive an upper bound for the term $\left| \frac{c\beta_{ij} + \alpha_{ij} + \epsilon_{ij}^-}{2 \beta_{ij}} - \frac{c{\hat\beta}_{ij}^{d-1} + {\hat\alpha}_{ij}^{d-1} + \epsilon_{ij}^-}{2 {\hat\beta}_{ij}^{d-1}}\right|$.

We can easily derive the following relation:
\begin{align}
\nonumber
& \left| \frac{c\beta_{ij} + \alpha_{ij} + \epsilon_{ij}^-}{2 \beta_{ij}} - \frac{c{\hat\beta}_{ij}^{d-1} + {\hat\alpha}_{ij}^{d-1} + \epsilon_{ij}^-}{2 {\hat\beta}_{ij}^{d-1}}\right| \\
\nonumber
= & \frac{1}{2\beta_{ij}{\hat\beta}_{ij}^{d-1}} \left| {\hat\beta}_{ij}^{d-1} \left(\alpha_{ij} + \epsilon_{ij}^-\right) - \beta_{ij} \left({\hat\alpha}_{ij}^{d-1} + \epsilon_{ij}^-\right) \right| \\
\nonumber
\overset{(a)}{\le} & \frac{1}{2 \beta_{\min}^2} \left(\beta_{\max}\left|\alpha_{ij}-{\hat\alpha}_{ij}^{d-1}\right| + \alpha_{\max}\left|\beta_{ij}-{\hat\beta}_{ij}^{d-1}\right|\right)  \\
\le & \frac{\alpha_{\max}+\beta_{\max}}{2 \beta_{\min}^2} || {\bm \theta}_{ij}-{\hat{\bm \theta}}_{ij}^{d-1} ||_2.
\end{align}
Note that when deriving inequality (a), we have used the facts that $\left|x_1x_2-x_3x_4\right|=\left|x_1x_2-x_3x_2+x_3x_2-x_3x_4\right|\le \left|x_2\right|\left|x_1-x_3\right|+\left|x_3\right|\left|x_2-x_4\right|$ (where $x_1$, $x_2$, $x_3$, and $x_4$ are real numbers) and $\alpha_{ij}+ \epsilon_{ij}^- >\alpha_{\min}+{\underline\epsilon}\ge\beta_{\max}p_{\max}>0$.

{\bf (Step 2-B)} In this part, we derive an upper bound for the following term:
\begin{align}
\nonumber
& \left| \left(R_{jk}\left({\bm \beta}\right)\!-\!R_{ik}\left({\bm \beta}\right)\!-\!R_{jm}\left({\bm \beta}\right)\!+\!R_{im}\left({\bm \beta}\right)\right) \right.\\
\nonumber
& \left.- \left(R_{jk}\left({\hat{\bm \beta}}^{d-1}\right)\!-\!R_{ik}\left({\hat{\bm \beta}}^{d-1}\right)\!-\!R_{jm}\left({\hat{\bm \beta}}^{d-1}\right)\!+\!R_{im}\left({\hat{\bm \beta}}^{d-1}\right) \!\right) \right|.
\end{align}
We can see that the following relation holds:
\begin{align}
\nonumber
& \left| \left(R_{jk}\left({\bm \beta}\right)\!-\!R_{ik}\left({\bm \beta}\right)\!-\!R_{jm}\left({\bm \beta}\right)\!+\!R_{im}\left({\bm \beta}\right)\right) \right.\\
\nonumber
& \left.- \left(R_{jk}\left({\hat{\bm \beta}}^{d-1}\right)\!-\!R_{ik}\left({\hat{\bm \beta}}^{d-1}\right)\!-\!R_{jm}\left({\hat{\bm \beta}}^{d-1}\right)\!+\!R_{im}\left({\hat{\bm \beta}}^{d-1}\right) \!\right) \right|\\
\nonumber
& \le \left|R_{jk}\left({\bm \beta}\right) - R_{jk}\left({\hat{\bm \beta}}^{d-1}\right)\right| + \left|R_{ik}\left({\bm \beta}\right) - R_{ik}\left({\hat{\bm \beta}}^{d-1}\right)\right| \\
\nonumber
& {~~}{~~}{~~}+ \left|R_{jm}\left({\bm \beta}\right) - R_{jm}\left({\hat{\bm \beta}}^{d-1}\right)\right| + \left| R_{im}\left({\bm \beta}\right) - R_{im}\left({\hat{\bm \beta}}^{d-1}\right)\right| \\
& \le \frac{4}{\beta_{\min}^2}\sum_{{\tilde m}\in{\mathcal N}} \sum_{{\tilde n}\in{\mathcal N}\setminus \left\{{\tilde m}\right\}} \xi_{{\tilde m}{\tilde n}}|| {\hat{\bm \theta}}_{{\tilde m}{\tilde n}}^{d-1} - {\bm \theta}_{{\tilde m}{\tilde n}} ||_2.
\end{align}
The second inequality is based on (\ref{app:equ:gapRRresult}). 


{\bf (Step 2-C)} In this part, we derive an upper bound for the term $\left| \left(\alpha_{km} \!-\! c\beta_{km} \!-\!  \epsilon_{km}^- \right) - \left({\hat\alpha}_{km}^{d-1} \!-\! c{\hat\beta}_{km}^{d-1} \!-\!  \epsilon_{km}^-\right)\right|$. We can derive the upper bound as follows:
\begin{align}
\nonumber
& \left| \left(\alpha_{km} - c\beta_{km} -  \epsilon_{km}^- \right) - \left({\hat\alpha}_{km}^{d-1} - c{\hat\beta}_{km}^{d-1} -  \epsilon_{km}^-\right)\right| \\
\nonumber
\le & \left|\alpha_{km} - {\hat\alpha}_{km}^{d-1}\right| + c\left| \beta_{km} - {\hat\beta}_{km}^{d-1} \right| \\
\le & \left(1+c\right) || {\bm \theta}_{km} - {\hat{\bm \theta}}_{km}^{d-1} ||_2.
\end{align}

{\bf (Step 2-D)} We derive upper bounds for $\left| {\hat\alpha}_{km}^{d-1} \!-\! c{\hat\beta}_{km}^{d-1} \!-\!  \epsilon_{km}^- \right|$ and $\left| R_{jk}\left({\bm \beta}\right)\!-\!R_{ik}\left({\bm \beta}\right)\!-\!R_{jm}\left({\bm \beta}\right)\!+\!R_{im}\left({\bm \beta}\right) \right|$.

Recall that we assume that $\alpha_{\min} - \beta_{\max} p_{\max} + {\underline \epsilon} \ge0$ and $0<c<p_{\max}$. Therefore, we have ${\hat\alpha}_{km}^{d-1} - c{\hat\beta}_{km}^{d-1} -  \epsilon_{km}^- > \alpha_{\min} - p_{\max}\beta_{\max} -  \epsilon_{km}^- \ge \alpha_{\min} - p_{\max}\beta_{\max} \ge - {\underline \epsilon} >0$. We derive an upper bound for $\left| {\hat\alpha}_{km}^{d-1} \!-\! c{\hat\beta}_{km}^{d-1} \!-\!  \epsilon_{km}^- \right|$ as follows:
\begin{align}
\left| {\hat\alpha}_{km}^{d-1} \!-\! c{\hat\beta}_{km}^{d-1} \!-\!  \epsilon_{km}^- \right|
\le \alpha_{\max} + \left|\epsilon_{km}^-\right| \le \alpha_{\max}- {\underline \epsilon},
\end{align}
where the second inequality is based on $\epsilon_{km}^- = \int_{\underline \epsilon}^0 \epsilon_{km}^d d F_{km}\left(\epsilon_{km}^d\right) \in \left[{\underline \epsilon},0\right]$.

We derive a bound for $\left| R_{jk}\left({\bm \beta}\right)\!-\!R_{ik}\left({\bm \beta}\right)\!-\!R_{jm}\left({\bm \beta}\right)\!+\!R_{im}\left({\bm \beta}\right)\right|$ as follows:
\begin{align}
\nonumber 
& \left| R_{jk}\left({\bm \beta}\right)\!-\!R_{ik}\left({\bm \beta}\right)\!-\!R_{jm}\left({\bm \beta}\right)\!+\!R_{im}\left({\bm \beta}\right)\right| \\
\nonumber
\le & \left| R_{jk}\left({\bm \beta}\right)\!-\!R_{jm}\left({\bm \beta}\right) \right| + \left| R_{im}\left({\bm \beta}\right)\!-\!R_{ik}\left({\bm \beta}\right) \right| \\
\nonumber
\overset{(a)}{\le} & 2 R_{km} \left({\bm \beta}\right) \overset{(b)}{\le}  2 r_{km} \\
= & 2\frac{1}{\frac{\beta_{km}}{\xi_{km}}+\frac{\beta_{mk}}{\xi_{mk}}} \le 2\frac{1}{\frac{\beta_{\min}}{\xi_{km}}+\frac{\beta_{\min}}{\xi_{mk}}} \le \frac{2\xi_{km}}{\beta_{\min}}.
\end{align}
The inequality (a) is based on the triangle inequality for effective resistances, and the inequality (b) is based on the fact that the effective resistance between two nodes is no greater than the resistance of the resistor that directly connects the two nodes.  

{\bf (Step 2-E)} We combine the results in {\bf Steps 2-A, 2-B, 2-C, and 2-D}, and derive an upper bound for $\left| p_{ij}^*\left({\bm \theta}\right) - p_{ij}^*\left({\hat{\bm \theta}}^{d-1}\right) \right|$. The basic idea is to utilize the inequality that $\left|x_1x_2-x_3x_4\right|\le \left|x_2\right|\left|x_1-x_3\right|+\left|x_3\right|\left|x_2-x_4\right|$ (where $x_1$, $x_2$, $x_3$, and $x_4$ are real numbers). 

According to the expressions of $p_{ij}^*\left({\bm \theta}\right)$ and $p_{ij}^*\left({\hat{\bm \theta}}^{d-1}\right)$ in (\ref{app:equ:plong:1}) and (\ref{app:equ:plong:2}) and the results in {\bf Steps 2-A, 2-B, 2-C, and 2-D}, we can get the following relation:
\begin{align}
\nonumber
& \left| p_{ij}^*\left({\bm \theta}\right) - p_{ij}^*\left({\hat{\bm \theta}}^{d-1}\right) \right| \\
\nonumber
\le & \frac{\alpha_{\max}+\beta_{\max}}{2 \beta_{\min}^2} || {\bm \theta}_{ij}-{\hat{\bm \theta}}_{ij}^{d-1} ||_2 \\
\nonumber
& + \frac{1}{4\xi_{ij}} \sum_{k\in{\mathcal N}} \sum_{m\in{\mathcal N}\setminus\left\{k\right\}} \frac{2\xi_{km}}{\beta_{\min}}  \left(1+c\right) || {\bm \theta}_{km} - {\hat{\bm \theta}}_{km}^{d-1} ||_2 \\
\nonumber
& + \frac{1}{4\xi_{ij}} \!\sum_{k\in{\mathcal N}} \!\sum_{m\in{\mathcal N}\setminus\left\{k\right\}} \!\!\!\left(\alpha_{\max}-{\underline \epsilon}\right) \frac{4}{\beta_{\min}^2}\!\!\sum_{{\tilde m}\in{\mathcal N}} \!\!\sum_{{\tilde n}\in{\mathcal N}\setminus \left\{{\tilde m}\right\}} \!\!\!\xi_{{\tilde m}{\tilde n}}|| {\hat{\bm \theta}}_{{\tilde m}{\tilde n}}^{d-1} - {\bm \theta}_{{\tilde m}{\tilde n}} ||_2 \\
\nonumber
= & \frac{\alpha_{\max}+\beta_{\max}}{2 \beta_{\min}^2} || {\bm \theta}_{ij}-{\hat{\bm \theta}}_{ij}^{d-1} ||_2 \\
\nonumber
& + \frac{1+c}{2\xi_{ij}\beta_{\min}} \sum_{k\in{\mathcal N}} \sum_{m\in{\mathcal N}\setminus\left\{k\right\}} \xi_{km} || {\bm \theta}_{km} - {\hat{\bm \theta}}_{km}^{d-1} ||_2 \\
\nonumber
& + \frac{\alpha_{\max}-{\underline \epsilon}}{\xi_{ij} \beta_{\min}^2} N\left(N-1\right)   \sum_{k\in{\mathcal N}} \sum_{m\in{\mathcal N}\setminus\left\{k\right\}} \xi_{km}|| {\bm \theta}_{km} - {\hat{\bm \theta}}_{km}^{d-1} ||_2 \\
\nonumber
= & \frac{\alpha_{\max}+\beta_{\max}}{2 \beta_{\min}^2} || {\bm \theta}_{ij}-{\hat{\bm \theta}}_{ij}^{d-1} ||_2 \\
& \!\!\!+ \!\frac{1}{\xi_{ij} \beta_{\min}} \left(\frac{1\!+\!c}{2} \!+\! \frac{\alpha_{\max}\!-\!{\underline \epsilon}}{ \beta_{\min}} N\!\left(N\!-\!1\right)\right)\!\! \sum_{k\in{\mathcal N}} \!\sum_{m\in{\mathcal N}\setminus\left\{k\right\}} \!\!\!\xi_{km}|| {\bm \theta}_{km} \!-\! {\hat{\bm \theta}}_{km}^{d-1} ||_2.\label{app:equ:resultboundpp}
\end{align}
This completes our analysis in {\bf Step 2}. 

{\bf Step 3:} We analyze the gap between $\Pi\left({\bm p}^*\left({\bm \theta}\right),{\bm w}^*\left({\bm \theta}\right),{\bm \epsilon}^d\right)$ and $\Pi\left({\bm p}^d,{\bm w}^d,{\bm \epsilon}^d\right)$ (i.e., the payoff under our policy) when $d$ is odd and $d\ge5$.  

According to our policy, when $d$ is odd, the provider implements $p_{ij}^*\left({\hat{\bm \theta}}^{d-1}\right)$ as the pricing decision and $w_{ij}^*\left({\hat{\bm \theta}}^{d-1}\right)$ as the supply decision for each link $\left(i,j\right)$. 
The expressions of ${\mathbb E}_{{\bm \epsilon}^d}\left\{\Pi\left({\bm p}^*\left({\bm \theta}\right),{\bm w}^*\left({\bm \theta}\right),{\bm \epsilon}^d\right)\right\}$ and ${\mathbb E}_{{\bm \epsilon}^d}\left\{\Pi\left({\bm p}^d,{\bm w}^d,{\bm \epsilon}^d\right)\right\}$ are given as follows:
\begin{align}
\nonumber
& {\mathbb E}_{{\bm \epsilon}^d}\left\{\Pi\left({\bm p}^*\left({\bm \theta}\right),{\bm w}^*\left({\bm \theta}\right),{\bm \epsilon}^d\right)\right\} \\
\nonumber
= & \sum_{i\in{\mathcal N}} \sum_{j\in{\mathcal N}\setminus\left\{i\right\}} \xi_{ij} \left(\alpha_{ij} - \beta_{ij} p_{ij}^*\left({\bm \theta}\right) + \epsilon_{ij}^-\right) p_{ij}^*\left({\bm \theta}\right) \\
& - \sum_{i\in{\mathcal N}} \sum_{j\in{\mathcal N}\setminus\left\{i\right\}} \xi_{ij} \left(\alpha_{ij} - \beta_{ij} p_{ij}^*\left({\bm \theta}\right)\right) c,\label{app:equ:payoffgapo1}
\end{align}
{\scriptsize
\begin{align}
\nonumber
& {\mathbb E}_{{\bm \epsilon}^d}\left\{\Pi\left({\bm p}^d,{\bm w}^d,{\bm \epsilon}^d\right)\right\} \\
\nonumber
=& {\mathbb E}_{{\bm \epsilon}^d}\left\{\Pi\left({\bm p}^*\left({\hat{\bm \theta}}^{d-1}\right),{\bm w}^*\left({\hat{\bm \theta}}^{d-1}\right),{\bm \epsilon}^d\right)\right\} \\
\nonumber
= & \sum_{i\in{\mathcal N}} \sum_{j\in{\mathcal N}\setminus\left\{i\right\}} \xi_{ij} {\mathbb E}_{{\bm \epsilon}^d}\left\{\min\left\{\alpha_{ij} - \beta_{ij} p_{ij}^*\left({\hat{\bm \theta}}^{d-1}\right) + \epsilon_{ij}^d,{\hat \alpha}_{ij}^{d-1} - {\hat \beta}_{ij}^{d-1} p_{ij}^*\left({\hat{\bm \theta}}^{d-1}\right)\right\} \right\} p_{ij}^*\left({\hat{\bm \theta}}^{d-1}\right) \\
& - \sum_{i\in{\mathcal N}} \sum_{j\in{\mathcal N}\setminus\left\{i\right\}} \xi_{ij} \left({\hat \alpha}_{ij}^{d-1} - {\hat \beta}_{ij}^{d-1} p_{ij}^*\left({\hat{\bm \theta}}^{d-1}\right)\right) c.\label{app:equ:payoffgapo2}
\end{align}}
In order to bound $\left|{\mathbb E}_{{\bm \epsilon}^d}\left\{\Pi\left({\bm p}^*\left({\bm \theta}\right),{\bm w}^*\left({\bm \theta}\right),{\bm \epsilon}^d\right)\right\} - {\mathbb E}_{{\bm \epsilon}^d}\left\{\Pi\left({\bm p}^d,{\bm w}^d,{\bm \epsilon}^d\right)\right\}\right|$, we first prove some preliminary results. 

\begin{figure*}
{\small
\begin{align}
{\mathbb E}_{{\bm \epsilon}^d}\left\{\Pi\left({\bm p}^*\left({\bm \theta}\right),{\bm w}^*\left({\bm \theta}\right),{\bm \epsilon}^d\right)\right\} =  & \sum_{i\in{\mathcal N}} \sum_{j\in{\mathcal N}\setminus\left\{i\right\}} \xi_{ij} \left(\alpha_{ij} - \beta_{ij} p_{ij}^*\left({\bm \theta}\right) + \epsilon_{ij}^-\right) p_{ij}^*\left({\bm \theta}\right)  - \sum_{i\in{\mathcal N}} \sum_{j\in{\mathcal N}\setminus\left\{i\right\}} \xi_{ij} \left(\alpha_{ij} - \beta_{ij} p_{ij}^*\left({\bm \theta}\right)\right) c.\label{app:equ:even:payoff1}\\
\nonumber
{\mathbb E}_{{\bm \epsilon}^d}\left\{\Pi\left({\bm p}^d,{\bm w}^d,{\bm \epsilon}^d\right)\right\} = & \sum_{i\in{\mathcal N}} \sum_{j\in{\mathcal N}\setminus\left\{i\right\}} \xi_{ij} {\mathbb E}_{{\bm \epsilon}^d}\left\{\min\left\{\alpha_{ij} - \beta_{ij} \left(p_{ij}^*\left({\hat{\bm \theta}}^{d-2}\right)-\frac{\rho}{{\hat\beta}_{ij}^{d-2}}d^{-\eta}\right) + \epsilon_{ij}^d,{\hat \alpha}_{ij}^{d-2} - {\hat \beta}_{ij}^{d-2} p_{ij}^*\left({\hat{\bm \theta}}^{d-2}\right)+\rho d^{-\eta}\right\} \right\} \left(p_{ij}^*\left({\hat{\bm \theta}}^{d-2}\right)-\frac{\rho}{{\hat\beta}_{ij}^{d-2}}d^{-\eta}\right) \\
& - \sum_{i\in{\mathcal N}} \sum_{j\in{\mathcal N}\setminus\left\{i\right\}} \xi_{ij} \left({\hat \alpha}_{ij}^{d-2} - {\hat \beta}_{ij}^{d-2} p_{ij}^*\left({\hat{\bm \theta}}^{d-2}\right)+\rho d^{-\eta}\right) c.\label{app:equ:even:payoff2}
\end{align}
\hrule
\begin{align}
\nonumber
& \left|{\mathbb E}_{{\bm \epsilon}^d}\left\{\Pi\left({\bm p}^*\left({\bm \theta}\right),{\bm w}^*\left({\bm \theta}\right),{\bm \epsilon}^d\right)\right\}-{\mathbb E}_{{\bm \epsilon}^d}\left\{\Pi\left({\bm p}^d,{\bm w}^d,{\bm \epsilon}^d\right)\right\} \right|\\
\nonumber
\le & \sum_{i\in{\mathcal N}} \sum_{j\in{\mathcal N}\setminus\left\{i\right\}} \xi_{ij} \left(\alpha_{\max}+\beta_{\max}p_{\rm up} \right) \left( \left| p_{ij}^*\left({\bm \theta}\right) - p_{ij}^*\left({\hat{\bm \theta}}^{d-2}\right) \right| + \frac{\rho}{\beta_{\min}}d^{-\eta} \right)  + \sum_{i\in{\mathcal N}} \sum_{j\in{\mathcal N}\setminus\left\{i\right\}} \xi_{ij} p_{\rm up} \left(2\left(1+p_{\rm up}\right) || {\bm \theta}_{ij} - {\hat{\bm \theta}}_{ij}^{d-2} ||_2 + \beta_{\max} \left| p_{ij}^*\left({\bm \theta}\right) - p_{ij}^*\left({\hat{\bm \theta}}^{d-2}\right) \right| + \frac{\beta_{\max}}{\beta_{\min}}\rho d^{-\eta} \right) \\
\nonumber
& + \sum_{i\in{\mathcal N}} \sum_{j\in{\mathcal N}\setminus\left\{i\right\}} \xi_{ij} c\left(  \left(1+p_{\rm up}\right) || {\bm \theta}_{ij} - {\hat{\bm \theta}}_{ij}^{d-2} ||_2 + \beta_{\max} \left| p_{ij}^*\left({\bm \theta}\right) - p_{ij}^*\left({\hat{\bm \theta}}^{d-2}\right) \right| + \rho d^{-\eta}\right) \\
\!=\!\! & \sum_{i\in{\mathcal N}} \!\!\sum_{j\in{\mathcal N}\setminus\left\{i\right\}} \!\!\xi_{ij}\left(2p_{\rm up}+c\right)\left(1+p_{\rm up}\right)  || {\bm \theta}_{ij} - {\hat{\bm \theta}}_{ij}^{d-2} ||_2   \!+ \!\sum_{i\in{\mathcal N}} \!\sum_{j\in{\mathcal N}\setminus\left\{i\right\}} \xi_{ij} \left(\alpha_{\max}+2p_{\rm up}\beta_{\max} +c\beta_{\max}\right) \left| p_{ij}^*\left({\bm \theta}\right) - p_{ij}^*\left({\hat{\bm \theta}}^{d-2}\right) \right| \!+\! \sum_{i\in{\mathcal N}} \sum_{j\in{\mathcal N}\setminus\left\{i\right\}} \xi_{ij} \left(\frac{\alpha_{\max}}{\beta_{\min}}+2\frac{\beta_{\max}}{\beta_{\min}} p_{\rm up}+c\right) \rho d^{-\eta}.
\label{app:equ:even:gapresult}
\end{align}}
\hrule
\end{figure*}

{\bf (Step 3-A)} We analyze an upper bound for $|\alpha_{ij} - \beta_{ij} p_{ij}^*\left({\bm \theta}\right) + \epsilon_{ij}^- - {\mathbb E}_{{\bm \epsilon}^d}\left\{\min\left\{\alpha_{ij} - \beta_{ij} p_{ij}^*\left({\hat{\bm \theta}}^{d-1}\right) + \epsilon_{ij}^d,{\hat \alpha}_{ij}^{d-1} - {\hat \beta}_{ij}^{d-1} p_{ij}^*\left({\hat{\bm \theta}}^{d-1}\right)\right\} \right\}|$. We define a threshold $\epsilon_{\rm Th}$ as follows:
\begin{align}
\epsilon_{\rm Th} \triangleq {\hat \alpha}_{ij}^{d-1} - {\hat \beta}_{ij}^{d-1} p_{ij}^*\left({\hat{\bm \theta}}^{d-1}\right) - \left(\alpha_{ij} - \beta_{ij} p_{ij}^*\left({\hat{\bm \theta}}^{d-1}\right)\right).
\end{align}
Using the notation $\epsilon_{\rm Th}$, we can bound the term $|\alpha_{ij} - \beta_{ij} p_{ij}^*\left({\bm \theta}\right) + \epsilon_{ij}^- - {\mathbb E}_{{\bm \epsilon}^d}\left\{\min\left\{\alpha_{ij} - \beta_{ij} p_{ij}^*\left({\hat{\bm \theta}}^{d-1}\right) + \epsilon_{ij}^d,{\hat \alpha}_{ij}^{d-1} - {\hat \beta}_{ij}^{d-1} p_{ij}^*\left({\hat{\bm \theta}}^{d-1}\right)\right\} \right\}|$ as follows:
\begin{align}
\nonumber
& \left| \alpha_{ij} - \beta_{ij} p_{ij}^*\left({\bm \theta}\right) + \epsilon_{ij}^- \right.\\
\nonumber
& \left. - {\mathbb E}_{{\bm \epsilon}^d}\left\{\min\left\{\alpha_{ij} - \beta_{ij} p_{ij}^*\left({\hat{\bm \theta}}^{d-1}\right) + \epsilon_{ij}^d,{\hat \alpha}_{ij}^{d-1} - {\hat \beta}_{ij}^{d-1} p_{ij}^*\left({\hat{\bm \theta}}^{d-1}\right)\right\} \right\} \right| \\
\nonumber
= & \left|  \int_{\epsilon_{ij}^d<\epsilon_{\rm Th}}  \left( \alpha_{ij} - \beta_{ij} p_{ij}^*\left({\bm \theta}\right) - \left(\alpha_{ij} - \beta_{ij} p_{ij}^*\left({\hat{\bm \theta}}^{d-1}\right)\right) \right) d F_{ij}\left(\epsilon_{ij}^d\right) \right. \\
\nonumber
& + \epsilon_{ij}^- - \int_{\epsilon_{ij}^d<\epsilon_{\rm Th}}  \epsilon_{ij}^d   d F_{ij}\left(\epsilon_{ij}^d\right) \\
\nonumber
& \left. + \int_{\epsilon_{ij}^d\ge\epsilon_{\rm Th}} \left( \alpha_{ij} - \beta_{ij} p_{ij}^*\left({\bm \theta}\right) - \left({\hat \alpha}_{ij}^{d-1} - {\hat \beta}_{ij}^{d-1} p_{ij}^*\left({\hat{\bm \theta}}^{d-1}\right)\right)\right) d F_{ij}\left(\epsilon_{ij}^d\right) \right| \\
\nonumber
\le &  F_{ij}\left(\epsilon_{\rm Th}\right) \left| \alpha_{ij} - \beta_{ij} p_{ij}^*\left({\bm \theta}\right) - \left(\alpha_{ij} - \beta_{ij} p_{ij}^*\left({\hat{\bm \theta}}^{d-1}\right)\right)  \right|\\
\nonumber
&  + \left(1- F_{ij}\left(\epsilon_{\rm Th}\right)\right)  \left| \alpha_{ij} - \beta_{ij} p_{ij}^*\left({\bm \theta}\right) - \left({\hat \alpha}_{ij}^{d-1} - {\hat \beta}_{ij}^{d-1} p_{ij}^*\left({\hat{\bm \theta}}^{d-1}\right)\right) \right|\\
& + \left| \epsilon_{ij}^- - \int_{\epsilon_{ij}^d<\epsilon_{\rm Th}}  \epsilon_{ij}^d   d F_{ij}\left(\epsilon_{ij}^d\right)  \right|.\label{app:equ:fourA}
\end{align}
It is easy to verify that the following results hold:
\begin{align}
& \left|\alpha_{ij} - \beta_{ij} p_{ij}^*\left({\bm \theta}\right) - \left(\alpha_{ij} - \beta_{ij} p_{ij}^*\left({\hat{\bm \theta}}^{d-1}\right)\right)\right| \!\le\! \beta_{\max} \left| p_{ij}^*\left({\bm \theta}\right) - p_{ij}^*\left({\hat{\bm \theta}}^{d-1}\right) \right|,\label{app:equ:fourB}\\
\nonumber
& \left| \alpha_{ij} - \beta_{ij} p_{ij}^*\left({\bm \theta}\right) - \left({\hat \alpha}_{ij}^{d-1} - {\hat \beta}_{ij}^{d-1} p_{ij}^*\left({\hat{\bm \theta}}^{d-1}\right)\right) \right| \\
\nonumber
\le & \left|\alpha_{ij} - {\hat \alpha}_{ij}^{d-1} \right| + \left| {\hat \beta}_{ij}^{d-1} p_{ij}^*\left({\hat{\bm \theta}}^{d-1}\right) -  \beta_{ij} p_{ij}^*\left({\bm \theta}\right)\right|\\
\le & \left(1+p_{\rm up}\right) || {\bm \theta}_{ij} - {\hat{\bm \theta}}_{ij}^{d-1} ||_2 + \beta_{\max} \left| p_{ij}^*\left({\bm \theta}\right) - p_{ij}^*\left({\hat{\bm \theta}}^{d-1}\right) \right|.\label{app:equ:fourC}
\end{align}
Recall that $p_{\rm up}$ is a notation defined before (in {\bf Step 3} of Section \ref{app:sec:proof:the1}) and it satisfies $p_{\rm up}\ge\left| p_{ij}^*\left({\hat{\bm \theta}}^{d-1}\right) \right|, \left| p_{ij}^*\left({{\bm \theta}}\right)\right|$.

Next, we analyze an upper bound for $\left| \epsilon_{ij}^- - \int_{\epsilon_{ij}^d<\epsilon_{\rm Th}}  \epsilon_{ij}^d   d F_{ij}\left(\epsilon_{ij}^d\right)  \right|$. We can derive the following relation:
\begin{align}
\nonumber 
& \left| \epsilon_{ij}^- - \int_{\epsilon_{ij}^d<\epsilon_{\rm Th}}  \epsilon_{ij}^d   d F_{ij}\left(\epsilon_{ij}^d\right)  \right| \\
= & \left| \int_{\epsilon_{ij}^d<0}  \epsilon_{ij}^d   d F_{ij}\left(\epsilon_{ij}^d\right) - \int_{\epsilon_{ij}^d<\epsilon_{\rm Th}}  \epsilon_{ij}^d   d F_{ij}\left(\epsilon_{ij}^d\right)  \right|.
\end{align}
If $\epsilon_{\rm Th} = {\hat \alpha}_{ij}^{d-1} - {\hat \beta}_{ij}^{d-1} p_{ij}^*\left({\hat{\bm \theta}}^{d-1}\right) - \left(\alpha_{ij} - \beta_{ij} p_{ij}^*\left({\hat{\bm \theta}}^{d-1}\right)\right)>0$, we have
\begin{align}
\left| \epsilon_{ij}^- - \int_{\epsilon_{ij}^d<\epsilon_{\rm Th}}  \epsilon_{ij}^d   d F_{ij}\left(\epsilon_{ij}^d\right)  \right| =  \left| \int_{0\le\epsilon_{ij}^d<\epsilon_{\rm Th}}  \epsilon_{ij}^d   d F_{ij}\left(\epsilon_{ij}^d\right) \right| \le \left| \epsilon_{\rm Th} \right|.
\end{align}
If $\epsilon_{\rm Th}\le 0$, we have
\begin{align}
\left| \epsilon_{ij}^- - \int_{\epsilon_{ij}^d<\epsilon_{\rm Th}}  \epsilon_{ij}^d   d F_{ij}\left(\epsilon_{ij}^d\right)  \right| =  \left| \int_{\epsilon_{\rm Th}\le\epsilon_{ij}^d<0}  \epsilon_{ij}^d   d F_{ij}\left(\epsilon_{ij}^d\right) \right| \le \left| \epsilon_{\rm Th} \right|.
\end{align}
Therefore, we can conclude that 
\begin{align}
\nonumber
& \left| \epsilon_{ij}^- - \int_{\epsilon_{ij}^d<\epsilon_{\rm Th}}  \epsilon_{ij}^d   d F_{ij}\left(\epsilon_{ij}^d\right)  \right|  \le \left| \epsilon_{\rm Th} \right| \\
\nonumber
= & \left| {\hat \alpha}_{ij}^{d-1} - {\hat \beta}_{ij}^{d-1} p_{ij}^*\left({\hat{\bm \theta}}^{d-1}\right) - \left(\alpha_{ij} - \beta_{ij} p_{ij}^*\left({\hat{\bm \theta}}^{d-1}\right)\right) \right| \\
\le & \left(1+p_{\rm up}\right) || {\bm \theta}_{ij} - {\hat{\bm \theta}}_{ij}^{d-1} ||_2.\label{app:equ:fourD}
\end{align}

Combining our results in (\ref{app:equ:fourA}), (\ref{app:equ:fourB}), (\ref{app:equ:fourC}), and (\ref{app:equ:fourD}), we can get the following relation:
\begin{align}
\nonumber
& \left| \alpha_{ij} - \beta_{ij} p_{ij}^*\left({\bm \theta}\right) + \epsilon_{ij}^- \right.\\
\nonumber
& \left. - {\mathbb E}_{{\bm \epsilon}^d}\left\{\min\left\{\alpha_{ij} - \beta_{ij} p_{ij}^*\left({\hat{\bm \theta}}^{d-1}\right) + \epsilon_{ij}^d,{\hat \alpha}_{ij}^{d-1} - {\hat \beta}_{ij}^{d-1} p_{ij}^*\left({\hat{\bm \theta}}^{d-1}\right)\right\} \right\} \right| \\
\le &  2\left(1+p_{\rm up}\right) || {\bm \theta}_{ij} - {\hat{\bm \theta}}_{ij}^{d-1} ||_2 + \beta_{\max} \left| p_{ij}^*\left({\bm \theta}\right) - p_{ij}^*\left({\hat{\bm \theta}}^{d-1}\right) \right|.
\end{align}
Here, we have used the fact that the convex combination of two real numbers is no greater than each of the two numbers. 

\begin{figure*}
{\scriptsize
\begin{align}
\nonumber 
& {\mathbb E} \left\{  \sum_{d=5}^D \Bigg( \Pi\left({\bm p}^*\left({\bm \theta}\right),{\bm w}^*\left({\bm \theta}\right),{\bm \epsilon}^d\right)  - \Pi\left({\bm p}^d,{\bm w}^d,{\bm \epsilon}^d\right) \Bigg) \right\} \\
\nonumber 
\le & {\mathbb E} \left\{ \sum_{s=2}^{\lfloor\frac{D-1}{2}\rfloor} \Bigg( \left|\Pi\left({\bm p}^*\left({\bm \theta}\right),{\bm w}^*\left({\bm \theta}\right),{\bm \epsilon}^{2s+1}\right)  - \Pi\left({\bm p}^{2s+1},{\bm w}^{2s+1},{\bm \epsilon}^{2s+1}\right)\right|+ \left|\Pi\left({\bm p}^*\left({\bm \theta}\right),{\bm w}^*\left({\bm \theta}\right),{\bm \epsilon}^{2s+2}\right)  - \Pi\left({\bm p}^{2s+2},{\bm w}^{2s+2},{\bm \epsilon}^{2s+2}\right) \right| \Bigg) \right\} \\
\nonumber
\le & {\mathbb E} \left\{ \sum_{s=2}^{\lfloor\frac{D-1}{2}\rfloor} \left( \sum_{i\in{\mathcal N}} \sum_{j\in{\mathcal N}\setminus\left\{i\right\}} \xi_{ij}2\left(2p_{\rm up}+c\right)\left(1+p_{\rm up}\right)  || {\bm \theta}_{ij} - {\hat{\bm \theta}}_{ij}^{2s} ||_2  + \sum_{i\in{\mathcal N}} \sum_{j\in{\mathcal N}\setminus\left\{i\right\}} \xi_{ij} 2\left(\alpha_{\max}+2p_{\rm up}\beta_{\max} +c\beta_{\max}\right) \left| p_{ij}^*\left({\bm \theta}\right) - p_{ij}^*\left({\hat{\bm \theta}}^{2s}\right) \right|\right)   \right\} \\
\nonumber
& + {\mathbb E} \left\{ \sum_{s=2}^{\lfloor\frac{D-1}{2}\rfloor} \left(\! \sum_{i\in{\mathcal N}} \sum_{j\in{\mathcal N}\setminus\left\{i\right\}} \xi_{ij} \left(\frac{\alpha_{\max}}{\beta_{\min}}+2\frac{\beta_{\max}}{\beta_{\min}} p_{\rm up}+c\right) \rho {\left(2s+2\right)}^{-\eta}\right)   \right\} \\
\nonumber
\le & {\mathbb E} \left\{ \sum_{s=2}^{\lfloor\frac{D-1}{2}\rfloor} \left( \sum_{i\in{\mathcal N}} \sum_{j\in{\mathcal N}\setminus\left\{i\right\}} \xi_{ij}2\left(2p_{\rm up}+c\right)\left(1+p_{\rm up}\right)  || {\bm \theta}_{ij} - {\hat{\bm \theta}}_{ij}^{2s} ||_2  + \sum_{i\in{\mathcal N}} \sum_{j\in{\mathcal N}\setminus\left\{i\right\}} \xi_{ij} \left(\frac{\alpha_{\max}}{\beta_{\min}}+2\frac{\beta_{\max}}{\beta_{\min}} p_{\rm up}+c\right) \rho {\left(2s+2\right)}^{-\eta} \right)   \right\} \\
\nonumber
& + {\mathbb E} \left\{ \sum_{s=2}^{\lfloor\frac{D-1}{2}\rfloor} \left(
\sum_{i\in{\mathcal N}} \sum_{j\in{\mathcal N}\setminus\left\{i\right\}} \xi_{ij} 2\left(\alpha_{\max}+2p_{\rm up}\beta_{\max} +c\beta_{\max}\right) 
\left(\frac{\alpha_{\max}+\beta_{\max}}{2 \beta_{\min}^2} || {\bm \theta}_{ij}-{\hat{\bm \theta}}_{ij}^{2s} ||_2 + \!\frac{1}{\xi_{ij} \beta_{\min}} \left(\frac{1\!+\!c}{2} \!+\! \frac{\alpha_{\max}\!-\!{\underline \epsilon}}{ \beta_{\min}} N\!\left(N\!-\!1\right)\right)\!\! \sum_{k\in{\mathcal N}} \!\sum_{m\in{\mathcal N}\setminus\left\{k\right\}} \!\!\!\xi_{km}|| {\bm \theta}_{km} \!-\! {\hat{\bm \theta}}_{km}^{2s} ||_2\right)
\right)   \right\}\\
\nonumber
= & {\mathbb E} \left\{ \sum_{s=2}^{\lfloor\frac{D-1}{2}\rfloor} \left( \sum_{i\in{\mathcal N}} \sum_{j\in{\mathcal N}\setminus\left\{i\right\}} \xi_{ij} \left(\frac{\alpha_{\max}}{\beta_{\min}}+2\frac{\beta_{\max}}{\beta_{\min}} p_{\rm up}+c\right) \rho {\left(2s+2\right)}^{-\eta} \right)   \right\}\\
\nonumber
& + {\mathbb E} \left\{ \sum_{s=2}^{\lfloor\frac{D-1}{2}\rfloor} \left(2\left(2p_{\rm up}+c\right)\left(1+p_{\rm up}\right) +  2\left(\alpha_{\max}+2p_{\rm up}\beta_{\max} +c\beta_{\max}\right) 
\frac{\alpha_{\max}+\beta_{\max}}{2 \beta_{\min}^2} + 2\left(\alpha_{\max}+2p_{\rm up}\beta_{\max} +c\beta_{\max}\right) \frac{1}{\beta_{\min}} \left(\frac{1\!+\!c}{2} \!+\! \frac{\alpha_{\max}\!-\!{\underline \epsilon}}{ \beta_{\min}} N\!\left(N\!-\!1\right)\right)N\left(N-1\right) \right) \sum_{i\in{\mathcal N}} \sum_{j\in{\mathcal N}\setminus\left\{i\right\}} \xi_{ij} || {\bm \theta}_{ij} - {\hat{\bm \theta}}_{ij}^{2s} ||_2   \right\}\\
\nonumber
\le &  \left( \sum_{i\in{\mathcal N}} \sum_{j\in{\mathcal N}\setminus\left\{i\right\}} \xi_{ij} \right)\rho\left(\frac{\alpha_{\max}}{\beta_{\min}}+2\frac{\beta_{\max}}{\beta_{\min}} p_{\rm up}+c\right) \sum_{s=2}^{\lfloor\frac{D-1}{2}\rfloor}  {\left(2s+2\right)}^{-\eta}   \\
& +\! \left( \sum_{i\in{\mathcal N}}\! \sum_{j\in{\mathcal N}\setminus\left\{i\right\}} \!\xi_{ij} \right) \left(\!2\left(2p_{\rm up}+c\right)\left(1+p_{\rm up}\right) \!+\!  2\left(\alpha_{\max}+2p_{\rm up}\beta_{\max} +c\beta_{\max}\right) 
\frac{\alpha_{\max}+\beta_{\max}}{2 \beta_{\min}^2} \!+\! 2\left(\alpha_{\max}+2p_{\rm up}\beta_{\max} +c\beta_{\max}\right) \frac{1}{\beta_{\min}} \left(\frac{1\!+\!c}{2} \!+\! \frac{\alpha_{\max}\!-\!{\underline \epsilon}}{ \beta_{\min}} N\!\left(N\!-\!1\right)\right)N\left(N-1\right) \right) \sqrt{\Phi_1\left(\rho,\eta\right)} \sum_{s=2}^{\lfloor\frac{D-1}{2}\rfloor}  \sqrt{\frac{\ln \left(2s\right)}{\left(2s\right)^{1-2\eta}}}.\label{app:equ:longestgap}
\end{align}} 
\hrule
\end{figure*}

{\bf (Step 3-B)} We derive an upper bound for $|{\mathbb E}_{{\bm \epsilon}^d}\left\{\Pi\left({\bm p}^*\left({\bm \theta}\right),{\bm w}^*\left({\bm \theta}\right),{\bm \epsilon}^d\right)\right\} - {\mathbb E}_{{\bm \epsilon}^d}\left\{\Pi\left({\bm p}^d,{\bm w}^d,{\bm \epsilon}^d\right)\right\}|$. 
The basic idea is still to utilize the inequality that $\left|x_1x_2-x_3x_4\right|\le \left|x_2\right|\left|x_1-x_3\right|+\left|x_3\right|\left|x_2-x_4\right|$ (where $x_1$, $x_2$, $x_3$, and $x_4$ are real numbers). Based on (\ref{app:equ:payoffgapo1}) and (\ref{app:equ:payoffgapo2}), we can get the bound as follows:
{\small
\begin{align}
\nonumber
& \left|{\mathbb E}_{{\bm \epsilon}^d}\left\{\Pi\left({\bm p}^*\left({\bm \theta}\right),{\bm w}^*\left({\bm \theta}\right),{\bm \epsilon}^d\right)\right\} -{\mathbb E}_{{\bm \epsilon}^d}\left\{\Pi\left({\bm p}^d,{\bm w}^d,{\bm \epsilon}^d\right)\right\} \right|\\
\nonumber
\le &  \sum_{i\in{\mathcal N}} \sum_{j\in{\mathcal N}\setminus\left\{i\right\}} \xi_{ij} p_{\rm up} \left(2\left(1+p_{\rm up}\right) || {\bm \theta}_{ij} - {\hat{\bm \theta}}_{ij}^{d-1} ||_2 + \beta_{\max} \left| p_{ij}^*\left({\bm \theta}\right) - p_{ij}^*\left({\hat{\bm \theta}}^{d-1}\right) \right|\right) \\
\nonumber
& + \sum_{i\in{\mathcal N}} \sum_{j\in{\mathcal N}\setminus\left\{i\right\}} \xi_{ij} \left(\alpha_{\max}+\beta_{\max}p_{\rm up} \right) \left| p_{ij}^*\left({\bm \theta}\right) - p_{ij}^*\left({\hat{\bm \theta}}^{d-1}\right) \right| \\
\nonumber
& + \sum_{i\in{\mathcal N}} \sum_{j\in{\mathcal N}\setminus\left\{i\right\}} \xi_{ij} c\left(  \left(1+p_{\rm up}\right) || {\bm \theta}_{ij} - {\hat{\bm \theta}}_{ij}^{d-1} ||_2 + \beta_{\max} \left| p_{ij}^*\left({\bm \theta}\right) - p_{ij}^*\left({\hat{\bm \theta}}^{d-1}\right) \right|\right) \\
\nonumber
= & \sum_{i\in{\mathcal N}} \sum_{j\in{\mathcal N}\setminus\left\{i\right\}} \xi_{ij}\left(2p_{\rm up}+c\right)\left(1+p_{\rm up}\right)  || {\bm \theta}_{ij} - {\hat{\bm \theta}}_{ij}^{d-1} ||_2  \\
& + \sum_{i\in{\mathcal N}} \sum_{j\in{\mathcal N}\setminus\left\{i\right\}} \xi_{ij} \left(\alpha_{\max}+2p_{\rm up}\beta_{\max} +c\beta_{\max}\right) \left| p_{ij}^*\left({\bm \theta}\right) - p_{ij}^*\left({\hat{\bm \theta}}^{d-1}\right) \right|.
\end{align}}
Recall that in (\ref{app:equ:resultboundpp}), we give an upper bound for $\left| p_{ij}^*\left({\bm \theta}\right) - p_{ij}^*\left({\hat{\bm \theta}}^{d-1}\right) \right|$. We complete our analysis in {\bf Step 3}.

{\bf Step 4:} We analyze the gap between $\Pi\left({\bm p}^*\left({\bm \theta}\right),{\bm w}^*\left({\bm \theta}\right),{\bm \epsilon}^d\right)$ and $\Pi\left({\bm p}^d,{\bm w}^d,{\bm \epsilon}^d\right)$ (i.e., the payoff under our policy) when $d$ is even and $d\ge6$.  

According to our policy, when $d$ is even, the provider implements $p_{ij}^*\left({\hat{\bm \theta}}^{d-2}\right)-\frac{\rho}{{\hat\beta}_{ij}^{d-2}}d^{-\eta}$ as the pricing decision and $w_{ij}^*\left({\hat{\bm \theta}}^{d-2}\right)+\rho d^{-\eta}$ as the supply decision for each link $\left(i,j\right)$. 
Based on our definition of $p_{\rm up}$ in Section \ref{app:sec:proof:the1}, we have $p_{\rm up}\ge\left|p_{ij}^*\left({\hat{\bm \theta}}^{d-2}\right)-\frac{\rho}{{\hat\beta}_{ij}^{d-2}}d^{-\eta}\right|$. The expressions of ${\mathbb E}_{{\bm \epsilon}^d}\left\{\Pi\left({\bm p}^*\left({\bm \theta}\right),{\bm w}^*\left({\bm \theta}\right),{\bm \epsilon}^d\right)\right\}$ and ${\mathbb E}_{{\bm \epsilon}^d}\left\{\Pi\left({\bm p}^d,{\bm w}^d,{\bm \epsilon}^d\right)\right\}$ are given in (\ref{app:equ:even:payoff1}) and (\ref{app:equ:even:payoff2}). 
We can use an approach that is similar to the one used in {\bf Step 3} to bound $|{\mathbb E}_{{\bm \epsilon}^d}\left\{\Pi\left({\bm p}^*\left({\bm \theta}\right),{\bm w}^*\left({\bm \theta}\right),{\bm \epsilon}^d\right)\right\}-{\mathbb E}_{{\bm \epsilon}^d}\left\{\Pi\left({\bm p}^d,{\bm w}^d,{\bm \epsilon}^d\right)\right\}|$, and we show the result in (\ref{app:equ:even:gapresult}).

{\bf Step 5:} In this step, we combine our results in the steps above, and bound $ {\mathbb E} \!\left\{ \frac{1}{D} \sum_{d=5}^D \!\Bigg( \Pi\left({\bm p}^*\left({\bm \theta}\right),{\bm w}^*\left({\bm \theta}\right),{\bm \epsilon}^d\right) \! -\! \Pi\left({\bm p}^d,{\bm w}^d,{\bm \epsilon}^d\right) \Bigg) \right\}$. Note that in {\bf Step 3} and {\bf Step 4}, we analyze the upper bound for ${\mathbb E}_{{\bm \epsilon}^d}\left\{\Pi\left({\bm p}^*\left({\bm \theta}\right),{\bm w}^*\left({\bm \theta}\right),{\bm \epsilon}^d\right)-\Pi\left({\bm p}^d,{\bm w}^d,{\bm \epsilon}^d\right)\right\}$, where the expectation is taken with respect to ${\bm \epsilon}^d$. Considering the randomness of ${\bm \epsilon}^1,{\bm \epsilon}^2,\ldots,{\bm \epsilon}^D$, we can utilize Theorem \ref{theorem:estimate} and the results in {\bf Step 2}, {\bf Step 3}, and {\bf Step 4} to derive (\ref{app:equ:longestgap}).

From (\ref{app:equ:longestgap}), we can see that we need to bound $\sum_{s=2}^{\lfloor\frac{D-1}{2}\rfloor}  {\left(2s+2\right)}^{-\eta}$ and $\sum_{s=2}^{\lfloor\frac{D-1}{2}\rfloor}  \sqrt{\frac{\ln \left(2s\right)}{\left(2s\right)^{1-2\eta}}}$. We first derive the upper bound of the term $\sum_{s=2}^{\lfloor\frac{D-1}{2}\rfloor}  {\left(2s+2\right)}^{-\eta}$ as follows:
\begin{align}
\nonumber
& \sum_{s=2}^{\lfloor\frac{D-1}{2}\rfloor}  {\left(2s+2\right)}^{-\eta} = \sum_{s=2}^{\lfloor\frac{D-1}{2}\rfloor} \int_{s}^{s+1} {\left(2s+2\right)}^{-\eta} dz \\
\nonumber
\le & \sum_{s=2}^{\lfloor\frac{D-1}{2}\rfloor} \int_{s}^{s+1} {\left(2z\right)}^{-\eta} dz = \int_{2}^{\lfloor\frac{D-1}{2}\rfloor+1} {\left(2z\right)}^{-\eta} dz \\
\nonumber
= & \frac{1}{2} \int_{4}^{2\lfloor\frac{D-1}{2}\rfloor+2} {\left({\tilde z}\right)}^{-\eta} d{\tilde z} < \frac{1}{2} \frac{1}{1-\eta} \left(2\left(\frac{D-1}{2}\right)+2\right)^{1-\eta}\\
= &  \frac{1}{2} \frac{1}{1-\eta} \left(D+1\right)^{1-\eta}.
\end{align}
It is easy to see that the following relation holds for any $D\ge5$ (recall that $0<\eta<\frac{1}{2}$):
\begin{align}
\left(\frac{D+1}{D}\right)^{1-\eta} \le \left(\frac{6}{5}\right)^{1-\eta} < \frac{6}{5}.
\end{align}
Therefore, we can bound $\sum_{s=2}^{\lfloor\frac{D-1}{2}\rfloor}  {\left(2s+2\right)}^{-\eta}$ as follows:
\begin{align}
\sum_{s=2}^{\lfloor\frac{D-1}{2}\rfloor}  {\left(2s+2\right)}^{-\eta} < \frac{3}{5} \frac{1}{1-\eta} D^{1-\eta}.\label{app:equ:sumbound1}
\end{align}
Next, we derive the upper bound of the term $\sum_{s=2}^{\lfloor\frac{D-1}{2}\rfloor}  \sqrt{\frac{\ln \left(2s\right)}{\left(2s\right)^{1-2\eta}}}$. We can easily prove that $\sqrt{\frac{\ln \left(z\right)}{\left(z\right)^{1-2\eta}}}$ increases with $z$ when $0< z\le e^{\frac{1}{1-2\eta}}$ and decreases with $z$ when $z> e^{\frac{1}{1-2\eta}}$. When $D>4+e^{\frac{1}{1-2\eta}}$, since $D$ is an integer, we can prove that $\lfloor\frac{D-1}{2}\rfloor\ge{\lfloor{\frac{1}{2}e^{\frac{1}{1-2\eta}}}\rfloor}+2$. Then, we can bound $\sum_{s=2}^{\lfloor\frac{D-1}{2}\rfloor}  \sqrt{\frac{\ln \left(2s\right)}{\left(2s\right)^{1-2\eta}}}$ as follows:
\begin{align}
\nonumber
& \sum_{s=2}^{\lfloor\frac{D-1}{2}\rfloor}  \sqrt{\frac{\ln \left(2s\right)}{\left(2s\right)^{1-2\eta}}} < \sum_{s=1}^{\lfloor\frac{D-1}{2}\rfloor}  \sqrt{\frac{\ln \left(2s\right)}{\left(2s\right)^{1-2\eta}}} \\
\nonumber
= & \sum_{s=1}^{\lfloor{\frac{1}{2}e^{\frac{1}{1-2\eta}}}\rfloor+1}  \sqrt{\frac{\ln \left(2s\right)}{\left(2s\right)^{1-2\eta}}} + \sum_{s={\lfloor{\frac{1}{2}e^{\frac{1}{1-2\eta}}}\rfloor}+2}^{\lfloor\frac{D-1}{2}\rfloor}  \sqrt{\frac{\ln \left(2s\right)}{\left(2s\right)^{1-2\eta}}} \\
\nonumber
= & \sum_{s=1}^{\lfloor{\frac{1}{2}e^{\frac{1}{1-2\eta}}}\rfloor+1}  \sqrt{\frac{\ln \left(2s\right)}{\left(2s\right)^{1-2\eta}}} + \sum_{s={\lfloor{\frac{1}{2}e^{\frac{1}{1-2\eta}}}\rfloor}+2}^{\lfloor\frac{D-1}{2}\rfloor} \int_s^{s+1} \sqrt{\frac{\ln \left(2s\right)}{\left(2s\right)^{1-2\eta}}} dz \\
\nonumber
\le & \sum_{s=1}^{\lfloor{\frac{1}{2}e^{\frac{1}{1-2\eta}}}\rfloor+1}  \sqrt{\frac{\ln \left(2s\right)}{\left(2s\right)^{1-2\eta}}} + \sum_{s={\lfloor{\frac{1}{2}e^{\frac{1}{1-2\eta}}}\rfloor}+2}^{\lfloor\frac{D-1}{2}\rfloor} \int_s^{s+1} \sqrt{\frac{\ln \left(2\left(z-1\right)\right)}{\left(2\left(z-1\right)\right)^{1-2\eta}}} dz \\
= & \sum_{s=1}^{\lfloor{\frac{1}{2}e^{\frac{1}{1-2\eta}}}\rfloor+1}  \sqrt{\frac{\ln \left(2s\right)}{\left(2s\right)^{1-2\eta}}} + \frac{1}{2}\int_{2{\lfloor{\frac{1}{2}e^{\frac{1}{1-2\eta}}}\rfloor}+2}^{2\lfloor\frac{D-1}{2}\rfloor} \sqrt{\frac{\ln \left({\tilde z}\right)}{\left({\tilde z}\right)^{1-2\eta}}} d{\tilde z}.
\end{align}
Since we have the following relation:
\begin{align}
\frac{d \left(\frac{1}{\left(\eta+0.5\right)}\sqrt{\ln z} z^{\eta+0.5}\right) }{d z} = \frac{1}{2\left(\eta+0.5\right)}\frac{z^{\eta-0.5}}{\sqrt{\ln z}} +  z^{\eta-0.5} \sqrt{\ln z},
\end{align}
we can bound $\frac{1}{2}\int_{2{\lfloor{\frac{1}{2}e^{\frac{1}{1-2\eta}}}\rfloor}+2}^{2\lfloor\frac{D-1}{2}\rfloor} \sqrt{\frac{\ln \left({\tilde z}\right)}{\left({\tilde z}\right)^{1-2\eta}}} d{\tilde z}$ as follows:
\begin{align}
\nonumber
& \frac{1}{2}\int_{2{\lfloor{\frac{1}{2}e^{\frac{1}{1-2\eta}}}\rfloor}+2}^{2\lfloor\frac{D-1}{2}\rfloor} \sqrt{\frac{\ln \left({\tilde z}\right)}{\left({\tilde z}\right)^{1-2\eta}}} d{\tilde z} \\
\nonumber 
< & \frac{1}{2} \int_{2{\lfloor{\frac{1}{2}e^{\frac{1}{1-2\eta}}}\rfloor}+2}^{2\lfloor\frac{D-1}{2}\rfloor} \sqrt{\frac{\ln \left({z}\right)}{\left({z}\right)^{1-2\eta}}} d{z} + \frac{1}{2} \int_{2{\lfloor{\frac{1}{2}e^{\frac{1}{1-2\eta}}}\rfloor}+2}^{2\lfloor\frac{D-1}{2}\rfloor} \frac{1}{2\left(\eta+0.5\right)}\frac{z^{\eta-0.5}}{\sqrt{\ln z}} dz \\
\nonumber
= & \frac{1}{2}  \int_{2{\lfloor{\frac{1}{2}e^{\frac{1}{1-2\eta}}}\rfloor}+2}^{2\lfloor\frac{D-1}{2}\rfloor} \left(\sqrt{\frac{\ln \left({z}\right)}{\left({z}\right)^{1-2\eta}}} + \frac{1}{2\left(\eta+0.5\right)}\frac{z^{\eta-0.5}}{\sqrt{\ln z}} \right) dz \\
\nonumber
< & \frac{1}{2} \left(\frac{1}{\left(\eta+0.5\right)}\sqrt{\ln z} z^{\eta+0.5}\right)|_{z=2\lfloor\frac{D-1}{2}\rfloor} \\
\nonumber
= & \frac{1}{2} \frac{1}{\left(\eta+0.5\right)}\sqrt{\ln \left(2\lfloor\frac{D-1}{2}\rfloor\right)} \left(2\lfloor\frac{D-1}{2}\rfloor\right)^{\eta+0.5}\\
\le & \frac{1}{\left(2\eta+1\right)}\sqrt{\ln D} \left(D\right)^{\eta+0.5}.
\end{align}
Therefore, we have derived an upper bound for $\sum_{s=2}^{\lfloor\frac{D-1}{2}\rfloor}  \sqrt{\frac{\ln \left(2s\right)}{\left(2s\right)^{1-2\eta}}}$ as follows:
\begin{align}
\sum_{s=2}^{\lfloor\frac{D-1}{2}\rfloor}  \sqrt{\frac{\ln \left(2s\right)}{\left(2s\right)^{1-2\eta}}} <  \sum_{s=1}^{\lfloor{\frac{1}{2}e^{\frac{1}{1-2\eta}}}\rfloor+1}  \sqrt{\frac{\ln \left(2s\right)}{\left(2s\right)^{1-2\eta}}} +  \frac{1}{2\eta+1}\sqrt{\ln D} \left(D\right)^{\eta+0.5}.\label{app:equ:sumbound2}
\end{align}

\begin{figure*}
{\scriptsize
\begin{align}
\nonumber 
& {\mathbb E} \left\{  \sum_{d=5}^D \Bigg( \Pi\left({\bm p}^*\left({\bm \theta}\right),{\bm w}^*\left({\bm \theta}\right),{\bm \epsilon}^d\right)  - \Pi\left({\bm p}^d,{\bm w}^d,{\bm \epsilon}^d\right) \Bigg) \right\} \\
\nonumber
< & \left( \sum_{i\in{\mathcal N}} \sum_{j\in{\mathcal N}\setminus\left\{i\right\}} \xi_{ij} \right)\rho\left(\frac{\alpha_{\max}}{\beta_{\min}}+2\frac{\beta_{\max}}{\beta_{\min}} p_{\rm up}+c\right) \frac{3}{5} \frac{1}{1-\eta} D^{1-\eta}   \\
\nonumber
& \!\!\!\!\!+\! \left( \sum_{i\in{\mathcal N}}\! \sum_{j\in{\mathcal N}\setminus\left\{i\right\}} \!\xi_{ij} \right) \left(\!2\left(2p_{\rm up}+c\right)\left(1+p_{\rm up}\right) \!+\!  2\left(\alpha_{\max}+2p_{\rm up}\beta_{\max} +c\beta_{\max}\right) 
\frac{\alpha_{\max}+\beta_{\max}}{2 \beta_{\min}^2} \!+\! 2\left(\alpha_{\max}+2p_{\rm up}\beta_{\max} +c\beta_{\max}\right) \frac{1}{\beta_{\min}} \left(\frac{1\!+\!c}{2} \!+\! \frac{\alpha_{\max}\!-\!{\underline \epsilon}}{ \beta_{\min}} N\!\left(N\!-\!1\right)\right)N\left(N-1\right) \right) \sqrt{\Phi_1\left(\rho,\eta\right)} \sum_{s=1}^{\lfloor{\frac{1}{2}e^{\frac{1}{1-2\eta}}}\rfloor+1}  \sqrt{\frac{\ln \left(2s\right)}{\left(2s\right)^{1-2\eta}}}  \\
& \!\!\!\!\!+\! \left( \sum_{i\in{\mathcal N}}\! \sum_{j\in{\mathcal N}\setminus\left\{i\right\}} \!\xi_{ij} \right) \left(\!2\left(2p_{\rm up}+c\right)\left(1+p_{\rm up}\right) \!+\!  2\left(\alpha_{\max}+2p_{\rm up}\beta_{\max} +c\beta_{\max}\right) 
\frac{\alpha_{\max}+\beta_{\max}}{2 \beta_{\min}^2} \!+\! 2\left(\alpha_{\max}+2p_{\rm up}\beta_{\max} +c\beta_{\max}\right) \frac{1}{\beta_{\min}} \left(\frac{1\!+\!c}{2} \!+\! \frac{\alpha_{\max}\!-\!{\underline \epsilon}}{ \beta_{\min}} N\!\left(N\!-\!1\right)\right)N\left(N-1\right) \right) \sqrt{\Phi_1\left(\rho,\eta\right)} \frac{1}{2\eta+1}\sqrt{\ln D} \left(D\right)^{\eta+0.5}.\label{app:equ:payofffrom5}
\end{align}}
\hrule
{\tiny
\begin{align}
\nonumber
\!\!\!\!\!\!\!\!\!\!\!\!\!\!\!\!\!\!\!\!\!\!\!\!\!\!\!\!\!\!\!\!\!\!\!\!\!\!\!\!\!\!\!\!\!\!\!\! & \Delta_D^{\bm \pi} =  {\mathbb E} \!\left\{ \frac{1}{D} \sum_{d=1}^D \!\Bigg( \Pi\left({\bm p}^*\left({\bm \theta}\right),{\bm w}^*\left({\bm \theta}\right),{\bm \epsilon}^d\right) \! -\! \Pi\left({\bm p}^d,{\bm w}^d,{\bm \epsilon}^d\right) \Bigg) \right\} \\
\nonumber
\!\!\!\!\!\!\!\!\!\!\!\!\!\!\!\!\!\!\!\!\!\!\!\!\!\!\!\!\!\!\!\!\!\!\!\!\!\!\!\!\!\!\!\!\!\!\!\!\!\!\!\!\!\!\!\!\!\!\!\!\!\!\!\!\!\!\!\!\!\!\!\!< \! & \left( \! 8 N^2 \xi_{\max}\left(p_{\rm up}\!+\!c\right)\!\left(\alpha_{\max}\!+\!\beta_{\max}p_{\rm up}\right) \!+\! \left( \sum_{i\in{\mathcal N}}\! \sum_{j\in{\mathcal N}\!\setminus\left\{i\right\}} \!\!\xi_{ij} \right) \left(\!2\left(2p_{\rm up}\!+\!c\right)\!\left(1\!+\!p_{\rm up}\right) \!+\! \left(\alpha_{\max}\!+\!2p_{\rm up}\beta_{\max} \!+\!c\beta_{\max}\right) 
\frac{\alpha_{\max}+\beta_{\max}}{\beta_{\min}^2} \!+\! 2\left(\alpha_{\max}\!+\!2p_{\rm up}\beta_{\max} \!+\!c\beta_{\max}\right) \!\frac{1}{\beta_{\min}} \!\left(\frac{1\!+\!c}{2} \!+\! \frac{\alpha_{\max}\!-\!{\underline \epsilon}}{ \beta_{\min}} N\!\left(N\!-\!1\right)\right)N\left(N\!-\!1\right) \!\right)\! \sqrt{\Phi_1\left(\rho,\eta\right)} \!\sum_{s=1}^{\lfloor{\frac{1}{2}e^{\frac{1}{1-2\eta}}}\rfloor+1}\!  \sqrt{\frac{\ln \left(2s\right)}{\left(2s\right)^{1-2\eta}}} \right) D^{-1} \\
\nonumber
\!\!\!\!\!\!\!\!\!\!\!\!\!\!\!\!\!\!\!\!\!\!\!\!\!\!\!\!\!\!\!\!\!\!\!\!\!\!\!\!\!\!\!\!\!\!\!\!\!\!\!\!\!\!\!\!\!\!\!\!\!\!\!\!\!\!\!\!\! & \!\!\!\!\!+\! \left( \sum_{i\in{\mathcal N}}\! \sum_{j\in{\mathcal N}\setminus\left\{i\right\}} \!\xi_{ij} \right) \left(\!2\left(2p_{\rm up}+c\right)\left(1+p_{\rm up}\right) \!+\! \left(\alpha_{\max}+2p_{\rm up}\beta_{\max} +c\beta_{\max}\right) 
\frac{\alpha_{\max}+\beta_{\max}}{\beta_{\min}^2} \!+\! 2\left(\alpha_{\max}+2p_{\rm up}\beta_{\max} +c\beta_{\max}\right) \frac{1}{\beta_{\min}} \left(\frac{1\!+\!c}{2} \!+\! \frac{\alpha_{\max}\!-\!{\underline \epsilon}}{ \beta_{\min}} N\!\left(N\!-\!1\right)\right)N\left(N-1\right) \right) \sqrt{\Phi_1\left(\rho,\eta\right)} \frac{1}{2\eta+1}\sqrt{\ln D} \left(D\right)^{\eta-0.5} \\
\!\!\!\!\!\!\!\!\!\!\!\!\!\!\!\!\!\!\!\!\!\!\!\!\!\!\!\!\!\!\!\!\!\!\!\!\!\!\!\!\!\!\!\!\!\!\!\!\!\!\!\!\!\!\!\!\!\!\!\!\!\!\!\!\!\!\!\!\! & \!\!\!\!\! +\!\left( \sum_{i\in{\mathcal N}} \sum_{j\in{\mathcal N}\setminus\left\{i\right\}} \xi_{ij} \right)\rho\left(\frac{\alpha_{\max}}{\beta_{\min}}+2\frac{\beta_{\max}}{\beta_{\min}} p_{\rm up}+c\right) \frac{3}{5} \frac{1}{1-\eta} D^{-\eta}. \label{app:equ:deltabound}
\end{align}
}
\hrule
\end{figure*}

Considering (\ref{app:equ:longestgap}), (\ref{app:equ:sumbound1}), and (\ref{app:equ:sumbound2}), we can characterize an upper bound for ${\mathbb E} \left\{  \sum_{d=5}^D \Bigg( \Pi\left({\bm p}^*\left({\bm \theta}\right),{\bm w}^*\left({\bm \theta}\right),{\bm \epsilon}^d\right)  - \Pi\left({\bm p}^d,{\bm w}^d,{\bm \epsilon}^d\right) \Bigg) \right\}$ in inequality (\ref{app:equ:payofffrom5}). We can easily prove that the value of the expression ${\mathbb E} \left\{  \sum_{d=1}^4 \Bigg( \Pi\left({\bm p}^*\left({\bm \theta}\right),{\bm w}^*\left({\bm \theta}\right),{\bm \epsilon}^d\right)  - \Pi\left({\bm p}^d,{\bm w}^d,{\bm \epsilon}^d\right) \Bigg) \right\}$ is upper-bounded by a term that is independent of $D$, as shown below:
\begin{align}
\nonumber
& {\mathbb E} \left\{  \sum_{d=1}^4 \Bigg( \Pi\left({\bm p}^*\left({\bm \theta}\right),{\bm w}^*\left({\bm \theta}\right),{\bm \epsilon}^d\right)  - \Pi\left({\bm p}^d,{\bm w}^d,{\bm \epsilon}^d\right) \Bigg) \right\} \\
\le & 8 N^2 \xi_{\max}\left(p_{\rm up}+c\right)\left(\alpha_{\max}+\beta_{\max}p_{\rm up}\right).
\end{align}
Hence, we can characterize an upper bound for $\Delta_D^{\bm \pi}$ in (\ref{app:equ:deltabound}).

In (\ref{app:equ:deltabound}), we can see that the upper bound consists of three terms, and they are proportional to $D^{-1}$, $\left(\ln D\right)^{\frac{1}{2}} \left(D\right)^{\eta-\frac{1}{2}}$, and $D^{-\eta}$, respectively. We can let $\Phi_2\left(\rho,\eta\right)$, $\Phi_3\left(\rho,\eta\right)$, and $\Phi_4\left(\rho,\eta\right)$ be the coefficients (i.e., the parts that are independent of $D$) of these three terms. Then, we can rewrite (\ref{app:equ:deltabound}) as
\begin{align}
\Delta_D^{\bm \pi} < \Phi_2\left(\rho,\eta\right) D^{-1} + \Phi_3\left(\rho,\eta\right) {\left(\ln D\right)}^{\frac{1}{2}} D^{\eta-\frac{1}{2}} + \Phi_4\left(\rho,\eta\right) D^{-\eta}. \label{app:equ:lastbound}
\end{align}
This completes our proof for Theorem \ref{theorem:key}.

\section{Proof of Corollary \ref{corollary:only}}\label{app:sec:corollary}
In this section, we prove that $\lim_{D\rightarrow \infty} \Delta_D^{\bm \pi}=0$. According to Theorem \ref{theorem:key}, the upper bound of $\Delta_D^{\bm \pi}$ in (\ref{equ:regretbound}) consists of the terms that are proportional to $D^{-1}$, ${\left(\ln D\right)}^{\frac{1}{2}} D^{\eta-\frac{1}{2}}$, and $D^{-\eta}$. It is easy to see that $\lim_{D\rightarrow \infty} D^{-1} =0$ and $\lim_{D\rightarrow \infty} D^{-\eta} =0$. Furthermore, since $\eta\in\left(0,\frac{1}{2}\right)$, we have the following relations:
\begin{align}
\nonumber 
\lim_{D\rightarrow \infty} \frac{{\left(\ln D\right)}^{\frac{1}{2}}}{D^{\frac{1}{2}-\eta}} & = \lim_{D\rightarrow \infty} \frac{\frac{1}{2 {\left(\ln D\right)}^{\frac{1}{2}}} \frac{1}{D} }{\left(\frac{1}{2}-\eta\right) D^{-\frac{1}{2}-\eta}} \\
& = \lim_{D\rightarrow \infty} \frac{1}{\left(1-2\eta\right) D^{\frac{1}{2}-\eta} {\left(\ln D\right)}^{\frac{1}{2}}} =0.
\end{align}
Therefore, as $D$ goes to infinity, all the terms of the upper bound of $\Delta_D^{\bm \pi}$ approaches zero. This implies that $\lim_{D\rightarrow \infty} \Delta_D^{\bm \pi}=0$.

\end{document}